# Counterfactual Analysis in Empirical Games

Brendan Kline* and Elie Tamer[&]

October 2024

ABSTRACT. We address counterfactual analysis in empirical models of games with partially identified parameters, and multiple equilibria and/or randomized strategies, by constructing and analyzing the counterfactual predictive distribution set (CPDS). This framework accommodates various outcomes of interest, including behavioral and welfare outcomes. It allows a variety of changes to the environment to generate the counterfactual, including modifications of the utility functions, the distribution of utility determinants, the number of decision makers, and the solution concept. We use a Bayesian approach to summarize statistical uncertainty. We establish conditions under which the population CPDS is sharp from the point of view of identification. We also establish conditions under which the posterior CPDS is consistent if the posterior distribution for the underlying model parameter is consistent. Consequently, our results can be employed to conduct counterfactual analysis after a preliminary step of identifying and estimating the underlying model parameter based on the existing literature. Our consistency results involve the development of a new general theory for Bayesian consistency of posterior distributions for mappings of sets. Although we primarily focus on a model of a strategic game, our approach is applicable to other structural models with similar features.

**Keywords:** predictive distributions, partial identification, incomplete models, inference on games, inference with strategic interactions, counterfactuals.

**JEL classification:** C11, C35, C57

* DEPT. OF ECONOMICS, UNIVERSITY OF TEXAS AT AUSTIN
[&] DEPT. OF ECONOMICS, HARVARD UNIVERSITY
*E-mail addresses*: brendan.kline@utexas.edu, elietamer@fas.harvard.edu.
We thank Hadi Khalaf and Hanbin Yang for excellent research assistance and seminar participants at NU and SMU for comments. Any errors are ours.

1. INTRODUCTION

This paper considers counterfactual analysis in structural models with incompleteness (e.g., multiple equilibria) and/or partial identification of underlying model parameters. We focus on the case of game theory models involving strategic interactions among multiple decision makers (DMs), but our approach applies as a special case to models involving a single DM, where for example the parameters of a discrete choice model may be partially identified. Our running example is an entry game.

Identification and estimation of the underlying model parameters of such models (e.g., the parameters of the utility functions of the DMs) has received considerable attention in the recent literature. Accordingly, we take as given an identification and estimation result for the underlying model parameters. However, there is a critical gap between recovering the parameters of the model and conducting a counterfactual analysis.

This paper proposes a method for counterfactual analysis in such models by constructing and analyzing an object we name a *counterfactual predictive distribution set* (CPDS). This approach corresponds to the general idea that structural econometrics is about recovering "fundamental" parameters and then using them in counterfactual analysis (e.g., Blundell (2017) and Low and Meghir (2017)).

The CPDS accommodates a broad range of counterfactual analyses, as we formalize in Section 2.2. Counterfactual analysis involves specification of changes to the environment, and specification of the outcome of interest. Our CPDS is flexible on both. The CPDS accommodates counterfactual analyses including: changes to the observables, changes to the utility functions, changes to the set of DMs (e.g., the addition or subtraction of potential market entrants), and changes to the solution concept. Also, our CPDS accommodates counterfactual analyses involving a variety of specific counterfactual outcomes of interest, including the behavioral outcome (e.g., firm entry decisions) and welfare outcomes (e.g., realized firm profits, or social welfare).

The CPDS simultaneously accounts for *four distinct unknowns.* One unknown concerns the possible incompleteness of the model (e.g., Tamer (2003)). This refers to the fact that there may not be a unique predicted outcome of interest in these models, for instance because of the existence of multiple equilibria. In such cases, even for a given value of the underlying model parameter, it is only possible to conclude that the outcome of interest will be a realization from some *set* of possible outcomes. Accordingly, the CPDS concerns predicting "sets" of counterfactual outcomes of interest.

Another unknown concerns unobservable determinants of outcomes, including unobservable determinants of utility (e.g., unobservable heterogeneity). With such unobservables, a counterfactual analysis can generally only make *probabilistic* statements about outcomes of interest. Accordingly, combined with the above issue of incompleteness, the CPDS concerns a *distribution* over *sets* of counterfactual outcomes of interest.

A third unknown concerns partial identification of the underlying model parameters. Often, the model parameters concern the utility functions of the DMs, and the literature (see Section 1.1) has found that partial identification of the utility function is common. This paper does not contribute to the literature on identification of the underlying model parameters. Indeed, the CPDS *builds on* the existence of an identified set for the underlying model parameters. (Our results allow the econometrician to choose any such result from the existing literature.) Accordingly, combined with the above issue of incompleteness and unobservables, CPDS concerns a *set* of *distributions* over *sets* of counterfactual outcomes of interest. The first "set" in this phrase is induced by the identified set for the underlying model parameter. The construction of the population CPDS stops here.

The final unknown concerns the statistical uncertainty over the identified set for the underlying model parameters. Thus, counterfactual analysis must account for the known challenges with inference for the identified set for a partially identified parameter (e.g., reviews in Canay and Shaikh (2017) and Canay, Illanes, and Velez (2023), in addition to many of the papers on partial identification cited elsewhere). Our setting introduces a new complication relative to this literature, because there are two "layers" of partial identification in our setting: each value of the (partially identified) underlying model parameter maps to a *set* of counterfactual outcomes (or, e.g., a *distribution* over *sets* of counterfactual outcomes when there is unobserved heterogeneity).[1] Accordingly, combined with the above issues, statistical

---

[1]Therefore, as far as we know, this setup is not covered by the recent literature about inference on projections or specific kinds of real-valued ("ordinary") functions of partially identified parameters. (The "specific kind" can be linear, or smooth, or something else, depending on the specific result in the literature.) See for example Bugni, Canay, and Shi (2017), Chen, Christensen, and Tamer (2018), Kaido, Molinari, and Stoye (2019), Cho and Russell (2023), and Cox and Shi (2023). Another major difference is that literature works in

inference for the CPDS concerns inference statements about a *set* of *distributions* over *sets* of counterfactual outcomes of interest. We take a Bayesian approach, and correspondingly take as given that the econometrician has access to a posterior distribution for the identified set for the underlying model parameters (e.g., the setups developed in Kline and Tamer (2016), Liao and Simoni (2019), and Giacomini and Kitagawa (2021), among other possible approaches to Bayesian inference for the identified set in partially identified models).[2] In that sense, our approach directly builds on existing results for identification and inference for the underlying model parameter. This additional step in the construction results in a posterior distribution for the CPDS, shortened to "posterior CPDS."

We prove that the CPDSes have two theoretical properties. First, the population CPDS is sharp from the point of view of identification: the population CPDS says exactly what is possible to say about the counterfactual analysis, given population data. Second, we give conditions for the posterior CPDS to be consistent. The consistency result addresses a key question in Bayesian inference, especially in non-standard settings: given that the posterior distribution by construction summarizes the statistical uncertainty, does the posterior distribution eventually "concentrate" around the truth? One practical implication of consistency is that it implies that the role of the prior is asymptotically zero, and thus any two Bayesians would agree asymptotically. Particularly with this view, results like Poirier (1998) or Moon and Schorfheide (2012) show that consistency of Bayesian procedures in partially identified models can be subtle, as those results show situations where the prior plays a role even asymptotically. A key baseline assumption is that the posterior distribution of the identified set for the underlying model parameters is consistent.

The consistency result involves proving three results that may be of independent interest. First, in Section 6.1, we prove equivalences between alternative definitions of consistency of posterior distributions for sets. This relates different parts of the existing literature on estimating the underlying model parameters, and is useful in proofs. Second, in Section 6.2,

---

the frequentist perspective, while we work in the Bayesian perspective. See Appendix C for some further discussion of a practical advantage of the Bayesian approach.

[2]There are also setups for Bayesian inference directly on the partially identified parameter, as in Moon and Schorfheide (2012). See Remark 2. Some results in this literature assume existence of a standard parametric likelihood for the data in terms of the underlying model parameter; our setup falls outside the scope of such analysis, due to incompleteness. Correspondingly, we do not make any assumptions that either explicitly or implicitly require that the posterior distribution for the underlying model parameter results from an analysis that assumes such a likelihood.

we consider the general statistical problem of estimating a mapping $f(\cdot)$ of an estimated set $A_0$. For CPDSes, $A_0$ is the identified set for the underlying model parameter; and $f(\cdot)$ concerns the *set* of *distributions* over *sets* of counterfactual outcomes of interest, as discussed above. Although subtle in the notation, note again the important point that $f(\cdot)$ is *not* restricted to be a real-valued ("ordinary") function of the underlying partially identified model parameter. (Also see Footnote 1.) We provide general conditions for the consistency of the posterior distribution for $f(A_0)$, given that the posterior distribution for $A_0$ is consistent. Our results are sufficiently general to be potentially applied to other situations that involve mapping an estimated set to some other (complicated) object of interest. Our discussions and proofs of these results, particularly in Appendix A, point to various topological subtleties. Our results require a suitable form of *continuity* of $f(\cdot)$, but not *differentiability*. Third, in Section 6.3, we prove conditions under which the mappings relevant for CPDSes are continuous, despite the fact that the set of equilibrium outcomes of a game generically exhibits discontinuities as a mapping from the utility functions of the DMs (which is relevant, since the underlying model parameters generally concern the utility functions).

Throughout the paper we use an entry game as a running example, and we conclude with an empirical application to market-level entry decisions of air carriers. Our analysis explores a range of counterfactual market outcomes, like the probability that a market is not served by any carrier, or the expected number of carriers that serve a market. Our analysis explores a range of changes to the environment that generates the counterfactual, like a change to the set of carriers, or a change in the profit function of the carriers. Our results point, in particular, to an interesting difference between low-cost carriers and other carriers. We also find that the computational burden of our approach to counterfactual analysis is low, an issue we discuss at various points in the paper. Specifically, the entire set of computations needed for our counterfactual analysis took less than one hour using a mid-level consumer-grade CPU and GPU.

In Section 2, we provide the basic setup for constructing CPDSes. In Section 3, we actually construct the CPDSes. In Section 4, we prove the consistency result. In Section 5, we discuss the computational implementation. In Section 6, we develop the general asymptotic theory discussed above. In Section 7, we provide our empirical application to market-level entry decisions of air carriers.

1.1. **Related literature.** The literature on identification of games has seen considerable recent interest. See for example reviews in Tamer (2010), de Paula (2013, 2020), Molchanov and Molinari (2014), Bontemps and Magnac (2017), Canay and Shaikh (2017), Ho and Rosen (2017), Aradillas-López (2020), Chesher and Rosen (2020), Molinari (2020), Kline, Pakes, and Tamer (2021), and Kline and Tamer (2023). Our paper turns to the question of identification and estimation of counterfactuals, rather than identification and estimation of the underlying model parameter, as has been the main focus of this existing literature. The question of identification of counterfactuals has seen increasing recent interest in a variety of different non-standard settings.

Many papers have focused on dynamic discrete choice models (e.g., Aguirregabiria (2010), Aguirregabiria and Suzuki (2014), Norets and Tang (2014), Arcidiacono and Miller (2020), and Kalouptsidi, Scott, and Souza-Rodrigues (2021)). Our incomplete model setup is a key point of departure from this literature, beyond just working on a different set of models.

Canen and Song (2020), Jun and Pinkse (2020), and Gu, Russell, and Stringham (2022) take different approaches to counterfactuals in related settings involving an incomplete model. The first takes a decomposition approach, and provides conditions under which re-weighting of the observed distribution of behavioral outcomes can be used to answer some counterfactual questions of interest. The second considers approaches for directly developing justified "point predictions" of the outcome of a two-player entry game under pure strategy Nash equilibrium, when some of the observables/unobservables are set to possibly counterfactual values. The third provides a method for discretizing the distribution of unobservables in a way that allows for directly partially identifying some counterfactual parameters of interest, with favorable computational implications. Thus, overall, the papers above collectively differ from ours in terms of the goals, setup, and results. In the next couple paragraphs, we describe some specific differences that may not be obvious from the summary of the papers.

Our method is intended to be used after the application of an existing identification result for the underlying model parameter. By contrast, any "self-contained" approach to identification of counterfactuals is tied to the specific set of assumptions used in those specific results, whereas our approach can accommodate any combination of assumptions for which there is an identification result for the underlying model parameter, so that empirical researchers do not need to "re-do" their identification analysis when conducting the counterfactual analysis.

Further, our setup accommodates the study of counterfactual outcomes other than just the behavioral outcome, including welfare outcomes. Another subtle but substantive difference is that other papers restrict attention to a certain class of counterfactuals, which seems to be less flexible than accommodated by our approach.[3] Then, in this setup, we provide a posterior distribution for the counterfactual predictions that accounts for the statistical uncertainty about the underlying model parameter.

Our setup does not fit into one particular framework sometimes used in the partial identification literature (e.g., Beresteanu, Molchanov, and Molinari (2011, Appendix D.2), Chesher and Rosen (2017), Chesher and Rosen (2020), Molinari (2020, Section 3.2.2), and Gu, Russell, and Stringham (2022)), whereby the observed outcome $Y$ is an element of a set of outcomes $\mathcal{Y}_\theta(X, \epsilon)$ for given $(X, \epsilon)$. This framework accommodates multiple *pure* strategy outcomes of the model. It is important that we accommodate *randomized* strategies (e.g., mixed strategies or correlated equilibrium), whereby the observed outcome is a realization from a distribution that is an element of a set of distributions. In the case of Nash equilibrium, perhaps the most fundamental reason to allow for mixing is that a pure strategy Nash equilibrium may not exist; and each counterfactual analysis provides "more" opportunities for this to happen, regardless of what happened in the data.[4] But even if a pure strategy

---

[3]These papers require the counterfactual involves only certain averages (or other linear functional) of the *behavioral outcome* (e.g., entry decisions), see e.g., Canen and Song (2020, Equation 2), Jun and Pinkse (2020, Section 2.2), and Gu, Russell, and Stringham (2022, Equation 2.9) and surrounding text for the specifics. Further, all focus attention on specific changes of the environment to generate the counterfactuals, restricted to involve some combination of changing the observed (or unobserved) explanatory variables, fixing the behavioral outcomes of some DMs, and/or changing the action spaces, depending on the specific paper. As we have already discussed, our approach allows for flexible specification of counterfactuals, including other outcomes (e.g., welfare outcomes) and other changes to generate the counterfactual (e.g., perhaps most notably, changing a DM's utility function, as in a subsidy/tax for taking certain actions).

[4]In particular, games of strategic substitutes (which overlaps with "entry games") are understood to often not have pure strategy Nash equilibria, except under special circumstances. See for example Vives (1999, page 43). Specifically with linear utility functions, Kline (2015, Supplement, Remark 2.2) observes that symmetric competitive effects suffices for existence of a pure strategy Nash equilibrium. On the other hand, one can fairly easily construct examples of entry games with asymmetric competitive effects among more than two potential entrants that have no pure strategy Nash equilibrium. One simple example (that can be generalized) is an odd number $M$ of DMs, where DM $i$ finds it profitable to enter the market exactly when DM $i+1$ does not enter the market (where DM $M+1$ is understood to be DM 1). By tracing through the necessary implications of DM 1 entering the market (or not entering the market) in a candidate PSNE, on the behavior of the other DMs, it can be easily seen that DM 1 entering (or not) the market would imply that DM 1 doesn't (or does) enter the market, so actually no PSNE can exist. This example is not requiring that DM $i$'s profit is only impacted by DM $i+1$. Thus, allowing randomized strategies is important to avoid making potentially undesirable assumptions on the nature of the competitive effects in an entry game.

Nash equilibrium does exist, it may be desirable to allow the use of randomized strategies (e.g., mixed strategy Nash equilibrium, or other solution concepts involving randomization). In particular, even if it is possible to show that the data could have been generated under the assumption of pure strategy Nash equilibrium, a counterfactual environment could plausibly involve the use of a randomized strategy especially if we use correlated equilibrium, which fundamentally involves randomized strategies. There are arguments directly in favor of randomized strategies in counterfactual analysis, as we discuss in Section 2.2.5 and Remark 1.

## 2. Counterfactual predictions

2.1. **Setup.** We consider an environment consisting of $K \geq 1$ decision makers (DMs). This could be a particular market based on geography or product, a particular social setting like a school or neighborhood, or something else. DMs are indexed by a subscripted $i$, and $K$ does not necessarily need to correspond to an observed number of DMs in the data.

Our approach to counterfactual analysis takes as given an identification result for the underlying fundamental model parameter $\theta$. The identified set for $\theta$ is $\Theta_{I,0}$. We allow our counterfactual analysis to depend on $\Theta_{I,0}$. This approach corresponds to the general idea that structural econometrics is about recovering "fundamental" parameters and then using them in counterfactual analysis (e.g., Blundell (2017) and Low and Meghir (2017)). In most applications of our approach, $\theta$ will concern the utility functions of the decision makers, and possibly other quantities (e.g., the distribution of unobservable determinants of utility). This is illustrated in Example 1, our running example of a binary (entry) game.

Our approach to counterfactual analysis avoids making strong statements about $\theta$ and $\Theta_{I,0}$. We can accommodate any combination of assumptions for which there is an identification result for $\theta$. Our results do not require the use of any particular identification result for $\theta$. We do not require $\theta$ to be finite-dimensional and allow $\theta$ to be point or partially identified.

The role of $\theta$ in the counterfactual analysis is that any link "to the data" in the counterfactual analysis must be via $\theta$ and $\Theta_{I,0}$. We allow the use of the parameter $\theta$ in various steps of the specification of the counterfactual analysis. For example, we allow the utility functions in the counterfactual analysis to have some specified relationship to the utility functions in the observed data (including being equal), by specifying that the utility functions in the counterfactual analysis depends on $\theta$ in such a way to achieve that relationship. We illustrate that in full detail in the following sections.

**Example 1** (Application to Binary Games). As a running example, consider the application to a structural model of a discrete game with two players ($M = 2$), two actions (so the action space for player $i$ is $\mathcal{A}_i = \{0, 1\}$), and complete information. We use this simple example for clarity; our approach to counterfactual analysis can handle more complicated setups, including setups with more than two players and more than two actions.

In normal form, with a parametric (linear) specification for the utility functions with the simplifying assumption of equality of parameters across players, the payoffs are as in Table 1.

|  | $a_2 = 0$ | $a_2 = 1$ |
|---|---|---|
| $a_1 = 0$ | $0, 0$ | $0, x_2\beta + \epsilon_2$ |
| $a_1 = 1$ | $x_1\beta + \epsilon_1, 0$ | $x_1\beta + \Delta + \epsilon_1, x_2\beta + \Delta + \epsilon_2$ |

TABLE 1. Parametric utility functions in a two player, two action game.

In this example, we assume this setup describes the environment in the observed data. The utility functions appearing in Table 1 can be written for player $i$,

$$u_i(a_1, a_2, x_i, \epsilon_i, \theta) = a_i[x_i\beta + \Delta a_{-i} + \epsilon_i], \qquad (1)$$

where $\theta$ includes $(\beta, \Delta)$ and perhaps other quantities, discussed below. As usual, $(\epsilon_1, \epsilon_2)$ is observed by the players but not the econometrician. The players know the utility functions exactly. The parameters $\beta$ and $\Delta$ are unknown to the econometrician and must be estimated from observed data. Further, the distribution of $\epsilon$ may be unknown. In such cases, the distribution of $\epsilon = (\epsilon_1, \epsilon_2)$ is often specified to be part of a parametric family of distributions, with generic parameterization $\theta_\epsilon$. For instance, $\theta_\epsilon$ could be the covariance of a joint normal distribution for $\epsilon$. Generally, some or all components of $\theta = (\beta, \Delta, \theta_\epsilon)$ are partially identified.

Our approach to counterfactual analysis takes as given an identification result for $\theta$, summarized by the identified set $\Theta_{I,0}$.

Our approach also allows for a non-parametric specification for the utility functions.

|         | $a_2 = 0$            | $a_2 = 1$                                  |
|---------|----------------------|--------------------------------------------|
| $a_1 = 0$ | $0, 0$             | $0, u_2(0, 1, x_2, \epsilon_2)$            |
| $a_1 = 1$ | $u_1(1, 0, x_1, \epsilon_1), 0$ | $u_1(1, 1, x_1, \epsilon_1), u_2(1, 1, x_2, \epsilon_2)$ |

TABLE 2. Non-parametric utility functions in a two player, two action game.

In a non-parametric specification, $\theta$ would include the utility *functions* $u_i(\cdot)$ displayed in Table 2. In that case, we would assume an identification result for those utility functions, again just resulting in an identified set $\Theta_{I,0}$. □

## 2.2. Specification of the counterfactual analysis.

**Definition 1** (Counterfactual analysis). A *counterfactual analysis* consists of the following elements, chosen by the econometrician and possibly depending on $\theta$:

(1) **Number of DMs.** There is a certain number $K$ of DMs.
(2) **Action Space.** Each DM chooses an action from a finite set $\tilde{\mathcal{A}}_i$.
(3) **Utility Function.** The utility function for DM $i$ is a function $\tilde{u}_i(a_i, a_{-i}, x_i, \epsilon_i, \theta)$ where $a_i \in \tilde{\mathcal{A}}_i$ is the action taken by DM $i$ and $a_{-i} \in \prod_{j \neq i} \tilde{\mathcal{A}}_j$ are the actions taken by the other DMs. $\tilde{u}(\cdot) = (\tilde{u}_1(\cdot), \tilde{u}_2(\cdot), \ldots) \in \mathcal{U}$, where $\mathcal{U}$ is the space of utility functions. The utility function may depend on $\theta$.
(4) **Distribution of Observables and Unobservables.** The distribution of $(x, \epsilon)$ is $\tilde{F}_\theta(x, \epsilon) = \tilde{F}_\theta(\epsilon|x)\tilde{F}_\theta(x)$. The distribution may depend on $\theta$.
(5) **Solution Concept.** The solution concept is a *correspondence* $\mathcal{S}(u) : \mathcal{U} \to 2^{\Delta(\prod \tilde{\mathcal{A}}_i)}$ where $2^{\Delta(\prod \tilde{\mathcal{A}}_i)}$ is the *set of distributions* over actions. An element $s \in \mathcal{S}(u)$ for given $u$ is called a solution for that $u$.
(6) **Outcome of Interest.** The outcome of interest is $t(u(\cdot), s)$, a function of the utility functions $u(\cdot)$ and a corresponding solution $s$.

**Comments on the specification of a counterfactual analysis**

2.2.1. *Number of DMs.* The number of DMs does not need to be the same as observed in the data. For instance, in an application to an entry game, the counterfactual can involve more or fewer potential entrants as compared to in the data. Generally, this will involve a corresponding change in the utility functions of the DMs, as discussed in Section 2.2.3.

2.2.2. *Action space.* Each DM $i$ chooses an action from a finite set $\tilde{\mathcal{A}}_i$. The action space $\tilde{\mathcal{A}}_i$ does not need to be the same as the action space in the observed data. For example, $\tilde{\mathcal{A}}_i$ might eliminate certain actions that were available in the observed data, corresponding to a counterfactual analysis that prohibits those actions. If $\tilde{\mathcal{A}}_i$ is a singleton, then the counterfactual involves specifying the action of DM $i$; this is closely related to the empirical analysis of best response functions, as in Kline and Tamer (2012).

We do not restrict the dimension of $\tilde{\mathcal{A}}_i$. $\tilde{\mathcal{A}}_i$ can have dimension 1 as in standard applications to the "econometrics of games" literature where the action represents, for instance, market entry decisions. $\tilde{\mathcal{A}}_i$ can have dimension greater than 1, for instance in applications to network formation models, where the actions are decisions to "link" to other DMs.

2.2.3. *Utility functions.* Generically, a counterfactual specification of the utility functions can be written as $\tilde{u}_i(a_i, a_{-i}, x_i, \epsilon_i, \theta)$. With further discussion in Section 2.2.4, $x_i$ are observed determinants of utility and $\epsilon_i$ are unobserved determinants of utility. Also, $a_i \in \tilde{\mathcal{A}}_i$ is the action taken by DM $i$ and $a_{-i} \in \prod_{j \neq i} \tilde{\mathcal{A}}_j$ are the actions taken by the other DMs. The dependence on $a_{-i}$ is the source of the interaction among the DMs. $x_i$ may include components of (or may be identical to) $x_j$, reflecting that observable determinants that influence DM $i$ also influence DM $j$. The dimension of $\epsilon_i$ is unrestricted, and $\epsilon_i$ may be correlated with $\epsilon_j$. As standard, we assume that all of $(x_i, \epsilon_i)$ is finite-dimensional. The space of utility functions $u(\cdot) = (u_1(\cdot), u_2(\cdot), \ldots)$ is denoted by $\mathcal{U}$.

A change in $x_i$ is accommodated in Section 2.2.4. A change in the utility function is accommodated in this section. The counterfactual utility functions can depend on $\theta$, which allows for many kinds of counterfactual analyses:

(1) The counterfactual utility *function* can be specified to be exactly equal to the "real world" utility *function* from the observed data. For instance, this arises when the counterfactual analysis concerns changing the value of an observable, but not the utility functions directly. This requires that $\theta$ contains sufficient information to construct the utility functions $u_i(a_i, a_{-i}, x_i, \epsilon_i)$ from the observed data. Then, $\tilde{u}_i(a_i, a_{-i}, x_i, \epsilon_i, \theta) \equiv$ $u_i(a_i, a_{-i}, x_i, \epsilon_i)$, interpreted to follow from the condition that $\theta$ includes the utility function $u_i(a_i, a_{-i}, x_i, \epsilon_i)$ from the observed data.

With a parametric/semi-parametric specification of the utility functions, this means that $\theta$ must contain the parameters of the utility functions. In the running example from Example 1, $\theta$ would need to contain $(\beta, \Delta)$, from which the utility function can be reconstructed via Equation 1. With a non-parametric specification, $\theta$ would need to contain the functions $u_i(a_i, a_{-i}, x_i, \epsilon_i)$ displayed in Table 2.

(2) Alternatively, the counterfactual utility functions can involve some change of the utility functions from the observed data. For instance, this could involve subsidizing/taxing certain actions, with $\tilde{u}_i(a_i, a_{-i}, x_i, \epsilon_i, \theta) \equiv u_i(a_i, a_{-i}, x_i, \epsilon_i) + S(a_i)$, where $S(a_i) > 0$ for any action $a_i$ that is subsidized and $S(a_i) < 0$ for any action $a_i$ that is taxed. In an application to market entry, this could be a subsidy/tax for entry.

(3) If the counterfactual analysis involves a change of the number of DMs, this would be reflected in both the functional dependence of the utility functions on $a_{-i}$ and the overall dimension of the space of utility functions. For instance, a slight generalization of the utility functions in Equation 1 is $u_i(a_1, a_2, \ldots, a_K, x_i, \epsilon_i, \theta) = a_i[x_i\beta + \Delta \sum_{j \neq i}^{K} a_j + \epsilon_i]$ for a certain number $K$ of DMs. In this case, there would be an "obvious" way to "extrapolate" to another number $K'$ of DMs by setting $u_i(a_1, a_2, \ldots, a_{K'}, x_i, \epsilon_i, \theta) = a_i[x_i\beta + \Delta \sum_{j \neq i}^{K'} a_j + \epsilon_i]$.

As just illustrated, the main requirement is that the counterfactual utility functions can be written in terms of $\theta$. We also require a regularity condition:

**Assumption 1.** *For each $i$, and each value of $(a_i, a_{-i}, \theta)$, $\tilde{u}_i(a_i, a_{-i}, x_i, \epsilon_i, \theta)$ is a measurable function of $(x_i, \epsilon_i)$.*

2.2.4. *Distribution of observables and unobservables.* The counterfactual analysis involves a possibly counterfactual specification of the observable and unobservable components of the utility functions. This consists of a specification $\tilde{F}_\theta(x, \epsilon)$ for the joint distribution of $(x, \epsilon)$, possibly depending on the parameter $\theta$ with $\tilde{F}_\theta(x, \epsilon) = \tilde{F}_\theta(\epsilon|x)\tilde{F}_\theta(x)$. In many cases, the distribution of $x$ is unrelated to $\theta$ and/or the distribution of $\epsilon|x$ is unrelated to $\theta$. This allows for many kinds of counterfactual analyses:

(1) In some cases, $\tilde{F}_\theta(x)$ does not depend on $\theta$. In particular, $\tilde{F}_\theta(x)$ can be a point mass at a particular value $\tilde{x}$, so that the counterfactual analysis concerns setting the observable components of utility to $\tilde{x}$. Alternatively, $\tilde{F}_\theta(x)$ can be non-degenerate, so that the counterfactual analysis concerns a distribution of observables. In both cases, this allows for *counterfactual* values of $x$ that are not directly observed in the data.

(2) In other cases, $\tilde{F}_\theta(x)$ depends on $\theta$, particularly when $\theta$ includes information about the distribution of the observed data. It could be that $\tilde{F}_\theta(x)$ is some transformation of the distribution of $x$ in the observed data. This could be a policy evaluation that involves increasing/decreasing the value of the observables relative to what is observed in the data. For example, it could be that $\tilde{F}_\theta(x) = G(x - c)$ where $G(\cdot)$ is the distribution of $x$ in the observed data, and $c$ has the same dimension as $x$ and reflects the addition/subtraction of values to each component of $x$.

In such cases, $G(\cdot)$ must be an element of $\theta$, in the sense that some elements of $\theta$ correspond to a (possibly fully non-parametric) model for the distribution of $x$, in which case our CPDS will account for the estimation[5] of the distribution of $x$. In some cases, the econometrician may want to abstract away from the uncertainty in the estimation of the distribution of $x$, e.g., if the sample in the observed data is equal to the population of interest. In those cases, the population distribution of $x$, $G(x)$, can be taken as a known quantity, equal to the sample distribution.

(3) Similar remarks apply to $\tilde{F}_\theta(\epsilon|x)$. $\tilde{F}_\theta(\epsilon|x)$ can be a point mass at a particular value $\tilde{\epsilon}$, so that the counterfactual analysis concerns setting the unobservable components of utility to $\tilde{\epsilon}$. Alternatively, $\tilde{F}_\theta(\epsilon|x)$ can be a non-degenerate distribution of unobservables. This could be assumed known by the econometrician. In that case, $\tilde{F}_\theta(\epsilon|x)$ would not depend on $\theta$, and the CPDS would account for the distribution of outcomes induced by the randomness in $\epsilon$, without uncertainty about what that distribution is.

Alternatively, $\tilde{F}_\theta(\epsilon|x)$ could be an estimated distribution of unobservables. In that case, components of $\theta$ correspond to the distribution of the unobservables. In that case, the CPDS would need to further account for the uncertainty about the distribution of the unobservables. Further, $\tilde{F}_\theta(\epsilon|x)$ could be some transformation of the estimated distribution of unobservables, parallel to the treatment of $x$ discussed above.

---

[5]It is straightforward to estimate the distribution of $x$ including from a non-parametric approach using Bayesian methods, reviewed for instance in Müller and Quintana (2004), Hjort, Holmes, Müller, and Walker (2010), Müller, Quintana, Jara, and Hanson (2015), and Ghosal and Van der Vaart (2017).

(4) We allow for the possibility that $\tilde{F}_\theta(\epsilon|x)$ depends on $x$, while recognizing that identification of the parameter $\theta$ corresponding to such a setup may be difficult. Thus, in most applications, $\epsilon \perp x$ would be assumed.

As notation, let $\tilde{\tilde{u}}_i(a_i, a_{-i}) \equiv \tilde{u}_i(a_i, a_{-i}, x_i, \epsilon_i, \theta)$ be the utility function that actually is relevant for the DMs, only depending on factors under the control of the DMs. The counterfactual analysis involves a *distribution* $\tilde{F}_\theta(\tilde{\tilde{u}})$ over $\tilde{\tilde{u}}$ that is induced by the counterfactual distribution $\tilde{F}_\theta(x, \epsilon)$. Note that $\tilde{\tilde{u}}_i$ can be represented as a vector in $\mathbb{R}^{\prod |\tilde{\mathcal{A}}_i|}$.

2.2.5. *Solution concept.* The solution concept provides the connection between the utility functions and the actions taken by the DMs. The solution concept used in the counterfactual analysis can be different from the solution concept that is used to identify $\theta$.

A solution concept[6] is a set-valued correspondence $\mathcal{S}(u) : \mathcal{U} \to 2^{\Delta(\prod \tilde{\mathcal{A}}_i)}$. A solution concept provides a *set* of *distributions* of actions for any given specification of the utility functions. Sometimes, such a distribution $s \in \Delta(\prod \tilde{\mathcal{A}}_i)$ is known as a *solution* or a *strategy*. We allow that the solution concept maps into (sets of) *distributions* to allow randomized strategies, including mixed strategies and correlated strategies. A pure strategy is a special case of a mixed strategy. A correlated strategy is a *joint* distribution over the entire profile of actions, as in correlated equilibrium. Also, the solution concept maps to a *set* in order to accommodate multiple equilibria, or multiple solutions. If randomized strategies or multiple solutions are present, the solution concept does not uniquely predict the actions taken by the DMs. Such non-unique predictions is a feature of models with multiple DMs, and complicates the counterfactual prediction problem.

$\mathcal{S}(\cdot)$ could be the Nash equilibrium solution concept allowing for mixed strategies, or the Nash equilibrium solution concept restricted to pure strategies, or the rationalizability solution concept, or the correlated equilibrium solution concept, and so forth.

Because each $\tilde{\mathcal{A}}_i$ is finite, $s \in \Delta(\prod \tilde{\mathcal{A}}_i)$ is a probability vector over $\prod \tilde{\mathcal{A}}_i$. By construction, each element of $s$ is the probability of a particular action profile actually arising in the game according to the solution $s$. For instance, in a game involving two DMs each with a binary action space, the possible action profiles are $(0,0)$, $(0,1)$, $(1,0)$, and $(1,1)$, where the first element is the action of the first DM and the second element is the action of the second DM. A corresponding solution $s$ would then be of the form $(p_{00}, p_{01}, p_{10}, p_{11})$ where each element gives the probability of the corresponding action profile.

---

[6]The solution concept mapping depends on various other parts of the specification of the counterfactual, like the action space and the number of DMs. We suppress this dependence in the notation, since those quantities are fixed for any given counterfactual analysis.

Our approach is not tied to the use of any particular solution concept. In particular, our approach *does not* require that the solution concept used in the identification of $\theta$ is the same as the solution concept used in the counterfactual analysis. For example, an econometrician might assume that the observed data is generated by Nash equilibrium, based on arguments that Nash equilibria are the outcomes of a long-run dynamic process as in Kalai and Lehrer (1993). The counterfactual analysis could be based on other solution concepts like rationalizability, correlated equilibrium, or some learning algorithm based on arguments that Nash equilibrium may not (initially) be reasonable in a novel policy environment. It might be more appealing to conduct the counterfactual analysis based on rationalizability, because it exhausts the content of "rationality" as in Bernheim (1984) and Pearce (1984). Or it might be more appealing to conduct the counterfactual analysis based on correlated equilibrium as in Aumann (1987), because it results from "Bayesian rationality" with a common prior. We return specifically to the advantages of correlated equilibrium in Remark 1.

Our results rely on a modest set of assumptions about the solution concept, thereby accommodating a wide range of solution concepts, including those that have not necessarily seen much theoretical investigation. The solution concept must satisfy the following assumption.

**Assumption 2.** *The solution concept $\mathcal{S}(\cdot)$ is non-empty, in the sense that $\mathcal{S}(u) \neq \emptyset$ for all $u$. Further, $\mathcal{S}(\cdot)$ is closed-valued, in the sense that $\mathcal{S}(u)$ is a closed set for all $u$. Further, $\mathcal{S}(\cdot)$ is upper hemi-continuous,[7] in the sense that $\{u : \mathcal{S}(u) \cap K \neq \emptyset\}$ is a closed set for every compact set $K$.*

---

[7]The definition of upper hemi-continuous follows usage as in Aliprantis and Border (2006), particularly the equivalence in Aliprantis and Border (2006, Lemma 17.4). Since the elements of $\mathcal{S}(u)$ are distributions over the finite set $\prod \tilde{\mathcal{A}}_i$ and therefore necessarily bounded, compactness is equivalent to closed-ness. Some sources like Rockafellar and Wets (2009, Theorem 5.7 and Chapter 5 generally) use the term "outer semi-continuous" for what we call "upper hemi-continuous." Further, in this setting, the upper hemi-continuity condition is equivalent to the condition that $\mathcal{S}(\cdot)$ has closed graph, sometimes known as the "Closed Graph Theorem" as in Aliprantis and Border (2006, Theorem 17.11) or Rockafellar and Wets (2009, Theorem 5.7). The closed graph condition is the condition that $\{(u, s) : s \in \mathcal{S}(u)\}$ is a closed set.

The condition that $\mathcal{S}(\cdot)$ is non-empty requires that at least one solution exists for each specification of the utility function. This rules out, for instance, the use of pure strategy Nash equilibrium if the game is such that a pure strategy Nash equilibrium does not exist for some $u$. Conversely, many solution concepts have solutions that exist for all $u$, as in mixed strategy Nash equilibrium (e.g., Fudenberg and Tirole (1991, Section 1.3.1)) or generalizations thereof, like correlated equilibrium or rationalizability.

The characterization of many solution concepts in terms of inequality conditions makes it clear they satisfy the closed-value condition. For example, this holds for Nash equilibrium and correlated equilibrium (e.g., Aumann (1987)). More generally, any solution concept such that $\mathcal{S}(\cdot)$ is finite-valued would satisfy the closed-value condition.

The sequential characterization of upper hemi-continuity makes it apparent that it is a common property of solution concepts.[8] See for example Fudenberg and Tirole (1991, Section 1.3.2) for a detailed discussion for Nash equilibrium. Similarly, correlated equilibrium clearly satisfies this condition (i.e., Aumann (1987, Proposition 2.3) or Hart and Schmeidler (1989)).

Further, the solution concept must satisfy a further continuity condition. This continuity condition requires special attention, and hence appears by itself.

**Assumption 3.** *There is a Borel set $\mathcal{U}^e$ such that: if $u \notin \mathcal{U}^e$, then $\mathcal{S}(\cdot)$ is lower hemi-continuous at $u$.*

It is not necessary to have an explicit representation for $\mathcal{U}^e$. The set $\mathcal{U}^e$ will depend on the choice of the solution concept. Since continuity is equivalent to the conjunction of upper hemi-continuity and lower hemi-continuity, when added to Assumption 2, Assumption 3 requires that $\mathcal{S}(\cdot)$ is continuous everywhere but for the *exceptional* set $\mathcal{U}^e$.

As formalized below, our consistency result requires a further condition: that $\mathcal{U}^e$ can be taken to be a "small" set, as foreshadowed by our use of the terminology "exceptional" set.[9] Essentially, the required condition is that $\mathcal{U}^e$ has probability zero under a relevant distribution of the utility functions, as further discussed in our consistency results.

---

[8]By Border (1985, Theorem 11.11), a sufficient condition for the upper hemi-continuity part of Assumption 2 is the condition: for any sequence of utility functions $u^{(\nu)} \to u$ with corresponding sequence $s^{(\nu)} \in \mathcal{S}(u^{(\nu)})$, it holds that there is a convergent subsequence of $s^{(\nu)}$ with limit point $s \in \mathcal{S}(u)$. Many solution concepts, like Nash equilibrium or correlated equilibrium, satisfy this condition because they concern inequality constraints on the utility functions and strategies that are respected by these limits.

[9]Assumption 3 can always be satisfied by taking $\mathcal{U}^e$ to be the entire space of utility functions.

Relevant results have been established in the economic theory literature. For instance, when the solution concept is Nash equilibrium, and for certain other solution concepts, Kohlberg and Mertens (1986), Schanuel, Simon, and Zame (1991), and Blume and Zame (1994) show that $\mathcal{U}^e$ can be taken to be a closed semi-algebraic set of lower dimension (compared to the entire space of utility functions). By similar arguments, the same result applies to correlated equilibrium and rationalizability (for both claims see Germano (2006)).

Critically, when $\mathcal{U}^e$ has those properties, it is indeed "small:" it is closed and has empty interior (e.g., Schanuel, Simon, and Zame (1991)) and therefore is a nowhere dense set, and therefore is a meagre set. Further, it is of Lebesgue measure 0. As discussed further below, these sorts of results can provide the "small"-ness needed for our consistency result.

**Example 1** (continuing from p. 9). Assumption 3 generalizes the approach from the econometrics of games where the space of utility functions are partitioned according to the corresponding set of Nash equilibria. For instance, in the Bresnahan and Reiss (1990, Figure 1)- Bresnahan and Reiss (1991, Figures 1-3) entry game with two players, it is common to partition the space of unobservables into five different "regions" according to the corresponding set of Nash equilibria. The set of Nash equilibria are obviously continuous within those regions (as a function of the unobservable), and so a discontinuity could arise only at the boundaries between regions: hence, $\mathcal{U}^e$ concerns the boundaries between the regions. And indeed, the set of Nash equilibria fails to be lower hemi-continuous at the boundary.[10] However, consistent with the interpretation of Assumption 3, the boundary between the regions is "small." Outside of particular examples like this, it might be difficult to explicitly characterize $\mathcal{U}^e$, and such an explicit characterization is not needed for Assumption 3.

In general, Assumption 3 is *not* true when $\mathcal{U}^e = \emptyset$, so the analysis must account for such discontinuities at the exceptional set. In other words, the analysis would be simpler if we could assume that $\mathcal{U}^e = \emptyset$, but we must allow the more complicated possibility that $\mathcal{U}^e \neq \emptyset$.[11]

---

[10]Consider the boundary between any two regions $A$ and $B$. For any two regions, there is a pure strategy Nash equilibrium $s$ in region $A$ that is not a pure strategy Nash equilibrium in region $B$. That strategy $s$ will also be a pure strategy Nash equilibrium along the boundary by upper hemi-continuity. However, for any utility function along a sequence of utility functions from region $B$ that approach the boundary, the strategy $s$ will not be a Nash equilibrium. And the Nash equilibrium from region $B$ will not be close to strategy $s$. This implies the Nash equilibrium solution concept is not lower hemi-continuous along this boundary.

[11]Intuitively, lower hemi-continuity fails for a specific utility function $u$ if there is a solution $s \in \mathcal{S}(u)$ and a sequence of utility functions that approach $u$, but yet $s$ cannot be approached as a limit of solutions

Because the convex hull of $\mathcal{S}$ becomes relevant in future steps of the counterfactual analysis, we report here a result on the properties of the convex hull of $\mathcal{S}$.

**Lemma 1.** *If $\mathcal{S}$ satisfies the conditions of Assumption 2, then co $\mathcal{S}$ satisfies the conditions of Assumption 2. If Assumption 3 holds, then the statement in Assumption 3 also holds with $\mathcal{S}(\cdot)$ replaced with co $\mathcal{S}(\cdot)$, i.e., with the same $\mathcal{U}^e$.*

2.2.6. *Outcome of interest.* The final part of the specification of a counterfactual analysis is the selection of an outcome of interest. We assume the outcome of interest can be written as a known continuous function $t(u(\cdot), s)$ of the entire vector of utility functions $u(\cdot)$ and a solution $s$, taking values in some finite-dimensional Euclidean space.[12]

This allows for many kinds of counterfactual analyses:

(1) Many outcomes of interest depend only on $s$. For example, $t(u, s) \equiv s$ means that the outcome of interest is the solution itself. Recall that $s$ is a distribution over the actions. In a model of market entry, this means that the counterfactual analysis concerns the distribution over market entry decisions. More generally, $t(u, s) \equiv t(s)$ is allowed. For example, summing components of $s$ that satisfy some specified condition, $t(s)$ can be the probability of an outcome with a specified property. For instance, in a model of market entry, $t(s)$ could sum components of $s$ with at least a certain number of entrants, or sum components of $s$ with a specified DM entering the market. For another example, if the counterfactual concerns a subsidy/tax, $t(s)$ could measure the fiscal cost of that subsidy/tax. Or, $t(s)$ could be a policy maker's "social welfare function" over the solution.

(2) Other outcomes of interest depend on the utility functions of the DMs. Specifically, many "welfare" measures concern the utility of the DMs. For example, $t(u, s)$ could be specified to be the linear combination of $u$ and $s$ giving the expected utility under

---

associated with the sequence of utility functions. In other words, lower hemi-continuity fails at $u$ if there are "discontinuously" more solutions at $u$ compared to some utility functions nearby $u$. Fudenberg and Tirole (1991, Section 1.3.2) shows simple examples where lower hemi-continuity fails for Nash equilibrium for some utility functions. As mentioned above, lower hemi-continuity also fails at the boundary between the regions of equilibria in the entry game.

[12]Many of our results would immediately generalize to cases where the outcome of interest is not in a Euclidean space, but since we are not aware of any such examples, we focus on a Euclidean outcome of interest.

the solution $s$, either for a particular DM or as a vector for all DMs. Alternatively, $t(u, s)$ could be specified to be the summed utility across the DMs. If utility is firm profits, then this is profits of the firms.

The dimension of the outcome of interest is unrestricted, so the counterfactual analysis can be specified to concern multiple of the above outcomes of interest. For instance, the outcome of interest could be both the fiscal cost of a subsidy/tax and the profits of the firms.

3. SPECIFICATION OF THE COUNTERFACTUAL PREDICTIVE DISTRIBUTION SET

We construct the counterfactual predictive distribution set in a sequence of steps. The first step considers a specific value of $\theta$, and derives a corresponding counterfactual predictive distribution set. This results in a distribution over sets of counterfactual outcomes, accounting both for the possibility of model incompleteness and unknown determinants of outcomes. The second step accounts for the possibility that $\theta$ is partially identified. This results in a set of distributions over sets of counterfactual outcomes. The resulting object is the "population CPDS." The last step accounts for estimation of the identified set for $\theta$. The resulting object is the "posterior CPDS."

3.1. **Step 1: Partial CPDS for a given value of $\theta$.** Consider the "partial" counterfactual analysis for a given value of the parameter $\theta$. Following the specification of the utility functions in Section 2.2.3, the counterfactual involves the distribution of outcomes induced by draws of utility functions from $\tilde{F}_\theta(\tilde{\tilde{u}})$. For each draw $\tilde{\tilde{u}}$, based on Section 2.2.5, the *solution* used will be an element of $\mathcal{S}(\tilde{\tilde{u}})$. Consequently,

$$P_{\tilde{\tilde{u}} \sim \tilde{F}_\theta(\tilde{\tilde{u}})}(\mathcal{S}(\tilde{\tilde{u}})) \tag{2}$$

is the *distribution of sets of solutions* arising in the counterfactual.

In the typical case of multiple solutions (e.g., multiple equilibria), $|\mathcal{S}(u)| > 1$ for some $u$. The determination of the action arising from the game involves two steps: first, a solution $s$ from $\mathcal{S}(u)$ is selected according to a "selection mechanism"; and second, a draw from the solution $s$ is taken to realize an actual action profile. This process implies that the action profile actually used by the DMs is a draw from a distribution in co $\mathcal{S}(u)$, the convex hull of $\mathcal{S}(u)$. The convex hull reflects the selection mechanism, which is an arbitrary mixture of solutions from $\mathcal{S}(u)$. Consequently,

$$P_{\tilde{\tilde{u}} \sim \tilde{F}_\theta(\tilde{\tilde{u}})}(\text{co } \mathcal{S}(\tilde{\tilde{u}})) \tag{3}$$

is the *distribution of sets of probabilities of the actions taken in the game* arising in the counterfactual.

Therefore, with $\mathcal{T}^{\text{co}}(\tilde{\tilde{u}}) = \{t(\tilde{\tilde{u}}, s) : s \in \text{co } \mathcal{S}(\tilde{\tilde{u}})\}$,

$$P_{\tilde{\tilde{u}} \sim \tilde{F}_\theta(\tilde{\tilde{u}})}(\mathcal{T}^{\text{co}}(\tilde{\tilde{u}})) \tag{4}$$

is the *distribution of sets of the outcome of interest* arising in the counterfactual. Another relevant quantity is $P_{\tilde{\tilde{u}} \sim \tilde{F}_\theta(\tilde{\tilde{u}})}(\mathcal{T}(\tilde{\tilde{u}}))$, with $\mathcal{T}(\tilde{\tilde{u}}) = \{t(\tilde{\tilde{u}}, s) : s \in \mathcal{S}(\tilde{\tilde{u}})\}$. The next result shows that these distributions are indeed valid distributions over sets.

**Lemma 2.** *Under Assumptions 1 and 2: $\mathcal{S}(\tilde{\tilde{u}})$ with $\tilde{\tilde{u}} \sim \tilde{F}_\theta(\tilde{\tilde{u}})$ is a random closed set. $\mathcal{T}(\tilde{\tilde{u}})$ with $\tilde{\tilde{u}} \sim \tilde{F}_\theta(\tilde{\tilde{u}})$ is a random closed set. co $\mathcal{S}(\tilde{\tilde{u}})$ with $\tilde{\tilde{u}} \sim \tilde{F}_\theta(\tilde{\tilde{u}})$ is a random closed set. $\mathcal{T}^{co}(\tilde{\tilde{u}})$ with $\tilde{\tilde{u}} \sim \tilde{F}_\theta(\tilde{\tilde{u}})$ is a random closed set.*

**Example 1** (continuing from p. 17)**.** For example, using the Nash equilibrium solution concept, consider an entry game involving two firms, and suppose that for a particular draw $\tilde{\tilde{u}}'$, there are three Nash equilibria: firm 1 enters as a monopolist, firm 2 enters as a monopolist, and the firms use (independent) mixed strategies with entry probabilities $p_1'$ and $p_2'$ respectively. Note that $p_1'$ and $p_2'$ would be known given $\tilde{\tilde{u}}'$. Continuing the same setup as in previous discussions of this running example, where a generic solution $s$ is of the form $(p_{00}, p_{01}, p_{10}, p_{11})$, and the subscript is a particular action profile, these three Nash equilibria convert to solutions $(0, 0, 1, 0)$, $(0, 1, 0, 0)$, and $((1 - p_1')(1 - p_2'), (1 - p_1')p_2', p_1'(1 - p_2'), p_1'p_2')$. The actual probabilities of entry arising from this particular $\tilde{\tilde{u}}'$ is the set co $\mathcal{S}(\tilde{\tilde{u}}')$, which is the set of convex combinations of these three solutions.

Alternatively, suppose that for another draw of $\tilde{\tilde{u}}''$, there is only one Nash equilibrium: both firms enter the market. In that case, co $\mathcal{S}(\tilde{\tilde{u}}'')$ is the single element $(0, 0, 0, 1)$, corresponding to $p_{11} = 1$. More generally, each value of $\tilde{\tilde{u}}$ would have a different corresponding co $\mathcal{S}(\tilde{\tilde{u}})$, inducing the *distribution* over co $\mathcal{S}(\tilde{\tilde{u}})$. This is a distribution over sets.

Our CPDSes can be used to answer (at least) two broad classes of counterfactual questions. These are defined below in Definition 2 and 3. Again, for now, we consider a given value of $\theta$. Later on, we accommodate partial identification of $\theta$ and estimation of the identified set for $\theta$.

**Definition 2** (*Counterfactual question: the probability that a set of values for the outcome of interest could arise, or must arise*)**.** Suppose $B$ is some interesting set of values for the outcome of interest. For instance, if the outcome of interest is the solution used by the DMs, $B$ could be a set of solutions satisfying some interesting property. Or, if the outcome of interest is welfare, $B$ could be an interesting set of welfare outcomes. For instance, if the welfare outcome is a "social welfare" that accounts for the cost of a policy intervention, $B$ could be the non-negative welfare outcomes. Then:

(1) $P_{\tilde{\tilde{u}} \sim \tilde{F}_\theta(\tilde{\tilde{u}})}(\mathcal{T}^{\text{co}}(\tilde{\tilde{u}}) \cap B \neq \emptyset)$ is the probability that some value in $B$ *could* arise.

(2) $P_{\tilde{\tilde{u}} \sim \tilde{F}_\theta(\tilde{\tilde{u}})}(\mathcal{T}^{\text{co}}(\tilde{\tilde{u}}) \cap B = \emptyset)$ is the probability that all values in $B$ *cannot* arise.

(3) $P_{\tilde{\tilde{u}} \sim \tilde{F}_\theta(\tilde{\tilde{u}})}(\mathcal{T}^{\text{co}}(\tilde{\tilde{u}}) \subseteq B)$ is the probability that some value in $B$ *must* arise.

Thus, using the CPDS, the econometrician could make counterfactual statements concerning whether some outcome could happen, or must happen, and so forth.

**Definition 3** (*Counterfactual question: averaged value of a function of the outcome of interest*)**.** Suppose the counterfactual question concerns a scalar-valued function $\omega(\cdot)$ of the outcome of interest. Many specifications of $\omega(\cdot)$ can be of interest:

(1) Suppose the counterfactual question concerns the probability of a particular outcome of the game. Suppose this particular outcome of the game corresponds to the $k$-th element of the solution profile. This could be, for example, a particular arrangement of entry decisions in an entry game. Then, specify that the outcome of interest is the solution, and $\omega(s) = s_k$. Or, rather than focus on the probability of a specific outcome of the game, $\omega(\cdot)$ could sum the probabilities of all outcomes of the game that satisfy a certain property (e.g., a certain firm enters the market, or a certain number of firms enter the market, etc.).

(2) Another counterfactual question concerns the *expected* outcome of the game (in cases where an "expected" outcome makes sense). For example, in an entry game, the object of interest could be the *expected* number of market entrants. Then, specify that the outcome of interest is the solution, and specify $\omega(s)$ to be the expected number of market entrants for given solution $s$.

(3) Alternatively, if the outcome of interest is welfare, then $\omega(\cdot)$ could be just the identity mapping, so that the counterfactual question concerns the average welfare outcome.

Then:

(1) The Choquet integral $E_{P_{\tilde{\tilde{u}} \sim \tilde{F}_\theta(\tilde{u})}(\mathcal{T}^{\text{co}}(\tilde{\tilde{u}}))}(\sup_{t \in \mathcal{T}^{\text{co}}(\tilde{\tilde{u}})} \omega(t))$ is the average of the "largest attainable" value of the function $\omega$ of the outcome of interest. The average concerns the distribution of the sets of the outcome of interest from Equation 4, so that the $t$ argument of $\omega(\cdot)$ is from the set $\mathcal{T}^{\text{co}}(\tilde{\tilde{u}})$ where $\tilde{\tilde{u}} \sim \tilde{F}_\theta(\tilde{u})$. Further, "largest attainable" concerns the supremum value of the function $\omega$ of the outcome of interest, for each draw of $\mathcal{T}^{\text{co}}(\tilde{\tilde{u}})$ from Equation 4.

(2) Similarly, $E_{P_{\tilde{\tilde{u}} \sim \tilde{F}_\theta(\tilde{u})}(\mathcal{T}^{\text{co}}(\tilde{\tilde{u}}))}(\inf_{t \in \mathcal{T}^{\text{co}}(\tilde{\tilde{u}})} \omega(t))$ is the "smallest attainable" value of the function $\omega$ of the outcome of interest.

Thus, using the CPDS, the econometrician could make counterfactual statements concerning the upper bound (or lower bound) on the probability of a particular outcome of the game, integrating over the distribution of the utility functions. Or, the econometrician could make counterfactual statements concerning the upper bound (or lower bound) of the welfare measure, integrating over the distribution of the utility functions.

We provide an empirical illustration of both kinds of counterfactual questions, respectively in Sections 7.2.2 and 7.2.1.

**Example 1** (continuing from p. 20)**.** Continue to consider the two realizations of utility $\tilde{\tilde{u}}'$ and $\tilde{\tilde{u}}''$. Suppose to simplify the example that these are the *only* possible values of $\tilde{\tilde{u}}$; however, our approach does not require any finiteness/discreteness restriction on the support of $\tilde{\tilde{u}}$. Next we discuss possible specific counterfactual questions from Definitions 2 and 3. We provide closed-form representations in this example, but such derivations are not part of our actual approach to counterfactual analysis, which does everything computationally. As a reminder, a solution $s$ in this example is of the form $(p_{00}, p_{01}, p_{10}, p_{11})$ where in the case that DM 1 enters with probability $p_1$ and DM 2 enters with probability $p_2$, this would be of the form $((1-p_1)(1-p_2), (1-p_1)p_2, p_1(1-p_2), p_1p_2)$.

(1) **Market entry outcomes:** Suppose the counterfactual question concerns the probability that *at least one firm* enters the market *as a pure strategy outcome*. Thus, the outcome of interest is the strategy, and $B = \{((1-p_1)(1-p_2), (1-p_1)p_2, p_1(1-p_2), p_1p_2) : p_1 = 1 \text{ or } p_2 = 1\}$, which is the event: "*at least one firm enters in pure strategies.*" Then, co $\mathcal{S}(\tilde{\tilde{u}}') \cap B \neq \emptyset$ and co $\mathcal{S}(\tilde{\tilde{u}}'') \cap B \neq \emptyset$. Thus,

$$P_{\tilde{\tilde{u}} \sim \tilde{F}_\theta(\tilde{\tilde{u}})}(\text{co } \mathcal{S}(\tilde{\tilde{u}}) \cap B \neq \emptyset) = 1, \tag{5}$$

in the sense that there is probability 1 that the outcome *could be* that at least one firm enters the market as a pure strategy outcome. Alternatively, co $\mathcal{S}(\tilde{\tilde{u}}') \not\subseteq B$ while co $\mathcal{S}(\tilde{\tilde{u}}'') \subseteq B$. Thus,

$$P_{\tilde{\tilde{u}} \sim \tilde{F}_\theta(\tilde{\tilde{u}})}(\text{co } \mathcal{S}(\tilde{\tilde{u}}) \subseteq B) = P_\theta(\tilde{\tilde{u}} = \tilde{\tilde{u}}''), \tag{6}$$

in the sense that there is probability $P_\theta(\tilde{\tilde{u}} = \tilde{\tilde{u}}'')$ that the outcome *must be* that at least one firm enters the market as a pure strategy outcome.

(2) **Market entry probabilities:** Suppose the counterfactual question concerns the probability that *at least one* firm enter the market. This might correspond to a planner's preference against having no entrants. Thus, $\omega(s) = s_{01} + s_{10} + s_{11}$, with the notation $s_a$ as the probability of the action profile $a$ according to the solution $s$. With different $\omega(s)$, a similar analysis could be conducted for the probability of other market outcomes (e.g., a monopoly market outcome, a duopoly market outcome).

First, consider the upper bound. When the utility is $\tilde{\tilde{u}}'$, we have $\sup \omega(\text{co } \mathcal{S}(\tilde{\tilde{u}}')) = 1$ since the solution $(0, 0, 1, 0)$ achieves this bound. And similarly when the utility is $\tilde{\tilde{u}}''$, $\sup \omega(\text{co } \mathcal{S}(\tilde{\tilde{u}}'')) = 1$. Therefore, the upper bound on the probability that at least one firm enters the market (integrating over the distribution of the utility functions) is

$$E_{P_{\tilde{\tilde{u}} \sim \tilde{F}_\theta(\tilde{\tilde{u}})}(\text{co } \mathcal{S}(\tilde{\tilde{u}}))}(\sup \omega(s)) = 1. \tag{7}$$

Even though this happens to be the "trivial" upper bound, note the logical steps here establishes that this upper bound is sharp: it is consistent with the model that there is probability 1 that at least one firm enters the market.

Now, consider the lower bound. It follows that $\inf \omega(\text{co } \mathcal{S}(\tilde{u}')) = 1 - (1-p'_1)(1-p'_2)$ since the mixed strategy achieves this bound and either of the pure strategies would have a greater value under $\omega(\cdot)$. And, $\inf \omega(\text{co } \mathcal{S}(\tilde{u}'')) = 1$. Therefore, the lower bound on the probability that at least one firm enters the market (integrating over the distribution of the utility functions) is

$$E_{P_{\tilde{u} \sim \tilde{F}_\theta(\tilde{u})}(\text{co } \mathcal{S}(\tilde{u}))}(\inf \omega(s)) = (1 - (1-p'_1)(1-p'_2))P_\theta(\tilde{u} = \tilde{u}') + P_\theta(\tilde{u} = \tilde{u}''). \tag{8}$$

Again, note this bound is sharp.

(3) **Expected entrants:** Suppose the counterfactual question concerns the expected number of entrants. Then, $\omega(s) = 0s_{00} + 1s_{01} + 1s_{10} + 2s_{11}$, and the rest of the analysis follows similarly to above.

First, consider the upper bound. It follows that $\sup \omega(\text{co } \mathcal{S}(\tilde{u}')) = \max\{1, (1-p'_1)p'_2 + p'_1(1-p'_2) + 2p'_1p'_2)\}$ since the maximum of a linear function over a convex set is achieved at an extreme point (again, $p'_1$ and $p'_2$ are known given $\tilde{u}'$). And, $\sup \omega(\text{co } \mathcal{S}(\tilde{u}'')) = 2$. Therefore, the upper bound on the expected number of entrants (integrating over the distribution of the utility functions) is

$$E_{P_{\tilde{u} \sim \tilde{F}_\theta(\tilde{u})}(\text{co } \mathcal{S}(\tilde{u}))}(\sup \omega(s)) = \max\{1, (1-p'_1)p'_2 + p'_1(1-p'_2) + 2p'_1p'_2)\}P_\theta(\tilde{u} = \tilde{u}') + 2P_\theta(\tilde{u} = \tilde{u}''). \tag{9}$$

Now, consider the lower bound. It follows that $\inf \omega(\text{co } \mathcal{S}(\tilde{u}')) = \min\{1, (1-p'_1)p'_2 + p'_1(1-p'_2) + 2p'_1p'_2)\}$ by the same reasoning. Again, $\inf \omega(\text{co } \mathcal{S}(\tilde{u}'')) = 2$. Therefore, the lower bound on the expected number of entrants (integrating over the distribution of the utility functions) is

$$E_{P_{\tilde{u} \sim \tilde{F}_\theta(\tilde{u})}(\text{co } \mathcal{S}(\tilde{u}))}(\inf \omega(s)) = \min\{1, (1-p'_1)p'_2 + p'_1(1-p'_2) + 2p'_1p'_2)\}P_\theta(\tilde{u} = \tilde{u}') + 2P_\theta(\tilde{u} = \tilde{u}''). \tag{10}$$

(4) **Profits:** Suppose the counterfactual question concerns the summed profits of the potential entrants. Then, $t(\tilde{\tilde{u}}, s) = (\tilde{\tilde{u}}_{1,00} + \tilde{\tilde{u}}_{2,00})s_{00} + (\tilde{\tilde{u}}_{1,01} + \tilde{\tilde{u}}_{2,01})s_{01} + (\tilde{\tilde{u}}_{1,10} + \tilde{\tilde{u}}_{2,10})s_{10} + (\tilde{\tilde{u}}_{1,11} + \tilde{\tilde{u}}_{2,11})s_{11}$, where $\tilde{\tilde{u}}_{i,a}$ is the utility of DM $i$ under action profile $a$. Then, the mapping $\omega(\cdot)$ would just be the identity mapping.

As a function of $s$, $\omega(t(\tilde{\tilde{u}}, s))$ is a linear function that is maximized/minimized over a convex set of solutions, so the max/min will be achieved at one of the extreme points. Thus, $\sup \mathcal{T}(\tilde{\tilde{u}}') = \max\{(\tilde{\tilde{u}}'_{1,10} + \tilde{\tilde{u}}'_{2,10}), (\tilde{\tilde{u}}'_{1,01} + \tilde{\tilde{u}}'_{2,01}), (\tilde{\tilde{u}}'_{1,00} + \tilde{\tilde{u}}'_{2,00})(1-p'_1)(1-p'_2) + (\tilde{\tilde{u}}'_{1,01} + \tilde{\tilde{u}}'_{2,01})(1-p'_1)p'_2 + (\tilde{\tilde{u}}'_{1,10} + \tilde{\tilde{u}}'_{2,10})p'_1(1-p'_2) + (\tilde{\tilde{u}}'_{1,11} + \tilde{\tilde{u}}'_{2,11})p'_1 p'_2\}$ for $\tilde{\tilde{u}}'$ and $\sup \mathcal{T}(\tilde{\tilde{u}}'') = (\tilde{\tilde{u}}''_{1,11} + \tilde{\tilde{u}}''_{2,11})$ for $\tilde{\tilde{u}}''$. Therefore, the upper bound on the average profits of the firms summed (integrating over the distribution of the utility functions) is

$$\max\{(\tilde{\tilde{u}}'_{1,10} + \tilde{\tilde{u}}'_{2,10}), (\tilde{\tilde{u}}'_{1,01} + \tilde{\tilde{u}}'_{2,01}),$$
$$(\tilde{\tilde{u}}'_{1,00} + \tilde{\tilde{u}}'_{2,00})(1-p'_1)(1-p'_2) + (\tilde{\tilde{u}}'_{1,01} + \tilde{\tilde{u}}'_{2,01})(1-p'_1)p'_2$$
$$+ (\tilde{\tilde{u}}'_{1,10} + \tilde{\tilde{u}}'_{2,10})p'_1(1-p'_2) + (\tilde{\tilde{u}}'_{1,11} + \tilde{\tilde{u}}'_{2,11})p'_1 p'_2\} P_\theta(\tilde{\tilde{u}} = \tilde{\tilde{u}}')$$
$$+ (\tilde{\tilde{u}}''_{1,11} + \tilde{\tilde{u}}''_{2,11}) P_\theta(\tilde{\tilde{u}} = \tilde{\tilde{u}}''). \tag{11}$$

The lower bound would simply switch max for min. □

Although our running example has found explicit expressions for the various counterfactual quantities, this is not needed for our approach to CPDSes, which does this computationally.

**Remark 1** (The value of the correlated equilibrium solution concept for counterfactual analysis). Our CPDSes accommodate a variety of solution concepts. In this remark, we provide a few arguments in favor of the *correlated equilibrium* solution concept.

First, the correlated equilibrium solution concept by construction contains the convex hull of the Nash equilibrium solution concept (e.g., Aumann (1987)).[13] In other words, any distribution of actions that could arise from a selection mechanism over multiple Nash equilibria is a correlated equilibrium. Further, the set of correlated equilibria is convex by construction, so $\operatorname{co} \mathcal{S} = \mathcal{S}$. Thus, using correlated equilibrium as the solution concept is a relaxation of using Nash equilibrium as the solution concept, *including accommodating for the possibility of multiple equilibria.*

---
[13]Indeed, the related solution concept sometimes known as "public correlated equilibrium" is exactly the convex hull of the Nash equilibrium solution concept. In general, the inclusion is strict, in that there are correlated equilibria that are not convex combinations of Nash equilibria (e.g., Aumann (1974)).

Second, using the set of correlated equilibria involves solving a relatively simple computational problem, because it is a linear programming problem (e.g., Hart and Schmeidler (1989), Gilboa and Zemel (1989), and Papadimitriou (2007)). Using the set of Nash equilibria can involve solving a substantially more complex computational problem. Thus, combined with the above, correlated equilibrium provides a *computationally attractive* solution to the problem of dealing with multiple Nash equilibria.

3.2. **Step 2: Population CPDS accounting for partial identification of $\theta$.** Partial identification of $\theta$ is another unknown in the counterfactual analysis. Suppose the identified set for the underlying model parameter $\theta$ is $\Theta_{I,0}$. As discussed above, we take this identified set as given, based on some existing identification strategy for $\theta$.

**Definition 4** (Population counterfactual predictive distribution set). Given the identified set $\Theta_{I,0}$, the *population counterfactual predictive distribution set* is

$$\{P_{\tilde{\tilde{u}} \sim \tilde{F}_\theta(\tilde{\tilde{u}})}(\mathcal{T}^{\text{co}}(\tilde{\tilde{u}})) : \theta \in \Theta_{I,0}\}. \tag{12}$$

This is the "population CPDS." This implies related constructions for each of the counterfactual questions described in the previous part. We demonstrate this below. From the point of view of identification, the set given in Display 12 is the sharp identified set.

**Theorem 3** (Sharpness of the population counterfactual predictive distribution set). *Suppose Assumptions 1 and 2 hold. Then:*

*(3.1) The set in Display 12 is the sharp identified set for the distribution of outcomes under the specification of the counterfactual given in Section 2.2, relative to $\Theta_{I,0}$. That is: a random variable $\tilde{T}$ can be generated as the distribution of the counterfactual outcomes under the setup of the counterfactual analysis described in Definition 1 if and only if $P(\tilde{T} \in K) \leq P_{\tilde{\tilde{u}} \sim \tilde{F}_{\theta^*}(\tilde{\tilde{u}})}(\mathcal{T}^{co}(\tilde{\tilde{u}}) \cap K \neq \emptyset)$ for all compact sets $K$, for some $\theta^* \in \Theta_{I,0}$.*

*(3.2) If $\Theta_{I,0}$ is the sharp identified set, then the set in Display 12 is the sharp identified set relative to the underlying data used to estimate $\theta$.*

The statement "sharp identified set for the distribution of outcomes under the specification of the counterfactual given in Section 2.2" means that any selection (in the sense of a selection from a random set) from one of the distributions in the set in Display 12 could be the distribution of counterfactual outcomes under the setup of the counterfactual analysis described in Definition 1, for some value of the parameter in $\Theta_{I,0}$. Since the counterfactual analysis can depend on the data only through $\Theta_{I,0}$ by assumption, if $\Theta_{I,0}$ is the sharp identified set, Display 12 is the sharp identified set relative to the setup used to estimate $\theta$.

**Example 1** (continuing from p. 22). Continue the running example, where now different values of $\theta$ correspond to different distributions of $\tilde{\tilde{u}}$. To simplify the example, we suppose the same $\tilde{\tilde{u}}'$ and $\tilde{\tilde{u}}''$ values as considered before are the only possible points in the support, but our approach does not require this restriction. For any of the counterfactual questions considered above for a given $\theta$, partial identification of $\theta$ introduces another unknown, per the expression in Equation 12 applied to the expressions derived previously.

**Market entry outcomes:** Since $P_{\tilde{\tilde{u}} \sim \tilde{F}_\theta(\tilde{\tilde{u}})}(\text{co } \mathcal{S}(\tilde{\tilde{u}}) \cap B \neq \emptyset) = 1$ for all values of $\theta$ per Equation 5, even with partial identification of $\theta$, the corresponding identified set for the probability that the outcome *could be* that at least one firm enters the market as a pure strategy outcome is the singleton set

$$\{P_{\tilde{\tilde{u}} \sim \tilde{F}_\theta(\tilde{\tilde{u}})}(\text{co } \mathcal{S}(\tilde{\tilde{u}}) \cap B \neq \emptyset) : \theta \in \Theta_{I,0}\} = \{1\}. \tag{13}$$

The overall probability with which the outcome *could be* that at least one firm enters the market as a pure strategy outcome is no larger than the supremum of this set. Although this is the trivial upper bound of 1 in this particular example, note this bound is sharp.

Since $P_{\tilde{\tilde{u}} \sim \tilde{F}_\theta(\tilde{\tilde{u}})}(\text{co } \mathcal{S}(\tilde{\tilde{u}}) \subseteq B) = P_\theta(\tilde{\tilde{u}} = \tilde{\tilde{u}}'')$ per Equation 6, the corresponding identified set for the probability that the outcome *must be* that at least one firm enters the market as a pure strategy outcome is

$$\{P_{\tilde{\tilde{u}} \sim \tilde{F}_\theta(\tilde{\tilde{u}})}(\text{co } \mathcal{S}(\tilde{\tilde{u}}) \subseteq B) : \theta \in \Theta_{I,0}\} = \{P_\theta(\tilde{\tilde{u}} = \tilde{\tilde{u}}'') : \theta \in \Theta_{I,0}\} \tag{14}$$

The overall probability with which the outcome *must be* that at least one firm enters the market as a pure strategy outcome is at least as large as the infimum of this set.

**Market entry probabilities:** Using Equation 7, the upper bound on the probability that at least one firm enters the market (integrating over the distribution of the utility functions) is 1 for every value of $\theta$. Thus, the corresponding identified set for the upper bound on the probability that at least one firm enters the market (integrating over the distribution of the utility functions) is the singleton set

$$\{E_{P_{\tilde{u} \sim \tilde{F}_\theta(\tilde{u})}(\operatorname{co} \mathcal{S}(\tilde{u}))}(\sup \omega(s)) : \theta \in \Theta_{I,0}\} = \{1\}. \tag{15}$$

The overall probability with which at least one firm enters the market is no larger than the supremum of this set. As above, although this is the trivial upper bound of 1 in this particular example, note this bound is sharp.

Using Equation 8, the corresponding identified set for the lower bound on the probability that at least one firm enters the market (integrating over the distribution of the utility functions) is

$$\{E_{P_{\tilde{u} \sim \tilde{F}_\theta(\tilde{u})}(\operatorname{co} \mathcal{S}(\tilde{u}))}(\inf \omega(s)) : \theta \in \Theta_{I,0}\} = \{(1-(1-p'_1)(1-p'_2))P_\theta(\tilde{\tilde{u}} = \tilde{\tilde{u}}') + P_\theta(\tilde{\tilde{u}} = \tilde{\tilde{u}}'') : \theta \in \Theta_{I,0}\} \tag{16}$$

The overall probability that at least one firm enters the market is at least as large as the infimum of this set.

**Expected entrants:** Using Equation 9, the corresponding identified set for the upper bound on the expected number of entrants is

$$\{\max\{1, (1-p'_1)p'_2 + p'_1(1-p'_2) + 2p'_1 p'_2\} P_\theta(\tilde{\tilde{u}} = \tilde{\tilde{u}}') + 2P_\theta(\tilde{\tilde{u}} = \tilde{\tilde{u}}'') : \theta \in \Theta_{I,0}\} \tag{17}$$

The overall expected number of entrants is no greater than as the supremum of this set.

Using Equation 10, the corresponding identified set for the lower bound on the expected number of entrants is

$$\{\min\{1, (1-p'_1)p'_2 + p'_1(1-p'_2) + 2p'_1 p'_2\} P_\theta(\tilde{\tilde{u}} = \tilde{\tilde{u}}') + 2P_\theta(\tilde{\tilde{u}} = \tilde{\tilde{u}}'') : \theta \in \Theta_{I,0}\} \tag{18}$$

The overall expected number of entrants is at least as large as the infimum of this set.

**Profits:** Using Equation 11, the identified set for the upper bound on the average profits of the firms summed (integrating over the distribution of the utility functions) is

$$\{\max\{(\tilde{\tilde{u}}'_{1,10} + \tilde{\tilde{u}}'_{2,10}), (\tilde{\tilde{u}}'_{1,01} + \tilde{\tilde{u}}'_{2,01}),$$

$$(\tilde{\tilde{u}}'_{1,00} + \tilde{\tilde{u}}'_{2,00})(1-p'_1)(1-p'_2) + (\tilde{\tilde{u}}'_{1,01} + \tilde{\tilde{u}}'_{2,01})(1-p'_1)p'_2 + (\tilde{\tilde{u}}'_{1,10} + \tilde{\tilde{u}}'_{2,10})p'_1(1-p'_2) +$$

$$(\tilde{\tilde{u}}'_{1,11} + \tilde{\tilde{u}}'_{2,11})p'_1 p'_2\} P_\theta(\tilde{\tilde{u}} = \tilde{\tilde{u}}') + (\tilde{\tilde{u}}''_{1,11} + \tilde{\tilde{u}}''_{2,11}) P_\theta(\tilde{\tilde{u}} = \tilde{\tilde{u}}'') : \theta \in \Theta_{I,0}\} \tag{19}$$

The overall average profits of the firms summed is no greater than as the supremum of this set. The lower bound would simply switch max for min.

As noted before, with regards to Equations 16, 17, 18, and 19, these expressions are substantially simplified by the assumptions used in this running example. For instance, in the general case allowed by our approach to counterfactual analysis, these expressions could not be written in terms of a discrete distribution of $\tilde{\tilde{u}}$.

Although our running example has found explicit expressions for the various counterfactual quantities, this is not needed for our approach to CPDSes, which does this computationally.

3.3. **Step 3: Posterior CPDS accounting for statistical uncertainty.** Statistical uncertainty about the identified set for $\theta$ is yet another unknown in the counterfactual analysis. Given a posterior distribution $\Pi^{(N)}(\Theta_I)$ for $\Theta_{I,0}$, which we take as given from the existing literature, we define next the overall posterior counterfactual predictive distribution set. In the interest of brevity, we generally suppress the dependence on "the data" in the notation for $\Pi^{(N)}(\Theta_I)$; another way of writing this would be $\Pi^{(N)}(\Theta_I|W^{(N)})$, for data $W^{(N)}$.

**Definition 5** (Posterior counterfactual predictive distribution set)**.** Given a posterior distribution $\Pi^{(N)}(\Theta_I)$ over the identified set $\Theta_{I,0}$, the *posterior counterfactual predictive distribution set* is

$$\Pi^{(N)}(\{P_{\tilde{\tilde{u}} \sim \tilde{F}_\theta(\tilde{\tilde{u}})}(\mathcal{T}^{\text{co}}(\tilde{\tilde{u}})) : \theta \in \Theta_I\}). \tag{20}$$

This is the "posterior CPDS." The posterior CPDS accounts for all of the unknowns in the counterfactual analysis described in the introduction. In particular, it accounts for statistical uncertainty. Using this posterior CPDS, it is possible to answer the counterfactual questions discussed above. As with Equation 12, this implies related constructions for each of the counterfactual questions.

**Example 1** (continuing from p. 27). Continue the running example, where now there is a posterior distribution for $\Theta_{I,0}$. Since the quantities previously discussed were mappings of a value for the identified set (e.g., Equation 14 or Equation 16 or Equation 17 or Equation 18 or Equation 19), a posterior distribution for $\Theta_{I,0}$ induces a posterior distribution over the quantities previously discussed in the obvious way given the expression in Equation 20.

4. Consistency of the posterior counterfactual predictive distribution set

We provide conditions under which consistency of the posterior distribution for $\Theta_{I,0}$ implies[14] consistency of the posterior CPDS. Particularly in light of the sharpness result in Theorem 3, the posterior CPDS summarizes the statistical uncertainty about the counterfactuals. The remaining question is whether the statistical uncertainty disappears in sufficiently large samples, in that the posterior CPDS concentrates on the true population CPDS (i.e., Display 12). As we noted in the introduction, consistency of Bayesian approaches in partially identified models can be quite subtle. Our setting is particularly complicated because we are dealing with non-standard objects such as sets of distributions over sets, and because the solution concept mappings are discontinuous in the underlying utilities. Our starting point is a consistent posterior distribution for $\Theta_{I,0}$ from the existing literature.

As with any approach to Bayesian asymptotics (even in standard point identified models), we need some underlying data generating process that results in $\Pi^{(N)}$, the posterior distribution over $\Theta_{I,0}$. We use the notation $P_0^{(N)}$ for this underlying data generating process in a sample of size $N$. We also use the notation $P_0^{(\infty)}$ for the sample paths for this underlying data generating process. Then, as with any approach to Bayesian asymptotics, we can ask about the limiting properties of $\Pi^{(N)}$ exactly in the same way that we can ask about the limiting properties of a classical estimator $\hat{\theta}_{(N)}$. Ultimately, both are random objects with a distribution induced by $P_0^{(N)}$ and sample paths induced by $P_0^{(\infty)}$. The limit notation $\to$ we use below is understood as a limit in probability with respect to $P_0^{(N)}$. In other words, the notation $\Pi^{(N)}(E) \to c$ for some event $E$ and some constant $c$ corresponds to the statement that: for any $\epsilon > 0$, $P_0^{(N)}(|\Pi^{(N)}(E) - c| > \epsilon) \to 0$ as $N \to \infty$. Next, we provide two definitions of consistency of a posterior distribution for a set.

---

[14] We do not use the terminology "continuous mapping theorem" despite what might appear to be the study of CMTs as our results are different. Consistency of the posterior distribution for the set $A_0$ concerns statements like: for any $\epsilon > 0$, $P_0^{(N)}(|\Pi^{(N)}(E) - c| > \epsilon) \to 0$ as $N \to \infty$, where for instance it could be that $E$ is the event $A \cap B \neq \emptyset$ and $c = 1[A_0 \cap B \neq \emptyset]$. I.e., $\Pi^{(N)}(E)$ converges in probability to $c$. Hence, a standard CMT would conclude $f(\Pi^{(N)}(E)) \to f(c)$ if $f(\cdot)$ is a continuous function. However, this is not the result that we want, since we are not interested in a function of the posterior distribution. We want $P_0^{(N)}(|\Pi^{(N)}(\tilde{E}) - \tilde{c}| > \epsilon) \to 0$ for *a different* event $\tilde{E}$ and *a different* $\tilde{c}$ (that is generically not $f(c)$), like $\tilde{E} = f(A) \cap B \neq \emptyset$ and $\tilde{c} = 1[f(A_0) \cap B \neq \emptyset]$.

**Definition 6** (Consistency in a general topological space). A sequence of posterior distributions for a set $A_0$ in an ambient topological space $S$ is consistent if:

(6.1) $\Pi^{(N)}(A \cap B \neq \emptyset) \to 1[A_0 \cap B \neq \emptyset]$ for any closed set $B \subseteq S$ such that bd $A_0 \cap B = \emptyset$, where bd $A_0$ is the topological boundary of $A_0$.

(6.2) $\Pi^{(N)}(A \neq \emptyset) \to 1$ if $A_0 \neq \emptyset$.

According to Definition 6, the posterior distribution is consistent when it assigns posterior probability approaching 1 (or 0) in large samples to certain true (or false) statements about the intersection of $A_0$ with other sets. For instance, if $A_0$ is a (sharp) identified set, and $B$ is a set of parameter values, $A_0 \cap B \neq \emptyset$ means that at least one value in $B$ could have generated the data. Part 6.2 deals with a nuance that arises when the posterior distribution might have empty realizations; it is irrelevant under the common circumstance that $\Pi^{(N)}(A \neq \emptyset) = 1$. In metric spaces, it can be useful to use a related definition.

**Definition 7** (Consistency in a metric space). A sequence of posterior distributions for a set $A_0$ in an ambient metric space $S$ is consistent if $\Pi^{(N)}(d_H(A, A_0) > \epsilon) \to 0$ for any $\epsilon > 0$, where $d_H$ is the usual Hausdorff distance.

This definition requires the metric in order to define Hausdorff distance. According to Definition 7, the posterior distribution is consistent when there is vanishing posterior probability that the Hausdorff distance between $A$ and $A_0$ exceeds any $\epsilon > 0$. In Section 6.1, under general conditions, we prove equivalences among the different definitions of consistency. In particular, those equivalences provide the result that (under suitable topological conditions) both definitions of consistency are equivalent to the standard definition of posterior consistency in the special case of a singleton $A_0$ and posterior distribution with singleton-valued $A$.

Our definitions of consistency refers to the ambient topological space. In everything that follows (unless indicated otherwise), we use the following standards: If $S$ is a Euclidean space, we use the standard Euclidean topology. If $S$ is a space of non-empty compact subsets of a metric space, we use the Hausdorff topology. If $S$ is a space of distributions of random sets of ordinary Euclidean objects (or other locally compact Hausdorff second countable space), we use the topology induced by the Lévy metric, since it is a metrization of weak convergence.

Theorem 4 is the main result of this section, and concerns consistency of the posterior CPDS. The proof is based on the development of a general set of results on posterior consistency in related settings, given in Section 6. It provides two sets of sufficient conditions, sharing a set of baseline conditions. The baseline conditions are the assumptions already discussed, and the following two assumptions.

**Assumption 4.** *Suppose the distribution of $(x, \epsilon)$ is a sequentially continuous function of $\theta$, in the sense that the sequence of distributions $\tilde{F}_{\theta^{(\nu)}}(x, \epsilon)$ converges weakly to $\tilde{F}_\theta(x, \epsilon)$ whenever $\theta^{(\nu)}$ is a sequence of parameter values that converges to $\theta$. Suppose the counterfactual utility functions $\tilde{u}_i(a_i, a_{-i}, x_i, \epsilon_i, \theta)$ are jointly continuous in $(x, \epsilon, \theta)$ for each $(a_i, a_{-i})$.*

This requires that the fundamentals of the model are suitably continuous.

**Assumption 5.** *Suppose that for all $\theta \in \Theta$, $P(\tilde{\tilde{u}}_\theta \in \mathcal{U}^e) = 0$ where $\tilde{\tilde{u}}_\theta \equiv \tilde{u}(\cdot, \cdot, x, \epsilon, \theta)$ with $(x, \epsilon) \sim \tilde{F}_\theta(x, \epsilon)$.*

This requires that $\mathcal{U}^e$ has probability zero under any attainable distribution of counterfactual utility. One simple sufficient condition is that $\mathcal{U}^e$ has Lebesgue measure zero (following the discussion of Assumption 3), and that every attainable distribution of counterfactual utility has an ordinary density with respect to Lebesgue measure (ignoring the components of utility normalized to 0). For instance, this would happen with the parametric (linear) specification for utility in Equation 1 when the unobservables are independent of the observables and have an ordinary density (e.g., Gut (2006, p 113)). The fact that Equation 1 involves location normalizing the profits to have $\tilde{\tilde{u}}_i(0, a_{-i}) \equiv 0$ means that $\tilde{\tilde{u}}$ overall does not have an ordinary density, but this does not change this result, for any solution concept where only differences in utility matter.[15] The justification of this claim is a bit intricate, and thus relegated to Appendix E.

**Theorem 4** (Consistency of the posterior counterfactual predictive distribution set)**.** *If:*

*(i) Assumptions 1, 2, 3, 4, and 5 hold.*

*(ii) $\Theta$ is a connected metric space.*

*(iii) The posterior distribution for $\Theta_{I,0}$ is consistent, in the sense of Definition 7.*

*(iv) $\Theta_{I,0} \neq \emptyset$, $\Theta_{I,0}^C \neq \emptyset$, $\Theta_{I,0}$ is compact, and the posterior for $\Theta_{I,0}$ respects these properties with posterior probability $1$.*

*Then:*

*(4.1) The posterior distribution for*

$$\left\{ P_{\tilde{\tilde{u}} \sim \tilde{F}_\theta(\tilde{u})}(\mathcal{T}^{co}(\tilde{\tilde{u}})) : \theta \in \Theta_{I,0} \right\}$$

*is consistent, in the sense of Definition 7.*

*If:*

*(v) Assumptions 1, 2, 3, 4, and 5 hold.*

*(vi) $\Theta$ is a $T_1$ space.*

*(vii) The posterior distribution for $\Theta_{I,0}$ is consistent, in the sense of Definition 6.*

*(viii) $\Theta_{I,0}$ is non-empty, connected, compact, either a singleton or regular closed, and the posterior distribution for $\Theta_{I,0}$ respects these properties with posterior probability $1$.*

*Then:*

*(4.2) For any set $B$ such that $[\mathcal{T}^{co}]^{-1}(B)$ is a Borel set, which includes all open or closed sets $B$, and such that either $P_{\tilde{\tilde{u}} \sim \tilde{F}_\theta(\tilde{u})}(\tilde{\tilde{u}} \notin [\mathcal{T}^{co}]^{-1}(int\ B), \tilde{\tilde{u}} \in [\mathcal{T}^{co}]^{-1}(bd\ B)) = 0$ for all $\theta \in \Theta$ or $P_{\tilde{\tilde{u}} \sim \tilde{F}_\theta(\tilde{u})}(\tilde{\tilde{u}} \in bd\ \mathcal{T}^{-1}(B)) = 0$ for all $\theta \in \Theta$, the posterior distribution for*

$$\sup \left\{ P_{\tilde{\tilde{u}} \sim \tilde{F}_\theta(\tilde{u})}(\mathcal{T}^{co}(\tilde{\tilde{u}}) \cap B \neq \emptyset) : \theta \in \Theta_{I,0} \right\}$$

*based on Definition 2 is consistent, in the sense of Definition 6.*

---

[15]Nash equilibrium, rationalizability, correlated equilibrium all satisfy this condition, and indeed "location normalizing" utility wouldn't be a normalization if using a solution concept that does not satisfy this condition.

(4.3) *For any set $B$ such that $[\mathcal{T}^{co}]^{-1}(B^C)$ is a Borel set, which includes all open or closed sets $B$, and such that either $P_{\tilde{\tilde{u}} \sim \tilde{F}_\theta(\tilde{u})}(\tilde{\tilde{u}} \notin [\mathcal{T}^{co}]^{-1}(int\ B^C), \tilde{\tilde{u}} \in [\mathcal{T}^{co}]^{-1}(bd\ B^C)) = 0$ for all $\theta \in \Theta$ or $P_{\tilde{\tilde{u}} \sim \tilde{F}_\theta(\tilde{u})}(\tilde{\tilde{u}} \in bd\ \mathcal{T}^{-1}(B^C)) = 0$ for all $\theta \in \Theta$, the posterior distribution for*

$$\inf\left\{P_{\tilde{\tilde{u}} \sim \tilde{F}_\theta(\tilde{u})}(\mathcal{T}^{co}(\tilde{\tilde{u}}) \subseteq B) : \theta \in \Theta_{I,0}\right\}$$

*based on Definition 2 is consistent, in the sense of Definition 6.*

(4.4) *For any continuous function $\omega(\cdot)$ such that*

$$\lim_{\alpha \to \infty} \sup_{\theta \in \Theta} E_{P_{\tilde{\tilde{u}} \sim \tilde{F}_\theta(\tilde{u})}}\left(\max\left\{\sup_{t \in \mathcal{T}(\tilde{u}) \cap K_\alpha^C} |\omega(t)|, 0\right\}\right) = 0,$$

*where $K_\alpha$ is the closed unit $L_1$-ball with radius $\alpha$, the posterior distribution for*

$$\sup\left\{E_{P_{\tilde{\tilde{u}} \sim \tilde{F}_\theta(\tilde{u})}(\mathcal{T}^{co}(\tilde{\tilde{u}}))}\left(\sup_{t \in \mathcal{T}^{co}(\tilde{\tilde{u}})} \omega(t)\right) : \theta \in \Theta_{I,0}\right\}$$

*based on Definition 3 is consistent, in the sense of Definition 6.*

(4.5) *For any continuous function $\omega(\cdot)$ such that*

$$\lim_{\alpha \to \infty} \sup_{\theta \in \Theta} E_{P_{\tilde{\tilde{u}} \sim \tilde{F}_\theta(\tilde{u})}}\left(\max\left\{\sup_{t \in \mathcal{T}(\tilde{u}) \cap K_\alpha^C} |\omega(t)|, 0\right\}\right) = 0,$$

*where $K_\alpha$ is the closed unit $L_1$-ball with radius $\alpha$, the posterior distribution for*

$$\inf\left\{E_{P_{\tilde{\tilde{u}} \sim \tilde{F}_\theta(\tilde{u})}(\mathcal{T}^{co}(\tilde{\tilde{u}}))}\left(\inf_{t \in \mathcal{T}^{co}(\tilde{\tilde{u}})} \omega(t)\right) : \theta \in \Theta_{I,0}\right\}$$

*based on Definition 3 is consistent, in the sense of Definition 6.*

*The same conclusions hold also when dropping the outermost* sup *or* inf*, or replacing the outermost* sup *(or* inf*, respectively) with* inf *(or* sup*, respectively).*

We discuss the assumption of consistency for the posterior distribution for $\Theta_{I,0}$ in Section 6.1. In short, existing Bayesian approaches for estimating $\Theta_{I,0}$ from the literature are compatible with this assumption, although it requires some proof.

In this result, and other places in the paper, we use phrases like "the posterior respects these properties with posterior probability 1" to mean that the posterior probability of having the stated property is 1.

Part 4.1 establishes consistency for the posterior distribution for the full CPDS. Consistency in this part means that the posterior distribution of the Hausdorff distance (using Lévy metric) between a "draw" of $\left\{P_{\tilde{u}\sim\tilde{F}_\theta(\tilde{u})}(\mathcal{T}^{\text{co}}(\tilde{\tilde{u}})) : \theta \in \Theta_I\right\}$, induced by the posterior distribution over $\Theta_I$, and the true value of $\left\{P_{\tilde{u}\sim\tilde{F}_\theta(\tilde{u})}(\mathcal{T}^{\text{co}}(\tilde{\tilde{u}})) : \theta \in \Theta_{I,0}\right\}$, approaches 0 in large samples.

Under different conditions, Parts 4.2-4.5 establish consistency for the posterior distribution for the specific quantities relevant to answering the counterfactual questions from Definitions 2 and 3. Consistency in these parts require that the posterior distribution for the identified set for a specific quantity "concentrates" on the corresponding true identified set, in the sense of Definition 6, which essentially requires that the posterior correctly assigns probability 1 (or 0) to true (or false) statements about the identified set (see also Theorem 16 for equivalent interpretations of consistency). Depending on whether the "outermost" sup or inf is kept in the statement of these parts, these parts concern either the posterior distribution for a scalar, or a posterior distribution for a set. Each of these parts requires a specific mild technical condition, which we describe in a separate remark at the end of this section.

**Corollary 5.** *Under the conditions used in Parts 4.2 and 4.3 of Theorem 4, the corresponding posterior distribution for*

$$\left[\inf\left\{P_{\tilde{u}\sim\tilde{F}_\theta(\tilde{u})}(\mathcal{T}^{co}(\tilde{\tilde{u}}) \subseteq B) : \theta \in \Theta_{I,0}\right\}, \sup\left\{P_{\tilde{u}\sim\tilde{F}_\theta(\tilde{u})}(\mathcal{T}^{co}(\tilde{\tilde{u}}) \cap B \neq \emptyset) : \theta \in \Theta_{I,0}\right\}\right]$$

*is consistent, in both the sense of Definition 6 and Definition 7. Under the conditions used in Parts 4.4 and 4.5 of Theorem 4, the corresponding posterior distribution for*

$$\left[\inf\left\{E_{P_{\tilde{u}\sim\tilde{F}_\theta(\tilde{u})}(\mathcal{T}^{co}(\tilde{\tilde{u}}))}\left(\inf_{t \in \mathcal{T}^{co}(\tilde{\tilde{u}})} \omega(t)\right) : \theta \in \Theta_{I,0}\right\}, \sup\left\{E_{P_{\tilde{u}\sim\tilde{F}_\theta(\tilde{u})}(\mathcal{T}^{co}(\tilde{\tilde{u}}))}\left(\sup_{t \in \mathcal{T}^{co}(\tilde{\tilde{u}})} \omega(t)\right) : \theta \in \Theta_{I,0}\right\}\right]$$

*is consistent, in both the sense of Definition 6 and Definition 7.*

The general asymptotic theory developed in Section 6 is sufficiently general to cover other settings that involve mapping an estimated set to some other object of interest.

**Remark 2** (With a posterior over $\theta$)**.** We can accommodate a setup with a posterior for $\theta$, since it is mechanically the special case of the above analysis where $\Theta_{I,0}$ is a singleton.

**Remark 3** (The conditions in Theorem 4). Parts 4.2 and 4.3 require conditions on $B$ related to how weak convergence of ordinary random variables concerns convergence of CDFs at points of continuity of the limiting distribution.[16] Parts 4.4 and 4.5 require a sort of "uniform integrability" condition.[17] The simplest sufficient condition for this condition is that $\mathcal{T}(\tilde{\tilde{u}})$ is uniformly bounded across $\tilde{\tilde{u}}$.[18] $\mathcal{T}(\tilde{\tilde{u}})$ is uniformly bounded across $\tilde{\tilde{u}}$ in any situation where $\mathcal{T} = \mathcal{S}$, since $\mathcal{S}(\tilde{\tilde{u}})$ is a set of distributions. This case is relevant for market entry probabilities or the expected number of entrants from the running Example 1. The outcome of interest can also be plausibly bounded in other cases. For example, welfare outcomes can be bounded under the assumption that the supports of the distribution of utility functions all share some common bound. The "uniform integrability" also accommodates situations with unbounded outcomes of interest. Concretely, Appendix D shows this condition holds for welfare outcomes (e.g., the profits of the firms) when $\omega(t) = t$, as long as $E_{\tilde{\tilde{u}} \sim \tilde{F}_\theta(\tilde{u})}(|\tilde{\tilde{u}}_i(a)|^{1+\epsilon})$ depends continuously on $\theta$ for some $\epsilon > 0$, with the assumption that $\Theta$ is compact.

## 5. Computational implementation

The computational steps to generate a numerical approximation to the posterior CPDS follows directly from the construction of the CPDSes:

(1) For a grid of values of $\tilde{\tilde{u}}$, pre-compute the relevant quantity for the counterfactual analysis: e.g., $\mathcal{T}^{\text{co}}(\tilde{\tilde{u}})$, or $\sup_{t \in \mathcal{T}^{\text{co}}(\tilde{\tilde{u}})} \omega(t)$, or $\inf_{t \in \mathcal{T}^{\text{co}}(\tilde{\tilde{u}})} \omega(t)$. Examples of closed-form expressions of these quantities are provided in Example 1, but in general this step can be done purely computationally. For example, when considering a counterfactual involving market entry probabilities as the outcome of interest (e.g., the example following Equations 7 and 8), this step would involve computing the maximum/minimum probability of those market outcomes, at each value of $\tilde{\tilde{u}}$ in a grid of values of $\tilde{\tilde{u}}$. There can be a range, even for a given $\tilde{\tilde{u}}$, because of model incompleteness.

---

[16]The condition is that there is probability 0 of $\tilde{\tilde{u}}$ that has a set of counterfactual outcomes that has non-empty intersection with bd $B$ but empty intersection with int $B$.

[17]The condition involves the max to correctly account for cases where $\mathcal{T}(\tilde{\tilde{u}}) \cap K_\alpha^C = \emptyset$.

[18]If so, $\alpha$ can be selected to be large enough that $\mathcal{T}(\tilde{\tilde{u}}) \cap K_\alpha^C = \emptyset$ for all $\tilde{\tilde{u}}$, so the limit is obviously 0.

The existing literature on computational game theory (e.g., von Stengel (2007)) has results that can be applied. Note that for many counterfactual analyses, it is not necessary to explicitly compute $\mathcal{S}(\tilde{\tilde{u}})$. For example, in some cases, it suffices to be able to conduct an optimisation over the space of solutions, with $\mathcal{S}(\tilde{\tilde{u}})$ as constraints, as in Equations 7 and 8. In some cases, some preliminary "pen and paper" calculations can increase computational speed. For example, in binary games, there are known expressions for the set of Nash equilibria as a mapping of the utility functions.

(2) For a grid of values of $\theta$, generate a sample of $\tilde{\tilde{u}}$ from $\tilde{F}_\theta(\tilde{\tilde{u}})$, and for each draw of $\tilde{\tilde{u}}$, look up the value from the previous step. Some graphical examples are provided in Appendix C. This results in $P_{\tilde{\tilde{u}} \sim \tilde{F}_\theta(\tilde{u})}(\mathcal{T}^{\text{co}}(\tilde{\tilde{u}}))$, or $E_{P_{\tilde{\tilde{u}} \sim \tilde{F}_\theta(\tilde{u})}(\mathcal{T}^{\text{co}}(\tilde{u}))}\left(\sup_{t \in \mathcal{T}^{\text{co}}(\tilde{u})} \omega(t)\right)$, or $E_{P_{\tilde{\tilde{u}} \sim \tilde{F}_\theta(\tilde{u})}(\mathcal{T}^{\text{co}}(\tilde{u}))}\left(\inf_{t \in \mathcal{T}^{\text{co}}(\tilde{u})} \omega(t)\right)$ for a grid of values of $\theta$.

(3) Draw many $\Theta_I^j$ from $\Pi^{(N)}(\Theta_I)$, following the existing literature. For each draw $\Theta_I^j$, using the pre-computed values from the previous step, construct the corresponding value of the identified set for the counterfactuals, based on Equation 12. This results in a numerical approximation to the posterior CPDS.

(4) Report summaries of the posterior CPDS as desired.

The computations are suitable for speedup by parallelism. Particularly, we have found significant speed-ups from a standard consumer-grade GPU-based computational implementation, although computation with a standard consumer-grade CPU can be quite fast.

## 6. General asymptotic theory

This section develops the asymptotic theory that leads up to Theorem 4. These results are sufficiently general to apply to other statistical problems involving mappings of an estimated set. The setup in this section starts with a posterior distribution for a set $A_0$. In our case, $A_0$ is the identified set $\Theta_{I,0}$ for the underlying model parameters $\theta$.

Section 6.1 establishes equivalences between different definitions of consistency of a posterior distribution for a set. This section allows for infinite-dimensional objects, as in an identified set for a semi-parametric/non-parametric model. This section has two consequences. First, alternative characterizations of consistency are useful in proofs. Second, this section establishes that multiple existing proposals for a posterior distribution for $\Theta_{I,0}$ are compatible with the assumptions we require in our results about CPDSes.

Section 6.2 establishes conditions under which consistency of the posterior distribution for $A_0$ implies consistency of the posterior distribution for $f(A_0) = \bigcup_{a \in A_0} f(a)$, where $f(\cdot)$ maps from elements of $A_0$ to some other space. The results cover cases where $f(\cdot)$ does not map to a Euclidean space; thus, $f(a)$ is *not* restricted to be a real-valued ("ordinary") function of the argument $a$. The results also cover cases where the domain of $f(\cdot)$ is not a Euclidean space. Oversimplifying, in the application to CPDSes, $f(\cdot)$ can be the mapping between a specification of the underlying model parameter and the CPDSes (e.g., Equation 4), though other specifications of $f(\cdot)$ are also important for CPDSes, as discussed in Section 6.2. A key condition is that $f(\cdot)$ has suitable continuity properties.

Section 6.3 establishes that the relevant CPDS mappings are suitably continuous. This addresses the issue of discontinuities discussed in the context of Assumption 3.

6.1. **Consistency of posterior distributions for sets.** This section provides general equivalences between different definitions of consistency of a posterior distribution for a set $A_0$. As an unknown, the set is $A$. $A$ and $A_0$ are sets of elements from a topological space $S$.

We assume that a posterior distribution exists and use the notation $\Pi^{(N)}(A)$ to denote that posterior distribution from a sample of size $N$. Generally, $\Pi^{(N)}(A) = \Pi(A|W^{(N)})$ where $W^{(N)}$ represents a realization of observed data from a sample of size $N$. However, we omit the explicit dependence on the data.[19]

**Remark 4** (Sense of convergence). Essentially, all the equivalence results in this section also hold (with effectively the same proofs, just with a different sense of limit used everywhere) if the limit notation $\to$ is understood as a limit almost surely with respect to $P_0^{(\infty)}$. In that case, $\Pi^{(N)}(E) \to c$ for some event $E$ and some constant $c$ corresponds to the statement that: $P_0^{(\infty)}(\lim_{N \to \infty} \Pi^{(N)}(E) = c) = 1$. It is standard in Bayesian asymptotics (as with classical asymptotics) to have parallel results for consistency "in probability" and consistency "almost surely". See for instance Ghosal and Van der Vaart (2017, Chapter 6). See in particular the proof of Lemma 12 for the details of this issue.

---

[19]We do not need to be specific about the data generating process for $W^{(N)}$ or how that data influences the posterior distribution. It might be that $W^{(N)}$ is an i.i.d. sample of realizations of $W_i$ for $i = 1, 2, \ldots, N$, but it does not need to be. The $\Pi^{(N)}$ notation represents a random object, in the same way that notation like $\hat{\theta}_{(N)}$ for a classical estimator represents a random object, even without explicit dependence on "the data."

There are equivalences between different ways to define posterior consistency. These equivalences are useful in proofs, and are useful to relate existing notions of consistency from the existing literature. We provide further results in Appendix A.

**Theorem 16** (Equivalences between definitions of consistency). *Suppose the space $S$ is a normal Hausdorff topological space (i.e., a $T_4$ space). Suppose $A_0$ is closed. The following two conditions are equivalent:*

(16.1) *The posterior for $A$ satisfies the condition that $\Pi^{(N)}(A \subseteq O) \to 1$ if $O$ is an open set such that $A_0 \subseteq O$. The posterior for $A$ satisfies the condition that $\Pi^{(N)}(x \in A) \to 1$ if $x \in int\ A_0$. And, $\Pi^{(N)}(A \neq \emptyset) \to 1$ if $A_0 \neq \emptyset$.*

(16.2) *The posterior for $A$ satisfies the condition that $\Pi^{(N)}(A \cap B \neq \emptyset) \to 1[A_0 \cap B \neq \emptyset]$ for any closed $B$ such that $bd\ A_0 \cap B = \emptyset$. And, $\Pi^{(N)}(A \neq \emptyset) \to 1$ if $A_0 \neq \emptyset$.*

*Further, suppose the space $S$ is Euclidean. Suppose $A_0$ either is a singleton or a convex body (i.e., a compact convex set with non-empty interior). Suppose the posterior for $A_0$ has convex values and non-empty compact values with posterior probability 1. The following condition is equivalent to either of the previous two conditions:*

(16.3) $\Pi^{(N)}(d_H(A, A_0) > \epsilon) \to 0$ *for any $\epsilon > 0$, where $d_H$ is the usual Hausdorff distance.*

Condition 16.1 requires the space $S$ is a normal Hausdorff topological space (i.e., a $T_4$ space). This includes all metrizable spaces. This way of defining consistency says that the posterior probability of two kinds of true statements approach 1 in large samples. In turn, note that Condition 16.1 is equivalent to the standard definition of posterior consistency in the special case that $A_0$ is a singleton and $A$ is a singleton with posterior probability 1. Condition 16.2 is Definition 6. Condition 16.3 is Definition 7, and requires the space is Euclidean and convexity of the sets.

Various approaches to Bayesian inference in partially identified models can be related to these conditions. For instance, Kline and Tamer (2016) establish their posterior for $\Theta_{I,0}$ satisfies condition 16.1 and Giacomini and Kitagawa (2021, Theorem 3) establish their posterior for $\Theta_{I,0}$ satisfies condition 16.3.

6.2. **Consistency of a mapping.** We give conditions under which consistency of the posterior distribution for $A_0$ implies consistency of the posterior distribution for $f(A_0)$, where $f(\cdot)$ is a mapping. As before, the ambient space for $A_0$ is a topological space $S$. Before giving the results, we describe some example applications of the results.

In some applications of the results, $S$ is the parameter space $\Theta$, and $A_0$ is the identified set $\Theta_{I,0}$ for the underlying model parameters $\theta$. And, $f(\cdot)$ is the mapping with domain $\Theta$, from a value of $\theta$ to the counterfactual prediction, per Equation 4 and the specific examples that follow Equation 4. For example, $f(\theta) = P_{\tilde{\tilde{u}} \sim \tilde{F}_\theta(\tilde{u})}(\mathcal{T}^{\text{co}}(\tilde{\tilde{u}}))$ is the partial CPDS for a given value of $\theta$. In this case, $f(\cdot)$ maps to the *space of distributions of random sets*. Alternatively, for example, $f(\theta) \equiv P_{\tilde{\tilde{u}} \sim \tilde{F}_\theta(\tilde{u})}(\mathcal{T}^{\text{co}}(\tilde{\tilde{u}}) \cap B \neq \emptyset)$ is the counterfactual probability that some value in $B$ *could* arise as the outcome, for a given value of $\theta$, from Definition 2.

In other applications of the results, $S$ is a space of *sets*, and $f(\cdot)$ is a mapping with domain $S$. In this case, $A$ is not viewed as a set of elements; rather, $A$ is viewed as a singleton in the space of sets. Thus, $f(A)$ does not map from "elements" of $A$; rather, $f(A)$ is a mapping of the entire set $A$. One example of this application concerns $f_{sup}(A) = \sup A$ and $f_{inf}(A) = \inf A$ with $A \subseteq \mathbb{R}$, as in Equation 12 and the specific examples that follow Equation 12.[20] Another example of this application concerns the interval of possible average values of a function $\omega$ of the outcome of interest, from Definition 3:

$$f(\Theta_I) = \left[ \inf \left\{ E_{P_{\tilde{\tilde{u}} \sim \tilde{F}_\theta(\tilde{u})}(\mathcal{T}^{\text{co}}(\tilde{\tilde{u}}))} \left( \inf_{t \in \mathcal{T}^{\text{co}}(\tilde{\tilde{u}})} \omega(t) \right) : \theta \in \Theta_I \right\}, \sup \left\{ E_{P_{\tilde{\tilde{u}} \sim \tilde{F}_\theta(\tilde{u})}(\mathcal{T}^{\text{co}}(\tilde{\tilde{u}}))} \left( \sup_{t \in \mathcal{T}^{\text{co}}(\tilde{\tilde{u}})} \omega(t) \right) : \theta \in \Theta_I \right\} \right].$$

And, it can be useful to *iteratively* apply the results. For example, we can start with a posterior distribution for a set $A_0$, and establish consistency for the posterior distribution of $f(A_0)$. Then, we can establish consistency for the posterior distribution of $g(f(A_0))$, viewing $g(\cdot)$ as a function on the space of *sets* relevant for the co-domain of $f(\cdot)$. See for example Lemma 8 and Corollary 9 for results that use that iterative logic.

---

[20]These specifications of $f(\cdot)$ are continuous when $S$ are non-empty compact sets of real numbers with the Hausdorff metric. It suffices to prove sequential continuity in this metric space. Suppose $A_n$ converges to $A$ in Hausdorff metric. Then, for any $\epsilon > 0$, $d_H(A_n, A) \leq \frac{\epsilon}{2}$ implies that $A_n \subseteq A + \epsilon \mathcal{B}$. Since $\sup A_n \in A_n$, it follows that $\sup A_n = a + b$ where $a \in A$ and $b \in \epsilon \mathcal{B}$. Thus, $\sup A_n \leq \sup A + \epsilon$. Also, $A \subseteq A_n + \epsilon \mathcal{B}$, so $\sup A \leq \sup A_n + \epsilon$. Thus, $\sup A_n \to \sup A$. The same proof applies to $f_{inf}$.

The first result allows that $S$ is any $T_1$ space. The restriction to a $T_1$ space requires only that distinct points in $S$ can be separated; basically any space that is used in practice is a $T_1$ space. This includes cases where $A_0$ is an identified set for an infinite-dimensional parameter. Also, this includes cases where $S$ is a space of *sets* as discussed above. For this first result, $f(\cdot)$ must map to a Euclidean space. In some applications, this is enough; in other applications, this is a building block in an iterative application of these results.

**Theorem 6.** *Suppose $S$ is a $T_1$ space, and suppose $A_0 \subseteq S$. Suppose the posterior distribution for $A_0$ satisfies Part 6.1 of Definition 6. Suppose $f(\cdot) : S \to \mathbb{R}^d$ is a continuous function. Suppose the posterior distribution for $A_0$ is such that $f(A)$ is convex with posterior probability 1. Suppose $A_0$ is either a singleton or regular closed (i.e., $A_0 = \operatorname{cl}(\operatorname{int} A_0)$).[21] Then $\Pi^{(N)}(f(A) \cap B \neq \emptyset) \to 1[f(A_0) \cap B \neq \emptyset]$ for any closed set $B$ such that $\operatorname{bd} f(A_0) \cap B = \emptyset$. Further suppose the posterior distribution for $A_0$ satisfies Part 6.2 of Definition 6. Then, the posterior distribution for $f(A_0)$ also satisfies Part 6.2 of Definition 6.*

This result requires continuity of $f(\cdot)$.

This result also requires that the posterior for $A_0$ is such that $f(A)$ is convex with posterior probability 1. In a setting with singleton-valued $A$, this convexity assumption is vacuously true. An important but subtle point is that "singleton-valued" must be understood relative to the ambient space $S$. The "obvious" example of "singleton-valued" is when $S$ is the parameter space $\Theta$, and $A_0$ is the single-element identified set for a point identified $\theta$. Another important case is when $S$ is a space of *sets*: in that case, $A_0$ could correspond to the identified set, which is a singleton *in the space of sets.*

Even if $A$ is not singleton-valued, convexity is only a mild restriction when $f(\cdot)$ is scalar-valued (i.e., $d = 1$), because then convexity would follow from the assumption that the posterior distribution for $A_0$ is such that $A$ is connected with posterior probability 1. That holds because continuous functions map connected sets to connected sets, and connected sets in $\mathbb{R}$ are the convex sets in $\mathbb{R}$.

In turn, note that convex sets in a Euclidean space are connected sets, and it is known that many identified sets for underlying model parameters are convex sets. In those cases with a convex $A_0$, the condition that the posterior for $A_0$ is such that $A$ is connected with posterior probability 1 requires only that the posterior for $A_0$ respects the convexity of $A_0$. Even without this convexity, it is possible to "enlarge" the set by taking a convex hull, thereby returning to a setting with convexity, albeit for a larger set. We cannot expect to be able to eliminate the convexity assumption, as the example in the footnote illustrates.[22]

---

[21] A closed set is not necessarily regular closed. In a $T_1$ space, every singleton set is closed. Thus, $A_0$ is closed under these conditions.

The second result restricts to settings where $S$ is a connected metric space. This is a stronger assumption than in the previous result, but still allows for infinite-dimensional unknowns when $A_0$ is an identified set. It also allows that $S$ is a space of non-empty compact subsets of some metric space, using Hausdorff metric. It allows that $f(\cdot)$ maps to any metric space. In particular, this allows that $f(\cdot)$ maps *to* (and perhaps *from*) a non-empty compact subset of a metric space (e.g., using the Hausdorff metric), or to a distribution of a random set of ordinary Euclidean objects (e.g., using the Lévy metric). This accommodates the case that $f(\theta) = P_{\tilde{u} \sim \tilde{F}_\theta(\tilde{u})}(\mathcal{T}^{\text{co}}(\tilde{u}))$. It also eliminates the convexity requirement from the previous result. Finally, unlike the previous result, this second result requires (and establishes) consistency in Hausdorff distance.

**Theorem 7.** *Suppose $S$ is a connected metric space, and suppose $A_0 \subseteq S$. Suppose the posterior for $A_0$ is consistent in Hausdorff distance, in the sense that $\Pi^{(N)}(d_H(A, A_0) > \epsilon) \to 0$ for any $\epsilon > 0$. Suppose $f(\cdot)$ is a continuous function from $S$ to some metric space. Suppose $A_0 \neq \emptyset$ and $A_0^C \neq \emptyset$. Suppose $A_0$ is compact. Suppose the posterior for $A_0$ respects those properties with posterior probability 1. Then $\Pi^{(N)}(d_H(f(A), f(A_0)) > \delta) \to 0$ for any $\delta > 0$.*

As illustrated by Corollary 5 of Theorem 4, a particular way to iteratively apply the results is based on the following.

---

[22]Suppose the posterior distribution for $A_0$ is degenerate, with $A_N = [\frac{1}{N}, 1] \times [\frac{1}{N}, 2\pi - \frac{1}{N}]$. This is consistent for $A_0 = [0, 1] \times [0, 2\pi]$, using Theorem 16, and the fact that every interior point of $A_0$ is contained in $A_N$ for large enough $N$. Let $f(\theta_1, \theta_2) = (\theta_1 \cos \theta_2, \theta_1 \sin \theta_2)$, which via the standard polar transform, is the point that is radius $\theta_1$ from the origin, at the angle $\theta_2$. Then, $f(A_0)$ is the unit disk (i.e., all points contained in a circle of radius 1). However, $f(A_N)$ is only a "ribbon" of points that are of distance between $\frac{1}{N}$ and 1 from the origin, and also of angle between $\frac{1}{N}$ and $2\pi - \frac{1}{N}$. In particular, regardless of sample size, $f(A_N)$ does not contain any point of the form $(x, 0)$ for $x \in [0, 1)$ because such points are at angle 0, but all such points are in the interior of $f(A_0)$. Thus, again using Theorem 16, $f(\Theta_N)$ is not consistent for $f(\Theta_0)$. However, as Theorem 7 also establishes, we should not expect the "magnitude of failure" of consistency to be too severe without convexity, since consistency in Hausdorff distance can hold even without convexity.

**Lemma 8.** *Suppose $S$ is a topological space, and $A_0 \subseteq S$. Suppose $A_0 \neq \emptyset$ and the posterior distribution for $A_0$ respects this with posterior probability $1$. Consider functions $f_L : 2^S \setminus \emptyset \to \mathbb{R}$ and $f_U : 2^S \setminus \emptyset \to \mathbb{R}$ such that $f_L(B) \leq f_U(B)$ for all $\emptyset \neq B \subseteq S$. Consider $f$ given by $f(B) = [f_L(B), f_U(B)] \subseteq \mathbb{R}$. Suppose the posterior distributions $L = f_L(A)$ and $U = f_U(A)$, for $L_0 = f_L(A_0)$ and $U_0 = f_U(A_0)$ respectively, are consistent in the sense of Definition 6. Then, the posterior distribution $f(A)$ for $f(A_0)$ has non-empty realizations with posterior probability $1$, and is consistent in both the sense of Definition 6 and Definition 7.*

There are many applications of these results to our CPDSes. One example follows:

**Corollary 9.** *The following holds:*

(9.1) *Suppose $\Theta$ is a $T_1$ space. Suppose $\Theta_{I,0}$ is connected, suppose $\Theta_{I,0}$ is either a singleton or regular closed, and suppose the posterior distribution for $\Theta_{I,0}$ respects these properties with posterior probability $1$. Suppose $f(\cdot)$ is a continuous function from $\Theta$ to $\mathbb{R}$. Suppose the posterior distribution for $\Theta_{I,0}$ is consistent per Definition 6. Then, the posterior distribution for $f(\Theta_{I,0})$ is consistent per Definition 6.*

(9.2) *Suppose the conditions from Part 9.1, and suppose further $\Theta_{I,0}$ is non-empty and compact and the posterior distribution for $\Theta_{I,0}$ respects this property with posterior probability $1$, and $g(\cdot)$ is a continuous function from non-empty compact sets of $\mathbb{R}$ to $\mathbb{R}$, then the posterior distribution for $g(f(\Theta_{I,0}))$ is consistent per Definition 6.*

6.3. **Continuity of counterfactuals.** This section establishes that the mappings that are relevant for our CPDSes are continuous in the sense required by the results in Section 6.2. To state the result, use this notation: For any set $B$, let $\mathcal{T}^{-1}(B) = \{\tilde{\tilde{u}} : \mathcal{T}(\tilde{\tilde{u}}) \cap B \neq \emptyset\}$.

**Theorem 10** (Limiting properties of the CPDS mappings). *Suppose $t(u, s)$ is a continuous function of a vector of utility functions $u$ and a solution $s$, and let $\mathcal{T}(\tilde{\tilde{u}}) = \{t(\tilde{\tilde{u}}, s) : s \in \mathcal{S}(\tilde{\tilde{u}})\}$. Under Assumptions 1, 2, and 3, if $\tilde{\tilde{u}}^{(\nu)} \sim F^{(\nu)}$ is a sequence of distributions of utility functions that converge weakly to the distribution of $\tilde{\tilde{u}} \sim F$, then:*

(10.1) *For any set $B$ such that $\mathcal{T}^{-1}(B)$ is a Borel set, which includes all open/closed sets $B$, $P(\mathcal{T}(\tilde{\tilde{u}}) \cap \text{int } B \neq \emptyset) - P(\tilde{\tilde{u}} \in \mathcal{U}^e) \leq \liminf_{\nu \to \infty} P(\mathcal{T}(\tilde{\tilde{u}}^{(\nu)}) \cap B \neq \emptyset) \leq \limsup_{\nu \to \infty} P(\mathcal{T}(\tilde{\tilde{u}}^{(\nu)}) \cap B \neq \emptyset) \leq P(\mathcal{T}(\tilde{\tilde{u}}) \cap \text{cl } B \neq \emptyset)$.*

(10.2) If $P(\tilde{\tilde{u}} \in \mathcal{U}^e) = 0$, then $\mathcal{T}(\tilde{\tilde{u}}^{(\nu)})$ converges weakly to $\mathcal{T}(\tilde{\tilde{u}})$ as random closed sets.

(10.3) For any set $B$ such that $\mathcal{T}^{-1}(B)$ is a Borel set, which includes all open/closed sets $B$, and such that $P(\tilde{\tilde{u}} \in \text{bd } \mathcal{T}^{-1}(B)) = 0$, $P(\mathcal{T}(\tilde{\tilde{u}}^{(\nu)}) \cap B \neq \emptyset) \to P(\mathcal{T}(\tilde{\tilde{u}}) \cap B \neq \emptyset)$. Further, if $B = \cup B_j$ where $B_1, \ldots, B_k$ are such that $\mathcal{T}^{-1}(B_j)$ is a closed convex set, and $\tilde{\tilde{u}}$ has an ordinary density with respect to Lebesgue measure, then $P(\mathcal{T}(\tilde{\tilde{u}}^{(\nu)}) \cap B \neq \emptyset) \to P(\mathcal{T}(\tilde{\tilde{u}}) \cap B \neq \emptyset)$.

(10.4) For any set $B$ such that $\mathcal{T}^{-1}(B^C)$ is a Borel set, which includes all open/closed sets $B$, $P(\mathcal{T}(\tilde{\tilde{u}}) \subseteq \text{int } B) \leq \liminf_{\nu \to \infty} P(\mathcal{T}(\tilde{\tilde{u}}^{(\nu)}) \subseteq B) \leq \limsup_{\nu \to \infty} P(\mathcal{T}(\tilde{\tilde{u}}^{(\nu)}) \subseteq B) \leq P(\mathcal{T}(\tilde{\tilde{u}}) \subseteq \text{cl } B) + P(\tilde{\tilde{u}} \in \mathcal{U}^e)$.

(10.5) For any set $B$ such that $\mathcal{T}^{-1}(B^C)$ is a Borel set, which includes all open/closed sets $B$, and such that $P(\tilde{\tilde{u}} \in \text{bd } \mathcal{T}^{-1}(B^C)) = 0$, $P(\mathcal{T}(\tilde{\tilde{u}}^{(\nu)}) \subseteq B) \to P(\mathcal{T}(\tilde{\tilde{u}}) \subseteq B)$.

(10.6) If $P(\tilde{\tilde{u}} \in \mathcal{U}^e) = 0$, then for any continuous function $f(\cdot)$ with compact support, $E \sup f(\mathcal{T}(\tilde{\tilde{u}}^{(\nu)})) \to E \sup f(\mathcal{T}(\tilde{\tilde{u}}))$ and $E \inf f(\mathcal{T}(\tilde{\tilde{u}}^{(\nu)})) \to E \inf f(\mathcal{T}(\tilde{\tilde{u}}))$.

(10.7) If $P(\tilde{\tilde{u}} \in \mathcal{U}^e) = 0$, then for any continuous function $f(\cdot)$ such that $\lim_{\alpha \to \infty} E \max\{\sup_{t \in \mathcal{T}(\tilde{\tilde{u}}) \cap K_\alpha^C} |f(t)|, 0\} = 0$ and $\lim_{\alpha \to \infty} \sup_\nu E \max\{\sup_{t \in \mathcal{T}(\tilde{\tilde{u}}^{(\nu)}) \cap K_\alpha^C} |f(t)|, 0\} = 0$, where $K_\alpha$ is the closed unit $L_1$-ball with radius $\alpha$, $E \sup f(\mathcal{T}(\tilde{\tilde{u}}^{(\nu)})) \to E \sup f(\mathcal{T}(\tilde{\tilde{u}}))$ and $E \inf f(\mathcal{T}(\tilde{\tilde{u}}^{(\nu)})) \to E \inf f(\mathcal{T}(\tilde{\tilde{u}}))$.

Also, the same result obtains with $\mathcal{T}^{co}$ in place of $\mathcal{T}$.

With regards to Part 10.3, note for example that the set of utility functions that support any particular strategy as a (Nash or correlated) equilibrium is indeed a closed convex set, as a direct consequence of the characterization in terms of inequality constraints.

**Lemma 11.** *Suppose the distribution of $(x, \epsilon)$ is a sequentially continuous function of $\theta$, in the sense that the sequence of distributions $\tilde{F}_{\theta^{(\nu)}}(x, \epsilon)$ converges weakly to $\tilde{F}_\theta(x, \epsilon)$ whenever $\theta^{(\nu)}$ is a sequence of parameter values that converges to $\theta$. Suppose the counterfactual utility functions $\tilde{u}_i(a_i, a_{-i}, x_i, \epsilon_i, \theta)$ are jointly continuous in $(x, \epsilon, \theta)$ for each $(a_i, a_{-i})$. Let $\tilde{\tilde{u}}^{(\nu)}$ be the distribution of $\tilde{u}(\cdot, \cdot, x^{(\nu)}, \epsilon^{(\nu)}, \theta^{(\nu)})$ with $(x^{(\nu)}, \epsilon^{(\nu)}) \sim \tilde{F}_{\theta^{(\nu)}}(x, \epsilon)$ and let $\tilde{\tilde{u}}$ be the distribution of $\tilde{u}(\cdot, \cdot, x, \epsilon, \theta)$ with $(x, \epsilon) \sim \tilde{F}_\theta(x, \epsilon)$, for a sequence $\theta^{(\nu)} \to \theta$. Then, $\tilde{\tilde{u}}^{(\nu)}$ converges weakly to $\tilde{\tilde{u}}$.*

Lemma 11 establishes under suitable continuity properties of the primitives that the induced distribution of $\tilde{\tilde{u}}$ corresponding to a sequence $\theta^{(\nu)}$ converges weakly to the distribution of $\tilde{\tilde{u}}$ corresponding to the limit $\theta$. Then, Theorem 10 can be applied. Overall, this results is the required continuity used to prove that the posterior CPDS is consistent.

7. Empirical application

7.1. **Background.** We apply our methods to the market-level entry decisions of air carriers in the U.S. domestic airline industry. This is an important setting for empirical industrial organization, including in Reiss and Spiller (1989), Berry (1990, 1992), Borenstein (1992), Ciliberto and Tamer (2009), Berry and Jia (2010), Ciliberto and Williams (2014) and Ciliberto, Murry, and Tamer (2021).

Our analysis builds directly on the setup from Kline and Tamer (2016). We specifically follow specification 3 of Kline and Tamer (2016), which involves the most explanatory variables of any specification considered there. Applying the methods of Kline and Tamer (2016) gives us the posterior distribution for the identified set for the utility function parameters.

Overall, the setup is standard relative to the related literature. Each "market" is defined as a route between a pair of airports, irrespective of intermediate stops. For the purposes of defining the DMs in the game, we aggregate air carriers to two kinds of air carriers: low cost carriers (LCC), and all other airlines (OA). The low cost carriers are defined as in the ICAO's "List of Low-Cost-Carriers (LCCs)."[23] Perhaps most notably, the set of LCCs includes Allegiant, Frontier, Jet Blue, Southwest, Spirit, Sun Country, and Virgin. Thus, we consider a two-player binary-action entry game. The utility functions have the linear functional form described in Equation 1. We use[24] the data from the fourth quarter 2023 Airline Origin and Destination Survey (DB1B).

The model involves two literature-standard explanatory variables. The first explanatory variable is market size, defined by the geometric mean of the populations of the two cities associated with the two airports in the market. We use Census Metropolitan and Micropolitan Statistical Areas Population Totals for 2023. The second explanatory variable is market presence (e.g., Berry (1992)), defined for a given DM-airport as the ratio of the number of airports served by that DM from that airport to the total number of airports served by any DM from that airport. Market presence for a given DM-market is the average of that DM's market presence at the two airports in the market. Specifically, we construct market presence based on flights between the 100 busiest airports, according to the DB1B data. We conduct the counterfactual analysis based on "low/mid-sized" markets, which we define to be those between the 25% and 50% quantile of market size, relative to all markets involving the 100 busiest airports. As in other similar applications, we discretize the explanatory variables; specifically, following the setup from Kline and Tamer (2016), we consider binary explanatory variables indicating above/below the median value.

---

[23]Available online: https://www.icao.int/sustainability/documents/lcc-list.pdf

[24]This is a similar data set as the one used in Kline and Tamer (2016) except that we obtain the version corresponding to the year 2023.

The underlying model parameters are estimated based on the assumption of pure strategy Nash equilibrium in the data; in our counterfactuals, we use both the solution concept of correlated equilibrium and pure strategy Nash equilibrium.

7.2. **Counterfactual analysis.** We experiment with five different counterfactuals. The first is the "real world" counterfactual that actually does not change anything about the environment. In many standard econometrics models, this would be expected to result in the same outcomes as observed in the data. However, as a consequence of incompleteness and partial identification, this is not the case in our setting. This baseline "counterfactual" concretely illustrates the empirical magnitude of this important feature of the model (mainly incompleteness and partial identification), even before turning to the question of changing the environment.

We then consider a variety of changes to the environment. The second and third counterfactual involves shutting down the entry of one of the two DMs. This turns the model into the decision of a single DM. The fourth and fifth involves increasing the profitability of entry for LCC and OA, respectively, without any changes for the other DM. Specifically, this involves increasing the profitability of entry by exactly the same as the upper bound of the estimated identified set for the impact on profits from entry of being in an above median-size market (defined relative to the baseline restriction to "low/mid-sized" markets). This can be viewed as an experiment into the results of the government subsidizing the entry of carriers. These counterfactuals help us understand the different determinants of outcomes, by investigating what happens when certain parts of the environment changes.

7.2.1. *Aggregate outcomes.* For each counterfactual, we consider a variety of outcomes. Table 3 shows the outcomes for the probability of various market outcomes, displayed in the columns. In the interest of space, we do not explicitly report the probability that at least one DM enters the market, since that is one minus the probability that no DM enters the market, which we do report. Table 4 shows the outcomes for the average number of entrants and average profits. These tables concern the counterfactual questions in Definition 3. As with many other applications, the scale of profits is not identified in the underlying analysis of $\theta$.[25] Consequently, as with similar applications, we can only consider relative profits across different counterfactuals. For sake of interpretation, we re-scale profits, by choosing the scale for each DM to be such that the maximum of the identified set for each DM's profits under the "shut down (other DM)" counterfactual is 1.[26]

We display the results based on the correlated equilibrium (CE) solution concept. We also computed results for the pure strategy Nash equilibrium (PSNE), and found that they are basically the same (up to one decimal place, if not exactly equally).

In Tables 3 and 4, for each cell (e.g., the probability that no DM enters in the "real world" counterfactual in the case of the upper-left cell of Table 3), the top most interval is the estimated identified set, and the second interval is the 95% credible set. We note that in some cases, the estimated identified set is the same (or nearly the same) as the 95% credible set. This makes sense as a general feature of this kind of analysis involving partial identification.[27]

Specifically, for the estimated identified set, we report the average of the posterior distribution for that identified set. Since each posterior distribution concerns an interval identified set, this boils down to reporting the average of the upper/lower bound of the interval identified set under the posterior distribution. For the 95% credible set, we find the narrowest 95% of draws from the posterior distribution, and then find the smallest interval that contains all of those intervals (e.g., the lower bound of the 95% credible set is the smallest lower bound among these narrowest 95% of draws). The bottom row in the Tables provide the *observed* frequency of the corresponding outcome. So in Table 3, the proportion of the markets in the data that are not served by any airline is 16.1.

---

[25] This concerns identification of the underlying model parameter, and thus is not directly the topic for this paper. Nevertheless, it is worth briefly mentioning that, in applications where "utility" is actually "profits," it is obviously possible to identify the scale of profits with adequate data. For example, an observable explanatory variable that contributes to costs by a known scale factor will have a known corresponding slope parameter.

[26] This profit level is also equal to the maximum of the identified set for each DM's profits in a counterfactual where there are no competitive effects.

[27] In particular, this can happen when the estimated minimum/maximum outcome (e.g., minimum/maximum probability that no DM enters) corresponds to a certain value of the underlying parameter $\theta^*$ that is contained within all realizations from the posterior distribution of the the identified set for $\theta$. In such a case, even though there is uncertainty about the identified set for $\theta$, there is no uncertainty about the corresponding minimum/maximum outcome. This is somewhat related to the fact that it is possible to have no uncertainty about statements like $\theta \geq 0$, even though there is uncertainty about $\theta$, when $\theta$ is far away from 0, so that all draws from the posterior for (the identified set for) $\theta$ are such that $\theta \geq 0$.

|  | Probability that ... enters ||||||| 
|---|---|---|---|---|---|---|---|
| Counterfactual | no DM | exactly 1 DM | both DMs | exactly LCC | exactly OA | LCC | OA |
| Real world | [6.2, 25.9] | [32.0, 57.0] | [27.0, 52.4] | [0.0, 21.0] | [24.6, 50.5] | [34.8, 61.0] | [61.3, 89.3] |
|  | [5.4, 26.1] | [31.5, 57.1] | [26.9, 52.6] | [0.0, 21.2] | [24.6, 50.6] | [34.3, 61.1] | [60.9, 90.9] |
| Shut down OA | [23.6, 64.0] | [36.0, 76.4] | [0.0, 0.0] | [36.0, 76.4] | [0.0, 0.0] | [36.0, 76.4] | [0.0, 0.0] |
|  | [19.1, 64.5] | [35.5, 80.9] | [0.0, 0.0] | [35.5, 80.9] | [0.0, 0.0] | [35.5, 80.9] | [0.0, 0.0] |
| Shut down LCC | [9.3, 34.7] | [65.3, 90.7] | [0.0, 0.0] | [0.0, 0.0] | [65.3, 90.7] | [0.0, 0.0] | [65.3, 90.7] |
|  | [7.9, 34.9] | [65.1, 92.1] | [0.0, 0.0] | [0.0, 0.0] | [65.1, 92.1] | [0.0, 0.0] | [65.1, 92.1] |
| More profitable LCC entry | [3.0, 17.4] | [20.2, 48.1] | [40.0, 71.4] | [4.6, 30.6] | [6.7, 30.7] | [58.1, 84.5] | [58.7, 88.9] |
|  | [2.5, 17.7] | [19.9, 49.3] | [38.4, 72.0] | [4.1, 32.0] | [3.6, 32.1] | [56.5, 85.6] | [57.6, 90.1] |
| More profitable OA entry | [1.4, 12.1] | [36.6, 61.9] | [32.8, 57.8] | [0.0, 7.3] | [34.7, 61.6] | [33.3, 60.6] | [85.9, 98.2] |
|  | [1.1, 12.5] | [36.5, 62.0] | [32.8, 57.9] | [0.0, 7.5] | [34.6, 61.9] | [33.2, 60.8] | [85.5, 98.4] |
| Obs. | 16.1 | 44.5 | 39.4 | 4.0 | 40.5 | 43.5 | 79.9 |

TABLE 3. Results of counterfactual analysis using CE. For each counterfactual, the first row displays the estimated identified set for the quantity in that column, and the second row displays the 95% credible set for the quantity in that column. Shading/underlining indicates that the real world observed value (displayed in the last row) does not fall within the set.

For every interval (except those relating to profits), we shade and underline those intervals that do not contain the corresponding observed real world quantity. These outcomes are predicted to be necessarily different from the observed real world quantity. We are unable to do so for profits, given we don't observe real world profits.

Based on the "real world" counterfactual, we see that there is a considerable range of outcomes consistent with the data and our model assumptions. This is jointly a consequence of model incompleteness and partial identification of the underlying utility functions. Both imply a range of outcomes, even without changing anything about the environment. These intervals go beyond just accounting for statistical uncertainty (due to sampling). The width of the intervals reflects both partial identification of the underlying model parameters and the strategic "uncertainty" that renders the model incomplete. Such strategic uncertainty is a fundamental property of environments with multiple DMs.

| Counterfactual | Avg. entrants | Avg. profit (LCC) | Avg. profit (OA) | Avg. profit (summed) |
|---|---|---|---|---|
| Real world | [1.05, 1.43] | [0.28, 0.79] | [0.39, 0.95] | [0.78, 1.46] |
|  | [1.05, 1.45] | [0.28, 0.85] | [0.39, 1.03] | [0.78, 1.51] |
| Shut down OA | [0.36, 0.76] | [0.31, 1.00] | [0.00, 0.00] | [0.31, 1.00] |
|  | [0.36, 0.81] | [0.31, 1.09] | [0.00, 0.00] | [0.31, 1.09] |
| Shut down LCC | [0.65, 0.91] | [0.00, 0.00] | [0.44, 1.00] | [0.44, 1.00] |
|  | [0.65, 0.92] | [0.00, 0.00] | [0.44, 1.07] | [0.44, 1.07] |
| More profitable LCC entry | [1.25, 1.66] | [0.65, 1.26] | [0.36, 0.94] | [1.13, 1.92] |
|  | [1.25, 1.68] | [0.64, 1.30] | [0.36, 1.01] | [1.12, 1.95] |
| More profitable OA entry | [1.24, 1.55] | [0.27, 0.79] | [0.73, 1.34] | [1.13, 1.85] |
|  | [1.24, 1.55] | [0.26, 0.84] | [0.73, 1.40] | [1.13, 1.89] |
| Obs. | 1.23 | n.a. | n.a. | n.a. |

TABLE 4. Results of counterfactual analysis using CE. For each counterfactual, the first row displays the estimated identified set for the quantity in that column, and the second row displays the 95% credible set for the quantity in that column. Shading/underlining indicates that the real world observed value (displayed in the last row) does not fall within the set.

Based on the second and third counterfactuals, we see the relatively more important role of OA in avoiding the outcome that no DM enters a given market. Specifically, shutting down LCC seems to have have relatively modest effects on the range of the probability of the no entry outcome, either when comparing with the "real world" counterfactual range or with the actual observed probability of no entry. On the other hand, when shutting down OA, we find that the 95% credible set for the probability of the no entry outcome excludes the observed value of the probability of the no entry outcome. Thus, the substantive result from this exercise is that within this model, and when we counterfactually shut down low cost carriers, "underserved" markets continue to receive service presumably because when these LCC are out, other airlines (OA) find it profitable to serve these markets, inducing entry. The story is reversed when we shut down OA, whereby the number of markets that become not served increases as evidenced by the increase in the range of "no DM" probabilities. This may be due to the inability of LCC to enter the markets where OAs were present.

Based on the fourth counterfactual, with the increased profitability of entry of LCC, there seems to be an increase in the range of probabilities that both DMs enter the market, and an increase in the range of the average number of entrants. Further, there seems to be a shift to the "exactly LCC" market outcome and away from the "exactly OA" market outcome. Thus, this kind of subsidy seems to have both an effect on determining who of LCC/OA serves a particular market, and an overall effect on the number of entrants. Based on the fifth counterfactual, with the increased profitability of entry of OA, there seems to be a reduction in the range of probabilities that no DM enters the market. Combining the results of the fourth and fifth counterfactuals, it seems that OA (can) play a more important role in providing service to "underserved" markets, compared to LCC. This is consistent with the above paragraph.

One interesting result about profits is that "real world" profits of OA could be as much as 95% of the maximum profits OA would realize with LCC shut down.[28] For LCC, that ratio is much lower, at only 79%. That suggests some evidence that competition has a more negative effect on the profits of LCC compared to OA.

Further, the counterfactual that involves making entry more profitable for a certain class of carrier (LCC or OA) has potentially substantial positive effects on the profits of that class of carrier. Combined with the above discussion of market outcomes, this suggests that the government subsidizing the entry of specifically low cost carriers may increase the profits of the subsidized carriers, while having little impact on the probability of carriers serving particular "underserved" markets. On the other hand, subsidizing the entry of OA may increase the profits of the subsidized carriers, while also having a meaningful impact on the probability of carriers serving particular "underserved" markets.

---

[28]It is correct that the identified set for the summed profits (across the 2 firms) is generally strictly included in the sum of the identified sets for the profits of the two DMs separately. In general, the identified set for the sum of two quantities is not the sum of the identified sets for the two quantities.

7.2.2. *Market-level outcomes.* Table 5 displays the bounds on the fraction of markets with a certain outcome. Each column concerns a different outcome, described in the header; for example, the first column concerns the outcome that LCC achieves at least 0.5 in average profits in a given market. Even a given market results in "average" profits for a given DM because of the possible use of randomized strategies. The lower bound of the interval displayed in a particular cell of Table 5 is the counterfactual probability that outcome *must* happen. The upper bound of the interval displayed in a particular cell of Table 5 is the counterfactual probability that outcome *could* happen. This concerns the counterfactual question in Definition 2.

This adds nuance to the analysis of overall aggregate outcomes from the Section 7.2.1. For example, consider the outcomes under the "real world" counterfactual compared to the "shut down OA" counterfactual. Table 4 shows the ranges of overall average profits for LCC in these counterfactuals. This aggregate outcome averages across markets. The maximum possible overall average profits in the "real world" is about 79% of the maximum possible overall average profits in the "shut down OA" counterfactual. But, one might ask how many individual markets end up with particular profit levels for LCC. Table 5 shows the fraction of markets that will be such that LCC achieves profits greater than 1 is about the same in the "real world" as compared to the "shut down OA" counterfactual (though the range is a bit higher under the "shut down OA" counterfactual). The same is true for profits greater than 1.5. On the other hand, Table 5 also shows the fraction of markets that will be such that LCC achieves profits greater than 0.5 is much lower in the "real world" as compared to the "shut down OA" counterfactual. In that sense, shutting down OA seems to have more of an effect on increasing the "lower tail" of LCC profits. This illustrates another example of the flexibility of the sorts of outcomes accommodated by our method for counterfactual analysis, in particular going beyond aggregate outcomes.

**Remark 5** (Computational speed)**.** Taking as given the draws from the posterior distribution for the underlying model parameter, the entire set of computations required to generate the results took less than one hour using a mid-level consumer-grade CPU and GPU. As discussed above, these results are based on taking care in the computational implementation. It is worth noting that a "naive" computational implementation would be orders of magnitude slower (based on our own experience). Thus, our approach to counterfactual analysis requires relatively little computational time.

| Counterfactual | Avg. profit (LCC)$\geq$0.5 | Avg. profit (LCC)$\geq$1 | Avg. profit (LCC)$\geq$1.5 | Avg. profit (OA)$\geq$0.5 | Avg. profit (OA)$\geq$1 | Avg. profit (OA)$\geq$1.5 | Avg. profit (summed) $\geq$0.5 | Avg. profit (summed) $\geq$1 | Avg. profit (summed)$\geq$1.5 |
|---|---|---|---|---|---|---|---|---|---|
| Real world | [17.8, 48.0] [17.6, 49.4] | [6.7, 35.8] [6.7, 38.5] | [1.6, 23.0] [1.5, 25.9] | [37.5, 74.7] [37.4, 76.7] | [3.9, 37.7] [3.7, 44.7] | [0.6, 22.7] [0.6, 31.1] | [53.6, 83.7] [53.2, 85.8] | [26.6, 61.2] [25.8, 67.5] | [17.1, 44.6] [17.1, 50.0] |
| Shut down OA | [19.8, 60.3] [19.7, 64.5] | [8.0, 43.4] [7.8, 47.5] | [2.2, 28.2] [2.1, 31.1] | [0.0, 0.0] [0.0, 0.0] | [0.0, 0.0] [0.0, 0.0] | [0.0, 0.0] [0.0, 0.0] | [19.8, 60.3] [19.7, 64.5] | [8.0, 43.4] [7.8, 47.5] | [2.2, 28.2] [2.1, 31.1] |
| Shut down LCC | [0.0, 0.0] [0.0, 0.0] | [0.0, 0.0] [0.0, 0.0] | [0.0, 0.0] [0.0, 0.0] | [41.6, 77.5] [41.2, 80.4] | [7.3, 40.6] [7.2, 46.8] | [1.5, 24.8] [1.5, 33.2] | [41.6, 77.5] [41.2, 80.4] | [7.3, 40.6] [7.2, 46.8] | [1.5, 24.8] [1.5, 33.2] |
| More profitable LCC entry | [42.5, 68.0] [42.3, 68.1] | [24.9, 53.3] [24.6, 53.6] | [10.6, 41.2] [10.3, 43.3] | [33.0, 74.4] [32.3, 76.3] | [3.6, 37.4] [3.6, 43.9] | [0.5, 22.2] [0.5, 30.6] | [66.3, 90.7] [66.1, 91.7] | [44.7, 76.5] [44.1, 79.0] | [31.0, 62.3] [30.7, 65.7] |
| More profitable OA entry | [17.5, 47.4] [17.3, 48.1] | [6.6, 35.5] [6.6, 38.5] | [1.6, 22.9] [1.5, 25.6] | [66.9, 92.2] [66.0, 93.2] | [22.8, 61.3] [22.8, 63.8] | [6.1, 42.4] [5.9, 48.3] | [74.5, 94.1] [74.4, 94.8] | [45.4, 77.1] [45.3, 79.4] | [26.7, 61.8] [26.2, 65.9] |
| Obs. | n.a. | n.a. | n.a. | n.a. | n.a. | n.a. | n.a. | n.a. | n.a. |

Table 5. Results of counterfactual analysis using CE. For each counterfactual, the first row displays the estimated identified set for the quantity in that column, and the second row displays the 95% credible set for the quantity in that column. Shading/underlining indicates that the real world observed value (displayed in the last row) does not fall within the set.

## 8. Conclusions

This paper is concerned with counterfactual analysis in models with incompleteness and/or partial identification, while accounting for statistical uncertainty. We take a Bayesian approach to the statistical uncertainty, and develop a computationally attractive approach to the counterfactual analysis. Consistency of the posterior counterfactual predictions is based on a set of new general results concerning the consistency of posterior distributions of mappings of estimated sets. We verified that our counterfactual predictions, under reasonable conditions, satisfy the assumptions of those results. Thus the posterior distribution of our counterfactual predictions is consistent given a consistent posterior distribution for the identified set for the underlying parameters of the utility functions. We take this estimate of the identified set for the underlying parameters from the existing literature, and thus our approach to counterfactual analysis is to directly build on the econometrician's choice of an identification and estimation result for the underlying model parameters. We also provide an empirical application using a real data set involving air carrier entry decisions, that in particular illustrates the various counterfactuals that can be accommodated and the empirical consequences of incompleteness and partial identification in counterfactual analysis.

The counterfactual framework that we propose is flexible. For example, we allow for changes in the utility functions, distributions of the observables or unobservables, the identity of the decision makers, and solution concept. Overall, based on having used the data to estimate a posterior distribution for the identified set for the underlying model parameters, we construct counterfactual predictions within a statistically coherent framework that accounts for incompleteness (e.g., multiple equilibria), partial identification, and statistical uncertainty.

Appendix A. Results leading to Theorem 16

Before developing the sequence of results leading up to our main result about equivalences of different definitions of consistency, we provide a remark about our definition.

**Remark 6** (On Definition 6). Definition 6 restricts attention to sets $B$ that satisfy bd $A_0 \cap B = \emptyset$. This is because it would be unreasonable to require the same limit of posterior probabilities holds for all sets $B$ such that bd $A_0 \cap B \neq \emptyset$. For instance, suppose $A_0$ is an interval of real numbers, perhaps degenerate, so $A_0 = [a_{L,0}, a_{U,0}]$ with $a_{L,0} \leq a_{U,0}$. Suppose the posterior distribution for $A$ has interval values with posterior probability 1. And let $B = \{a_{L,0}\}$. Extending Definition 6 to that $B$ would require $\Pi^{(N)}(a_{L,0} \in A) \to 1$, which is an unreasonably strong condition: Supposing that $A_0$ is a non-degenerate interval (i.e., $a_{L,0} < a_{U,0}$), and if $a_{L,0}$ viewed as ordinary scalar object has a posterior distribution that is approximately normal in large samples centered at the "true value" of $a_{L,0}$, we should expect that $\Pi^{(N)}(a_{L,0} \in A) \approx \frac{1}{2}$, since a draw from the posterior for the lower bound would have approximately $\frac{1}{2}$ posterior probability of being smaller the mean of the (limiting) posterior distribution.

A.1. **Equivalences in normal Hausdorff topological spaces.**

**Lemma 12.** *Suppose $A_0$ is closed. Consider the following two conditions:*

(1) *The posterior for $A$ satisfies the condition that $\Pi^{(N)}(A \subseteq O) \to 1$ if $O$ is an open set such that $A_0 \subseteq O$. The posterior for $A$ satisfies the condition that $\Pi^{(N)}(x \in A) \to 1$ if $x \in \text{int } A_0$.*

(2) *The posterior for $A$ satisfies the condition that $\Pi^{(N)}(A \cap B \neq \emptyset) \to 1[A_0 \cap B \neq \emptyset]$ for any closed $B$ such that bd $A_0 \cap B = \emptyset$.*

*In any normal Hausdorff topological space (i.e., a $T_4$ space), the two conditions are equivalent. In any $T_1$ space, the second condition implies the first condition.*

The first condition guarantees that finite sets $K \subseteq \text{int } A_0$ have $\Pi^{(N)}(K \subseteq A) \to 1$. With the addition of a convexity assumption, this can be extended to any convex polytope $K$.

**Corollary 13.** *Suppose the space $S$ is Euclidean. Suppose $A_0$ is closed. Suppose the posterior distribution for $A_0$ has the property in Part 6.1 of Definition 6. Suppose the posterior for $A_0$* has convex values with posterior probability 1. Let $K$ be a convex polytope (i.e., a compact and convex set with finitely many extreme points) with $K \subseteq \text{int } A_0$. Then, $\Pi^{(N)}(K \subseteq A) \to 1$.

A.2. **Equivalences in Hausdorff distance.** Anywhere we use the term "Euclidean space" we mean the topology induced by the Euclidean metric on $\mathbb{R}^d$ for some finite $d$. Even if the original set $A_0$ does not satisfy the convexity assumptions, it is possible to apply the following results on the convex hull, $\text{co} A_0$.[29]

**Lemma 14.** *Suppose the space $S$ is Euclidean. Suppose $A_0$ either is a singleton or a convex body (i.e., a compact convex set with non-empty interior). Suppose the posterior for $A_0$ has convex values and non-empty compact values with posterior probability 1. The following two conditions are equivalent:*

*(1) $\Pi^{(N)}(A \cap B \neq \emptyset) \to 1[A_0 \cap B \neq \emptyset]$ for any closed set $B \subseteq S$ such that $\text{bd } A_0 \cap B = \emptyset$.*

*(2) $\Pi^{(N)}(d_H(A, A_0) > \epsilon) \to 0$ for any $\epsilon > 0$.*

**Remark 7** (Consistency in Hausdorff distance without convexity)**.** Without convexity, we cannot generally expect that consistency in Hausdorff distance implies consistency per Definition 6. For an example of consistency in Hausdorff distance but not consistency per Definition 6, suppose the posterior distribution for $A_0$ at sample size $n$ is the degenerate distribution at the set $A_n = A_0 \cap \left(\cup \mathcal{B}_{\frac{1}{n}}(a_i)\right)^C$ where $\mathcal{B}_r(a)$ is the neighborhood of radius $r$ centered at $a$. Suppose there are finitely many $a_i$, and each has $a_i \in \text{int } A_0$. Thus, $A_n$ is $A_0$ with removed neighborhoods of radius $\frac{1}{n}$ centered at a finite set of points $a_i \in \text{int } A_0$. Obviously, $A_n$ generally is not convex. Then, it can be shown that $d_H(A_n, A_0) \to 0$ but clearly $a_i \notin A_n$ for any sample size, which contradicts the characterization of consistency in terms of covering points in the interior of $A_0$ with posterior probability approaching 1.[30] This counterexample is inspired by Rockafellar and Wets (2009, page 119).

---

[29]For instance, if $A_0$ is an identified set, $\text{co} A_0$ still contains all parameter values that could have generated the data, but possibly parameter values that could not have generated the data.

[30]Proof: Clearly, $A_n \subseteq A_0$. Also, $A_0 \subseteq A_n + \frac{4}{n}\mathcal{B}$ for large enough $n$, specifically such that $\frac{5}{n} < \min_{i \neq j} d(a_i, a_j)$ and $\mathcal{B}_{\frac{2}{n}}(a_i) \subseteq A_0$ for all $i$. The argument is as follows: if $a \in A_0$ and $a \in \left(\cup \mathcal{B}_{\frac{1}{n}}(a_i)\right)^C$ then obviously $a \in A_n$. If $a \in A_0$ and $a \in \left(\cup \mathcal{B}_{\frac{1}{n}}(a_i)\right)$ then $a \in \mathcal{B}_{\frac{1}{n}}(a_i)$ for some $a_i \in \text{int } A_0$. Since $n$ is large enough, $a$ can be in at most one of the $\mathcal{B}_{\frac{1}{n}}(a_i)$ sets. Further, $n$ is large enough, so $\mathcal{B}_{\frac{2}{n}}(a_i) \subseteq A_0$ so find such $b \in \mathcal{B}_{\frac{2}{n}}(a_i) \cap [\mathcal{B}_{\frac{1}{n}}(a_i)]^C$. Clearly, $b \in A_n$ since $n$ is large enough also that $\mathcal{B}_{\frac{2}{n}}(a_i)$ is disjoint from $\mathcal{B}_{\frac{1}{n}}(a_j)$ for $j \neq i$. And, by construction, $a \in \{b\} + \frac{4}{n}\mathcal{B}$, since $d(a, b) \leq d(a, a_i) + d(a_i, b) \leq \frac{1}{n} + \frac{2}{n} = \frac{3}{n}$. Thus, $a \in A_n + \frac{4}{n}\mathcal{B}$.

**Remark 8** (On the family of sets used in the definition of posterior consistency)**.** A posterior distribution that satisfies the definition of consistency per Part 6.1 of Definition 6 necessarily has certain limiting behavior for other sets $B$, as the following result establishes.

**Lemma 15.** *Suppose the space $S$ is a $T_1$ space. Suppose the posterior for $A_0$ satisfies Part 6.1 of Definition 6.*

(15.1) *Then, it also holds that $\Pi^{(N)}(A \cap B \neq \emptyset) \to 1[A_0 \cap B \neq \emptyset]$ for any closed set $B$ that satisfies int $A_0 \cap B \neq \emptyset$.*

(15.2) *If either (a) $A_0$ is a singleton and the posterior for $A_0$ is non-empty with posterior probability 1, or (b) $A_0$ is regular closed, it also holds that $\Pi^{(N)}(A \cap B \neq \emptyset) \to 1[A_0 \cap B \neq \emptyset]$ for any closed set $B$ that satisfies $A_0 \cap $ int $B \neq \emptyset$ or $A_0 \cap$ bd $B = \emptyset$.*

In the (frequentist) theory of random sets, weak convergence of a random set to a degenerate distribution (in a locally compact Hausdorff second countable space, like Euclidean spaces) requires that $P^{(N)}(A \cap B \neq \emptyset) \to 1[A_0 \cap B \neq \emptyset]$ for closed sets $B$ such that $A_0 \cap$ int $B \neq \emptyset$ or $A_0 \cap$ bd $B = \emptyset$, known as the continuity family. See for instance Molchanov (2017, Section 1.7). Thus, by the second claim, consistency of the posterior distribution per Definition 6 makes statements about every set in this continuity family.

However, the condition in Part 15.2 does not suffice for consistency per Definition 6. A counterexample is $A_0 = [0, 2]$ with degenerate posterior distribution of $A_N = [0, 1 - \frac{1}{N}] \cup [1 + \frac{1}{N}, 2]$. This satisfies the condition in Part 15.2.[31] But, this posterior distribution is not consistent per Definition 6, as $\Pi^{(N)}(A \cap \{1\}) = 0$ despite the fact that $\{1\} \in$ int $A_0$.

A.3. **Main equivalence result.** Summarizing, we have the following equivalence result.

**Theorem 16** (Equivalences between definitions of consistency)**.** *Suppose the space $S$ is a normal Hausdorff topological space (i.e., a $T_4$ space). Suppose $A_0$ is closed. The following two conditions are equivalent:*

---

[31] Suppose $A_0 \cap B = \emptyset$. Then since $A_N \subseteq A_0$, also $A_N \cap B = \emptyset$, so $\Pi^{(N)}(A \cap B \neq \emptyset) = 0$. Suppose $A_0 \cap B \neq \emptyset$. If $A_0 \cap$ bd $B = \emptyset$ then $A_0 \cap$ int $B \neq \emptyset$. Otherwise, if $A_0 \cap$ bd $B \neq \emptyset$ then $A_0 \cap$ int $B \neq \emptyset$ by the condition in Part 15.2. Thus, $A_0 \cap$ int $B \neq \emptyset$. Therefore, there is an interval in $A_0 \cap B$, which will have non-empty intersection with $A_N$ for large enough $N$, since that interval will contain at least one point that is not $\{1\}$.

(16.1) *The posterior for A satisfies the condition that $\Pi^{(N)}(A \subseteq O) \to 1$ if $O$ is an open set such that $A_0 \subseteq O$. The posterior for A satisfies the condition that $\Pi^{(N)}(x \in A) \to 1$ if $x \in int\ A_0$. And, $\Pi^{(N)}(A \neq \emptyset) \to 1$ if $A_0 \neq \emptyset$.*

(16.2) *The posterior for A satisfies the condition that $\Pi^{(N)}(A \cap B \neq \emptyset) \to 1[A_0 \cap B \neq \emptyset]$ for any closed $B$ such that bd $A_0 \cap B = \emptyset$. And, $\Pi^{(N)}(A \neq \emptyset) \to 1$ if $A_0 \neq \emptyset$.*

*Further, suppose the space S is Euclidean. Suppose $A_0$ either is a singleton or a convex body (i.e., a compact convex set with non-empty interior). Suppose the posterior for $A_0$ has convex values and non-empty compact values with posterior probability 1. The following condition is equivalent to either of the previous two conditions:*

(16.3) $\Pi^{(N)}(d_H(A, A_0) > \epsilon) \to 0$ *for any $\epsilon > 0$, where $d_H$ is the usual Hausdorff distance.*

## Appendix B. Proofs

*Proof of Lemma 1.* First claim: The convex hull of a non-empty set is non-empty. The convex hull of a compact set is closed by Aliprantis and Border (2006, Corollary 5.33). The convex hull of a compact-valued upper hemi-continuous mapping is upper hemi-continuous by Aliprantis and Border (2006, Thorem 17.35). Second claim: Follows immediately from Aliprantis and Border (2006, Theorem 17.36). □

*Proof of Lemma 2.* By Assumption 2, $\{\tilde{\tilde{u}} : \mathcal{S}(\tilde{\tilde{u}}) \cap K \neq \emptyset\}$ is a closed set for all compact sets $K$. Since $\tilde{\tilde{u}}$ is itself a random variable under Assumption 1, the event that $\tilde{\tilde{u}}$ is an element of a particular closed set is itself a measurable event, so $\mathcal{S}(\tilde{\tilde{u}})$ is a random closed set.

Now consider $\mathcal{T}(\tilde{\tilde{u}}) = \{t(\tilde{\tilde{u}}, s) : s \in \mathcal{S}(\tilde{\tilde{u}})\}$. By Assumption 2, $\mathcal{S}(\cdot)$ is compact-valued since it is necessarily bounded. Therefore, $\mathcal{T}(\tilde{\tilde{u}})$ is compact-valued, since the continuous image of a compact set is a compact set, e.g., Aliprantis and Border (2006, Theorem 2.34). By the same arguments as in the beginning of the proof of Theorem 10, $\mathcal{T}(\tilde{\tilde{u}})$ is upper hemi-continuous also. Therefore, by the same arguments as above, $\mathcal{T}(\tilde{\tilde{u}})$ is a random closed set.

The convex hull co $\mathcal{S}(\tilde{\tilde{u}})$ is a random closed set by Molchanov (2017, Theorem 1.3.25). The proof for $\mathcal{T}^{co}(\tilde{\tilde{u}}) = \{t(\tilde{\tilde{u}}, s) : s \in$ co $\mathcal{S}(\tilde{\tilde{u}})\}$ is essentially the same as the proof for $\mathcal{T}(\tilde{\tilde{u}})$. □

*Proof of Theorem 3.* By Molchanov (2017, Theorem 1.4.1), a measurable selection exists. By the analysis that led up to Equation 12 in the main text, any selection from $\mathcal{T}^{co}(\tilde{\tilde{u}})$ can be the outcome for that given $\tilde{\tilde{u}}$. Thus, letting $\tilde{T}$ be a random variable that is a measurable selection for some $\theta \in \Theta_{I,0}$ with associated $\tilde{F}_\theta(\tilde{\tilde{u}})$, the corresponding distribution of $\tilde{T}$ can be the distribution of outcomes under the setup of the counterfactual analysis described in Definition 1. The characterization of $\tilde{T}$ follows from Artstein's inequality, as in Artstein (1983) or Molchanov (2017, Corollary 1.4.11). □

*Proof of Theorem 4.* Consider a sequence of parameter values $\theta^{(\nu)}$ that converges to $\theta$. Let $\tilde{\tilde{u}}^{(\nu)}$ be the distribution of the vector $\tilde{u}(\cdot, \cdot, x^{(\nu)}, \epsilon^{(\nu)}, \theta^{(\nu)})$ with $(x^{(\nu)}, \epsilon^{(\nu)}) \sim \tilde{F}_{\theta^{(\nu)}}(x, \epsilon)$ and $\tilde{\tilde{u}}$ be the distribution of $\tilde{u}(\cdot, \cdot, x, \epsilon, \theta)$ with $(x, \epsilon) \sim \tilde{F}_\theta(x, \epsilon)$. Then, by Lemma 11, given the stated assumptions, $\tilde{\tilde{u}}^{(\nu)}$ converges weakly to $\tilde{\tilde{u}}$.

**Proof of Part 4.1:** Let $f(\theta) \equiv P_{\tilde{\tilde{u}} \sim \tilde{F}_\theta(\tilde{\tilde{u}})}(\mathcal{T}^{\text{co}}(\tilde{\tilde{u}}))$, viewed as a mapping from $\Theta$ to the space of distributions of random closed sets (with elements of those sets from the space of the outcome of interest), using the Lévy metric. By Part 10.2 of Theorem 10, $P_{\tilde{\tilde{u}} \sim \tilde{F}_{\theta^{(\nu)}}(\tilde{\tilde{u}})}(\mathcal{T}^{\text{co}}(\tilde{\tilde{u}}))$ converges weakly to $P_{\tilde{\tilde{u}} \sim \tilde{F}_\theta(\tilde{\tilde{u}})}(\mathcal{T}^{\text{co}}(\tilde{\tilde{u}}))$ as random closet sets, since $\tilde{\tilde{u}}^{(\nu)}$ converges weakly to $\tilde{\tilde{u}}$. Therefore, $f(\theta^{(\nu)}) \to f(\theta)$, since the Lévy metric is a metrization of weak convergence, e.g., Molchanov (2017, Corollary 1.7.35). Thus, $f(\cdot)$ is continuous, since $\Theta$ is a metric space and thus sequential continuity is continuity. Then, the result follows from Theorem 7.

**Proof of Parts 4.2 and 4.3:** Under the first condition, by Parts 10.1 and 10.4 of Theorem 10, given the stated assumptions, $f(\theta) \equiv P_{\tilde{\tilde{u}} \sim \tilde{F}_\theta(\tilde{\tilde{u}})}(\mathcal{T}^{\text{co}}(\tilde{\tilde{u}}) \cap B \neq \emptyset)$ and $f(\theta) \equiv P_{\tilde{\tilde{u}} \sim \tilde{F}_\theta(\tilde{\tilde{u}})}(\mathcal{T}^{\text{co}}(\tilde{\tilde{u}}) \subseteq B)$ are continuous. This uses $[\mathcal{T}^{\text{co}}]^{-1}(\text{cl } B) = [\mathcal{T}^{\text{co}}]^{-1}(\text{int } B) \cup [\mathcal{T}^{\text{co}}]^{-1}(\text{bd } B)$ since $\text{cl } B = \text{int } B \cup \text{bd } B$. Therefore, $P(\tilde{\tilde{u}} \in [\mathcal{T}^{\text{co}}]^{-1}(\text{cl } B)) = P(\tilde{\tilde{u}} \in [\mathcal{T}^{\text{co}}]^{-1}(\text{int } B) \cup [\mathcal{T}^{\text{co}}]^{-1}(\text{bd } B)) = P(\tilde{\tilde{u}} \in [\mathcal{T}^{\text{co}}]^{-1}(\text{int } B)) + P(\tilde{\tilde{u}} \in [\mathcal{T}^{\text{co}}]^{-1}(\text{bd } B)) - P(\tilde{\tilde{u}} \in [\mathcal{T}^{\text{co}}]^{-1}(\text{int } B) \cap [\mathcal{T}^{\text{co}}]^{-1}(\text{bd } B)) = P(\tilde{\tilde{u}} \in [\mathcal{T}^{\text{co}}]^{-1}(\text{int } B))$ if $P(\tilde{\tilde{u}} \notin [\mathcal{T}^{\text{co}}]^{-1}(\text{int } B), \tilde{\tilde{u}} \in [\mathcal{T}^{\text{co}}]^{-1}(\text{bd } B)) = P(\tilde{\tilde{u}} \in [\mathcal{T}^{\text{co}}]^{-1}(\text{bd } B)) - P(\tilde{\tilde{u}} \in [\mathcal{T}^{\text{co}}]^{-1}(\text{int } B) \cap [\mathcal{T}^{\text{co}}]^{-1}(\text{bd } B)) = 0$. Under the second condition, by Parts 10.3 and 10.5 of Theorem 10, given the stated assumptions, $f(\theta) \equiv P_{\tilde{\tilde{u}} \sim \tilde{F}_\theta(\tilde{\tilde{u}})}(\mathcal{T}^{\text{co}}(\tilde{\tilde{u}}) \cap B \neq \emptyset)$ and $f(\theta) \equiv P_{\tilde{\tilde{u}} \sim \tilde{F}_\theta(\tilde{\tilde{u}})}(\mathcal{T}^{\text{co}}(\tilde{\tilde{u}}) \subseteq B)$ are continuous. Then, the result follows from Corollary 9.

**Proof of Parts 4.4 and 4.5:** By Part 10.7 of Theorem 10, given the stated assumptions, $f(\theta) \equiv E_{P_{\tilde{\tilde{u}} \sim \tilde{F}_\theta(\tilde{\tilde{u}})}(\mathcal{T}^{\text{co}}(\tilde{\tilde{u}}))}(\sup_{t \in \mathcal{T}^{\text{co}}(\tilde{\tilde{u}})} \omega(t))$ and $f(\theta) \equiv E_{P_{\tilde{\tilde{u}} \sim \tilde{F}_\theta(\tilde{\tilde{u}})}(\mathcal{T}^{\text{co}}(\tilde{\tilde{u}}))}(\inf_{t \in \mathcal{T}^{\text{co}}(\tilde{\tilde{u}})} \omega(t))$ are continuous. Then, the result follows from Corollary 9. □

*Proof of Corollary 5.* Given the conclusions from Theorem 4, Lemma 8 applies. Relative to the conditions in Lemma 8, it is obvious that the lower bound is less than the upper bound for the second interval. For the first interval, note that $\mathcal{T}^{\text{co}}(\tilde{u}) \subseteq B$ implies that $\mathcal{T}^{\text{co}}(\tilde{u}) \cap B \neq \emptyset$ as long as $\mathcal{T}^{\text{co}}(\tilde{u}) \neq \emptyset$, which is true by Assumption 2. We also use here the assumptions that $\Theta_{I,0} \neq \emptyset$, and that the posterior distribution respects this with posterior probability 1. □

*Proof of Theorem 6.* The first claim is: Suppose $B$ is a closed set such that $f(A_0) \cap B = \emptyset$. Then $\Pi^{(N)}(f(A) \cap B \neq \emptyset) \to 0$. Proof: in this case, $A_0 \cap f^{-1}(B) = \emptyset$. Consequently, since $A_0$ is closed by assumption, and therefore contains its boundary, $\text{bd} A_0 \cap f^{-1}(B) = \emptyset$. Since $f$ is continuous and $B$ is a closed set, $f^{-1}(B)$ is a closed set. Therefore, $\Pi^{(N)}(f(A) \cap B \neq \emptyset) = \Pi^{(N)}(A \cap f^{-1}(B) \neq \emptyset) \to 0$.

The second claim is: Suppose $B$ is a closed set such that $f(A_0) \cap B \neq \emptyset$ and $\text{bd } f(A_0) \cap B = \emptyset$. Then $\Pi^{(N)}(f(A) \cap B \neq \emptyset) \to 1$. Proof: first note that if $A_0$ is a singleton then $f(A_0) \cap B \neq \emptyset$ implies that $\text{bd } f(A_0) \cap B \neq \emptyset$. Thus, $A_0$ cannot be a singleton, so $A_0$ must be regular closed. Let $x \in \text{int } f(A_0) \cap B$. Then, there is an open set $O$ containing $x$ such that $O \subseteq f(A_0)$. Now find a polytope $P$ inside of $O$ that also contains $x$ in the interior, so that $x \in \text{int } P \subseteq O \subseteq f(A_0)$. Let the extreme points be $p_1, p_2, \ldots, p_k$. There must be corresponding $a_1, a_2, \ldots, a_k \in A_0$ such that $f(a_i) = p_i$. Because $A_0$ is regular closed, each $a_i$ is a point of closure of $\text{int } A_0$. Therefore, any neighborhood of $a_i$ contains an element of $\text{int } A_0$. Therefore, even if $a_i \in \text{bd } A_0$, there is a point $a_i'$ within a small neighborhood of $a_i$ in the interior of $A_0$ such that $p_i' = f(a_i') \approx f(a_i) = p_i$. In particular, such $a_i'$ can be found such that $x \in \text{co } \{p_1', p_2', \ldots, p_k'\} = P'$. By posterior consistency, since all $a_i' \in \text{int } A_0$, and singletons are closed sets in the $T_1$ space, it follows that $\Pi^{(N)}(A \cap \{a_i'\} \neq \emptyset) \to 1$ and hence $\Pi^{(N)}(\{a_1', a_2', \ldots, a_k'\} \subseteq A) \to 1$, implying that $\Pi^{(N)}(\{p_1', p_2', \ldots, p_k'\} \subseteq f(A)) = \Pi^{(N)}(\{f(a_1'), f(a_2'), \ldots, f(a_k')\} \subseteq f(A)) \to 1$. Because $f(A)$ is convex, this implies that $\Pi^{(N)}(x \in P' \subseteq f(A)) \to 1$. Therefore, by construction of $x$, $\Pi^{(N)}(f(A) \cap B \neq \emptyset) \to 1$.

For the main result, combine the above two claims.

Further, suppose that $f(A_0) \neq \emptyset$. Then it must have been that $A_0 \neq \emptyset$. Thus, $\Pi^{(N)}(A \neq \emptyset) \to 1$ by Part 6.2 of Definition 6, which implies that $\Pi^{(N)}(f(A) \neq \emptyset) \to 1$. □

*Proof of Theorem 7.* Let $\delta > 0$. Let $O = f(A_0) + \delta \mathcal{B}$, where $\mathcal{B}$ is the unit ball centered at 0. We want to show that $f(A) \subseteq O$. Let $P = f^u(O)$ be the upper inverse of $O$ under $f(\cdot)$, which is an open set since $f$ is continuous. By construction, $A_0 \subseteq P$. Let $Q$ be an open set such that $A_0 \subseteq Q \subseteq \text{cl } Q \subseteq P$. This exists by Munkres (2000, Lemma 31.1), since this is a normal topological space (since it is a metric space). Moreover, $Q \neq A_0$ since the only clopen sets in a connected space are the empty set and the entire set. We know that $A_0 \subseteq \text{int}(\text{cl } Q)$, since $Q$ is open and therefore $Q \subseteq \text{int}(\text{cl } Q)$ given that $Q \subseteq \text{cl } Q$. Therefore, for any $x \in A_0$, $\mathcal{B}_\delta(x) \subseteq \text{int}(\text{cl } Q)$ for some $\delta$. And hence if $y \in [\text{int}(\text{cl } Q)]^C$ it must be that $d(x, y) \geq \delta$. Therefore, $d(x, [\text{int}(\text{cl } Q)]^C) \geq \delta$. Since the distance is continuous, it follows that the distance between $A_0$ and $[\text{int}(\text{cl } Q)]^C$ is strictly positive since $A_0$ is compact so the distance to $[\text{int}(\text{cl } Q)]^C$ is achieved at some point in $A_0$. Let $\delta$ be that distance. Now consider $x \in A_0 + \frac{\delta}{2}\mathcal{B}$. It follows that $x \notin [\text{int}(\text{cl } Q)]^C$ because of the distance between $A_0$ and $[\text{int}(\text{cl } Q)]^C$. So, $A_0 + \frac{\delta}{2}\mathcal{B} \subseteq \text{int}(\text{cl } Q) \subseteq P$. When $d_H(A, A_0) \leq \frac{\epsilon}{3}$ it follows that $A \subseteq A_0 + \frac{\epsilon}{2}\mathcal{B}$, by for example Aliprantis and Border (2006, Lemma 3.71). (The different fractions are intentional: $\frac{\epsilon}{3}$ is less than $\frac{\epsilon}{2}$, as required in order to be sure we can apply this result.) Thus, as long as $\epsilon < \delta$, it follows that $A \subseteq P$, and hence $f(A) \subseteq O = f(A_0) + \delta\mathcal{B}$ as required. Then, swapping $A$ and $A_0$ everywhere, it would follow that $f(A_0) \subseteq f(A) + \delta\mathcal{B}$. Thus, $d_H(A, A_0) \leq \frac{\frac{\delta}{2}}{3}$ implies that $d_H(f(A), f(A_0)) \leq \delta$. Therefore, $\Pi^{(N)}(d_H(f(A), f(A_0)) \leq \delta) \geq \Pi^{(N)}(d_H(A, A_0) \leq \frac{\frac{\delta}{2}}{3}) \to 1$. □

*Proof of Lemma 8.* Use the notation $L_0 = f_L(A_0)$, $U_0 = f_U(A_0)$, $L = f_L(A)$, and $U = f_U(A)$. By the condition $f_L \leq f_U$, the posterior distribution $f(A)$ has non-empty compact realizations with posterior probability 1. Given also $f(A)$ and $f(A_0)$ are interval subsets of $\mathbb{R}$, it is enough to establish Part 16.3 from Theorem 16. It holds that $d_H(f(A), f(A_0)) \leq \max\{|L - L_0|, |U - U_0|\}$. Since the posteriors for $L$ and U are consistent, $\Pi^{(N)}(L \in \mathcal{B}_\epsilon(L_0)) \to 1$ and $\Pi^{(N)}(U \in \mathcal{B}_\epsilon(U_0)) \to 1$. Consequently, $\Pi^{(N)}(d_H(f(A), f(A_0)) > \epsilon) \to 0$ as needed. □

*Proof of Corollary 9.* By Theorem 6, the posterior distribution for $f(\Theta_{I,0})$ is consistent per Definition 6. This uses the fact that $f(\Theta_I)$ is connected and hence convex, since continuous functions map connected sets to connected sets, and connected sets in $\mathbb{R}$ are convex. In particular, $f(\Theta_I)$ is compact, since continuous functions map compact sets to compact sets. And, $f(\Theta_{I,0}) \neq \emptyset$ since $\Theta_{I,0} \neq \emptyset$, and the posterior for $f(\Theta_{I,0})$ respects this with posterior probability 1 given that the posterior for $\Theta_{I,0}$ does. By another application of Theorem 6, the posterior distribution for $g(f(\Theta_{I,0}))$ is consistent per Definition 6. This uses the fact that $f(\Theta_{I,0})$ is a singleton in the space of non-empty compact sets, and similarly also the convexity condition in Theorem 6 is vacuous. $\square$

*Proof of Theorem 10.* By the sequential characterization of hemi-continuity, $\mathcal{T}(\tilde{\tilde{u}})$ is upper/lower hemi-continuous at any $\tilde{\tilde{u}}$ such that $\mathcal{S}(\tilde{\tilde{u}})$ is respectively upper/lower hemi-continuous. For upper hemi-continuity: suppose $x^{(\nu)} \in \mathcal{T}(\tilde{\tilde{u}}^{(\nu)})$, such that $x^{(\nu)} = t(\tilde{\tilde{u}}^{(\nu)}, s^{(\nu)})$ for $s^{(\nu)} \in \mathcal{S}(\tilde{\tilde{u}}^{(\nu)})$, and $\tilde{\tilde{u}}^{(\nu)} \to \tilde{\tilde{u}}$. Assuming that $\mathcal{S}$ is upper hemi-continuous at $\tilde{\tilde{u}}$, there is a subsequence $s^{(\nu_k)} \to s \in \mathcal{S}(\tilde{\tilde{u}})$. Therefore since $t(\cdot)$ is continuous, along this same subsequence, $x^{(\nu_k)} \to t(\tilde{\tilde{u}}, s) \in \mathcal{T}(\tilde{\tilde{u}})$. Therefore, $\mathcal{T}$ is upper hemi-continuous at $\tilde{\tilde{u}}$. For lower hemi-continuity: suppose $\tilde{\tilde{u}}^{(\nu)} \to \tilde{\tilde{u}}$ and $x \in \mathcal{T}(\tilde{\tilde{u}})$, such that $x = t(\tilde{\tilde{u}}, s)$ for $s \in \mathcal{S}(\tilde{\tilde{u}})$. Assuming that $\mathcal{S}$ is lower hemi-continuous at $\tilde{\tilde{u}}$, there is a subsequence $\tilde{\tilde{u}}^{(\nu_k)}$ with $s^{(\nu_k)} \in \mathcal{S}(\tilde{\tilde{u}}^{(\nu_k)})$ and $s^{(\nu_k)} \to s$. Therefore since $t(\cdot)$ is continuous, along this same subsequence, $x^{(\nu_k)} = t(\tilde{\tilde{u}}^{(\nu_k)}, s^{(\nu_k)}) \in \mathcal{T}(\tilde{\tilde{u}}^{(\nu_k)})$ and $x^{(\nu_k)} \to t(\tilde{\tilde{u}}, s) \in \mathcal{T}(\tilde{\tilde{u}})$. Therefore, $\mathcal{T}$ is lower hemi-continuous at $\tilde{\tilde{u}}$. The same arguments also apply to $\mathcal{T}^{co}$, since co $\mathcal{S}$ is upper/lower hemi-continuous at any argument that $\mathcal{S}$ is, by Aliprantis and Border (2006, Theorem 17.35 and Theorem 17.36).

**Proof of Part 10.1:** Because $\mathcal{S}(\cdot)$ is upper hemi-continuous by Assumption 2, $\mathcal{T}(\cdot)$ is also upper hemi-continuous, so $\mathcal{T}^{-1}(\text{cl } B)$ is a closed set. Therefore, via the "Portmanteau Theorem" as in van der Vaart (1998, Lemma 2.2) and by weak convergence of $\tilde{\tilde{u}}^{(\nu)}$ to $\tilde{\tilde{u}}$, $\limsup_{\nu \to \infty} P(\tilde{\tilde{u}}^{(\nu)} \in \mathcal{T}^{-1}(\text{cl } B)) \leq P(\tilde{\tilde{u}} \in \mathcal{T}^{-1}(\text{cl } B))$.

From Assumption 3, there is a set $\mathcal{U}^e$ such that $\mathcal{S}(\cdot)$ is lower hemi-continuous on $(\mathcal{U}^e)^C$. Define $\mathcal{S}_c(\tilde{\tilde{u}}) = \mathcal{S}(\tilde{\tilde{u}}) 1\left[\tilde{\tilde{u}} \notin \mathcal{U}^e\right]$. Thus, $\mathcal{S}_c(\tilde{\tilde{u}}) = \emptyset$ for $\tilde{\tilde{u}} \in \mathcal{U}^e$.

For any set $B$, $\mathcal{S}_c^{-1}(B) \subseteq \mathcal{S}^{-1}(B)$: if $\tilde{\tilde{u}} \in \mathcal{S}_c^{-1}(B)$ then $\mathcal{S}_c(\tilde{\tilde{u}}) \cap B \neq \emptyset$, which implies that $\tilde{\tilde{u}} \notin \mathcal{U}^e$ and $\mathcal{S}(\tilde{\tilde{u}}) \cap B \neq \emptyset$, so $\tilde{\tilde{u}} \in \mathcal{S}^{-1}(B)$. For any set $B$, $\mathcal{S}^{-1}(B) \subseteq \mathcal{S}_c^{-1}(B) \cup \mathcal{U}^e$: (1) if $\tilde{\tilde{u}} \in \mathcal{S}^{-1}(B)$ and $\tilde{\tilde{u}} \notin \mathcal{U}^e$ then $\mathcal{S}(\tilde{\tilde{u}}) \cap B \neq \emptyset$ and therefore $\mathcal{S}_c(\tilde{\tilde{u}}) \cap B \neq \emptyset$ so $\tilde{\tilde{u}} \in \mathcal{S}_c^{-1}(B)$; alternatively, (2), if $\tilde{\tilde{u}} \in \mathcal{U}^e$, then the set inclusion is by construction. Moreover, $\mathcal{S}_c^{-1}(B) \cap \mathcal{U}^e = \emptyset$ by the same arguments as above. Therefore, $\mathcal{S}_c^{-1}(B) \subseteq \mathcal{S}^{-1}(B) \subseteq \mathcal{S}_c^{-1}(B) \cup \mathcal{U}^e$.

By essentially the same arguments, $\mathcal{S}_c^{-1}(B) = \mathcal{S}^{-1}(B) \cap (\mathcal{U}^e)^C$: If $\tilde{\tilde{u}} \in \mathcal{S}_c^{-1}(B)$ then $\mathcal{S}_c(\tilde{\tilde{u}}) \cap B \neq \emptyset$, which implies that $\tilde{\tilde{u}} \notin \mathcal{U}^e$ and $\mathcal{S}(\tilde{\tilde{u}}) \cap B \neq \emptyset$, so $\tilde{\tilde{u}} \in \mathcal{S}^{-1}(B) \cap (\mathcal{U}^e)^C$. If $\tilde{\tilde{u}} \in \mathcal{S}^{-1}(B) \cap (\mathcal{U}^e)^C$, then $\mathcal{S}(\tilde{\tilde{u}}) \cap B \neq \emptyset$ so $\mathcal{S}_c(\tilde{\tilde{u}}) \cap B \neq \emptyset$ so $\tilde{\tilde{u}} \in \mathcal{S}_c^{-1}(B)$. This implies $\mathcal{S}_c(\cdot)$ is lower hemi-continuous. By the above, $\mathcal{S}_c^{-1}(O) = \mathcal{S}^{-1}(O) \cap (\mathcal{U}^e)^C$ in particular for any open set $O$. Because $\mathcal{S}$ is lower hemi-continuous on $(\mathcal{U}^e)^C$ by Assumption 3, $\mathcal{S}^{-1}(O) \cap (\mathcal{U}^e)^C$ is an open set. See for instance Rockafellar and Wets (2009, Theorem 5.7).

By using the arguments involving the sequential characterization of hemi-continuity, $\mathcal{T}_c(\tilde{\tilde{u}}) = \{t(\tilde{\tilde{u}}, s) : s \in \mathcal{S}_c(\tilde{\tilde{u}})\}$ is also lower hemi-continuous. Also, $\mathcal{T}_c^{-1}(B) \subseteq \mathcal{T}^{-1}(B) \subseteq \mathcal{T}_c^{-1}(B) \cup \mathcal{U}^e$, given the same result for $\mathcal{S}$ as above.

Therefore, $P(\tilde{\tilde{u}}^{(\nu)} \in \mathcal{T}_c^{-1}(B)) \leq P(\tilde{\tilde{u}}^{(\nu)} \in \mathcal{T}^{-1}(B)) \leq P(\tilde{\tilde{u}}^{(\nu)} \in \mathcal{T}_c^{-1}(B)) + P(\tilde{\tilde{u}}^{(\nu)} \in \mathcal{U}^e)$.

In particular, $P(\tilde{\tilde{u}}^{(\nu)} \in \mathcal{T}_c^{-1}(\text{int } B)) \leq P(\tilde{\tilde{u}}^{(\nu)} \in \mathcal{T}^{-1}(\text{int } B))$. Because $\mathcal{T}_c(\cdot)$ is lower hemi-continuous, and int $B$ is an open set, $\mathcal{T}_c^{-1}(\text{int } B)$ is an open set. Therefore, via the "Portmanteau Theorem" and by weak convergence of $\tilde{\tilde{u}}^{(\nu)}$ to $\tilde{\tilde{u}}$, $P(\tilde{\tilde{u}} \in \mathcal{T}_c^{-1}(\text{int } B)) \leq \liminf_{\nu \to \infty} P(\tilde{\tilde{u}}^{(\nu)} \in \mathcal{T}_c^{-1}(\text{int } B)) \leq \liminf_{\nu \to \infty} P(\tilde{\tilde{u}}^{(\nu)} \in \mathcal{T}^{-1}(\text{int } B))$.

Moreover, since $P(\tilde{\tilde{u}} \in \mathcal{T}^{-1}(\text{int } B)) - P(\tilde{\tilde{u}} \in \mathcal{U}^e) \leq P(\tilde{\tilde{u}} \in \mathcal{T}_c^{-1}(\text{int } B))$, it follows that $P(\tilde{\tilde{u}} \in \mathcal{T}^{-1}(\text{int } B)) - P(\tilde{\tilde{u}} \in \mathcal{U}^e) \leq \liminf_{\nu \to \infty} P(\tilde{\tilde{u}}^{(\nu)} \in \mathcal{T}^{-1}(\text{int } B)) \leq \liminf_{\nu \to \infty} P(\tilde{\tilde{u}}^{(\nu)} \in \mathcal{T}^{-1}(B)) \leq \limsup_{\nu \to \infty} P(\tilde{\tilde{u}}^{(\nu)} \in \mathcal{T}^{-1}(B)) \leq \limsup_{\nu \to \infty} P(\tilde{\tilde{u}}^{(\nu)} \in \mathcal{T}^{-1}(\text{cl } B)) \leq P(\tilde{\tilde{u}} \in \mathcal{T}^{-1}(\text{cl } B))$, where the last inequality was proved above.

Note that $\mathcal{T}^{-1}(B)$ is Borel in particular for any open/closed $B$, by (the arguments of) Lemma 2, and general results on measurability (e.g., Molchanov (2017, Proposition 1.1.2 or Theorem 1.3.3)).

**Proof of Part 10.2:** If $B$ is in the continuity family relative to $\mathcal{T}(\tilde{\tilde{u}})$, so that $P(\mathcal{T}(\tilde{\tilde{u}}) \cap \text{cl } B \neq \emptyset) = P(\mathcal{T}(\tilde{\tilde{u}}) \cap \text{int } B \neq \emptyset)$, and if $P(\tilde{\tilde{u}} \in \mathcal{U}^e) = 0$, the endpoints are equal in Part 10.1. Consequently, $\lim_{\nu \to \infty} P(\tilde{\tilde{u}}^{(\nu)} \in \mathcal{T}^{-1}(B)) = P(\tilde{\tilde{u}} \in \mathcal{T}^{-1}(B))$ for any such closed $B$. By Molchanov (2017, Theorem 1.7.7), this suffices for weak convergence.

**Proof of Part 10.3:** The first sub-claim is a direct implication of the Portmanteau Theorem. There is an intermediate claim: In particular, if $\mathcal{T}^{-1}(B)$ is a closed convex set and $\tilde{\tilde{u}}$ has an ordinary density with respect to Lebesgue measure, then $P(\mathcal{T}(\tilde{\tilde{u}}^{(\nu)}) \cap B \neq \emptyset) \to P(\mathcal{T}(\tilde{\tilde{u}}) \cap B \neq \emptyset)$. This claim follows since convex sets have a boundary with Lebesgue measure zero, e.g., Dudley (2014, Lemma 2.24). The second sub-claim follows since $\mathcal{T}^{-1}(B) = \cup \mathcal{T}^{-1}(B_j)$: if $u \in \mathcal{T}^{-1}(B)$ then $\mathcal{T}(u) \cap B \neq \emptyset$ and thus $\mathcal{T}(u) \cap B_j \neq \emptyset$ for some $j$; conversely, if $u \in \cup \mathcal{T}^{-1}(B_j)$ then $u \in \mathcal{T}^{-1}(B_j)$ for some $j$ and thus $\mathcal{T}(u) \cap B_j \neq \emptyset$ and thus $\mathcal{T}(u) \cap B \neq \emptyset$ and thus $u \in \mathcal{T}^{-1}(B)$. Further, the boundary of a finite union of sets is a subset of the union of the boundaries, and so the boundary of $\mathcal{T}^{-1}(B)$ has Lebesgue measure zero.

**Proof of Part 10.4:** Let $\mathcal{T}^l$ be the lower inverse of $\mathcal{T}$. Note that $\mathcal{T}^l(B) = [\mathcal{T}^{-1}(B^C)]^C$, as in Aliprantis and Border (2006, page 557). Therefore, $P(\mathcal{T}(\tilde{\tilde{u}}) \subseteq B) = P(\tilde{\tilde{u}} \in \mathcal{T}^l(B)) = P(\tilde{\tilde{u}} \notin \mathcal{T}^{-1}(B^C)) = 1 - P(\tilde{\tilde{u}} \in \mathcal{T}^{-1}(B^C))$, and the same would be true for any distribution of utility functions. Therefore, applying Part 10.1, $\limsup_{\nu\to\infty} P(\mathcal{T}(\tilde{\tilde{u}}^{(\nu)}) \subseteq B) = 1 - \liminf_{\nu\to\infty} P(\tilde{\tilde{u}}^{(\nu)} \in \mathcal{T}^{-1}(B^C)) \leq 1 - P(\mathcal{T}(\tilde{\tilde{u}}) \cap \text{int } B^C \neq \emptyset) + P(\tilde{\tilde{u}} \in \mathcal{U}^e) = P(\mathcal{T}(\tilde{\tilde{u}}) \subseteq [\text{int } B^C]^C) + P(\tilde{\tilde{u}} \in \mathcal{U}^e) = P(\mathcal{T}(\tilde{\tilde{u}}) \subseteq \text{cl } B) + P(\tilde{\tilde{u}} \in \mathcal{U}^e)$. And, $\liminf_{\nu\to\infty} P(\mathcal{T}(\tilde{\tilde{u}}^{(\nu)}) \subseteq B) = 1 - \limsup_{\nu\to\infty} P(\tilde{\tilde{u}}^{(\nu)} \in \mathcal{T}^{-1}(B^C)) \geq 1 - P(\mathcal{T}(\tilde{\tilde{u}}) \cap \text{cl} B^C \neq \emptyset) = P(\mathcal{T}(\tilde{\tilde{u}}) \subseteq [\text{cl } B^C]^C) = P(\mathcal{T}(\tilde{\tilde{u}}) \subseteq \text{int } B)$. Both results use the fact that $[\text{int } D]^C = \text{cl } D^C$ for any set $D$, as in Aliprantis and Border (2006, Lemma 2.4).

**Proof of Part 10.5:** This follows directly from Part 10.3.

**Proof of Part 10.6:** This follows from Part 10.2 and Molchanov (2017, Proposition 1.7.15). The result for inf follows since $\inf f = -\sup(-f)$.

**Proof of Part 10.7:** Write $f(t) = f(t)s(t, K_\alpha) + f(t)(1 - s(t, K_\alpha)) = f_{1,\alpha}(t) + f_{2,\alpha}(t)$, where $s(t, K_\alpha) = 1 - \min\{d_1(t, K_\alpha), 1\}$ is continuous in $t$, using the $L_1$ distance $d_1$. Since $f$ is continuous, $f_{1,\alpha}(t) = f(t)s(t, K_\alpha)$ is continuous. Also, $f_{1,\alpha}(t)$ has compact support since $s(t, K_\alpha) = 0$ for $t \notin K_{\alpha+1}$.

Because $f_{1,\alpha}(t) - |f_{2,\alpha}(t)| \leq f(t) = f_{1,\alpha}(t) + f_{2,\alpha}(t) \leq f_{1,\alpha}(t) + |f_{2,\alpha}(t)|$, it follows that $E \sup f_{1,\alpha}(\mathcal{T}(\tilde{\tilde{u}}^{(\nu)})) - E \sup |f_{2,\alpha}(\mathcal{T}(\tilde{\tilde{u}}^{(\nu)}))| \leq E \sup f(\mathcal{T}(\tilde{\tilde{u}}^{(\nu)})) \leq E \sup f_{1,\alpha}(\mathcal{T}(\tilde{\tilde{u}}^{(\nu)})) + E \sup |f_{2,\alpha}(\mathcal{T}(\tilde{\tilde{u}}^{(\nu)}))|$. The same holds for the limiting distribution $\tilde{\tilde{u}}$.

Therefore, $E \sup f_{1,\alpha}(\mathcal{T}(\tilde{\tilde{u}}^{(\nu)})) - E \sup |f_{2,\alpha}(\mathcal{T}(\tilde{\tilde{u}}^{(\nu)}))| - E \sup f_{1,\alpha}(\mathcal{T}(\tilde{\tilde{u}})) - E \sup |f_{2,\alpha}(\mathcal{T}(\tilde{\tilde{u}}))| \leq E \sup f(\mathcal{T}(\tilde{\tilde{u}}^{(\nu)})) - E \sup f(\mathcal{T}(\tilde{\tilde{u}})) \leq E \sup f_{1,\alpha}(\mathcal{T}(\tilde{\tilde{u}}^{(\nu)})) + E \sup |f_{2,\alpha}(\mathcal{T}(\tilde{\tilde{u}}^{(\nu)}))| - E \sup f_{1,\alpha}(\mathcal{T}(\tilde{\tilde{u}})) + E \sup |f_{2,\alpha}(\mathcal{T}(\tilde{\tilde{u}}))|$.

For any given $\alpha$, by Part 10.6, $E \sup f_{1,\alpha}(\mathcal{T}(\tilde{\tilde{u}}^{(\nu)})) \to E \sup f_{1,\alpha}(\mathcal{T}(\tilde{\tilde{u}}))$, as $\nu \to \infty$.

Since $f_{2,\alpha}(t) = 0$ for $t \in K_\alpha$, it follows that $E \sup |f_{2,\alpha}(\mathcal{T}(\tilde{\tilde{u}}^{(\nu)}))| = E \sup_{t \in \mathcal{T}(\tilde{\tilde{u}}^{(\nu)})} |f_{2,\alpha}(t)| = E \max\{\sup_{t \in \mathcal{T}(\tilde{\tilde{u}}^{(\nu)}), t \in K_\alpha^C} |f_{2,\alpha}(t)|, 0\} \leq E \max\{\sup_{t \in \mathcal{T}(\tilde{\tilde{u}}^{(\nu)}), t \in K_\alpha^C} |f(t)|, 0\}$. Note that the outer max is relevant to handle cases where $\mathcal{T}(\tilde{\tilde{u}}) \cap K_\alpha^C = \emptyset$. By assumption, for any given $\epsilon > 0$, for large enough $\alpha$, $E \max\{\sup_{t \in \mathcal{T}(\tilde{\tilde{u}}^{(\nu)}), t \in K_\alpha^C} |f(t)|, 0\} < \frac{\epsilon}{3}$. Thus, for large enough $\alpha$ and all $\nu$, $E \sup |f_{2,\alpha}(\mathcal{T}(\tilde{\tilde{u}}^{(\nu)}))| < \frac{\epsilon}{3}$. Furthermore, for large enough $\alpha$, $E \sup |f_{2,\alpha}(\mathcal{T}(\tilde{\tilde{u}}))| < \frac{\epsilon}{3}$.

Therefore, for any given $\epsilon > 0$, by choosing large enough $\alpha$, and then by choosing $\nu$ large enough, it follows that $E \sup f(\mathcal{T}(\tilde{\tilde{u}}^{(\nu)})) - E \sup f(\mathcal{T}(\tilde{\tilde{u}}))$ can be squeezed between $-\epsilon$ and $\epsilon$.

The result for inf follows similarly.

Finally, note that exactly the same arguments also apply to $\mathcal{T}^{\mathrm{co}}$. □

*Proof of Lemma 11.* By some abuse of notation, temporarily viewing $(x^{(\nu)}, \epsilon^{(\nu)})$ as deterministic sequences converging to some $(x^{(\infty)}, \epsilon^{(\infty)})$, $\tilde{u}_{(\nu)}(\cdot, \cdot, x^{(\nu)}, \epsilon^{(\nu)}) \equiv \tilde{u}(\cdot, \cdot, x^{(\nu)}, \epsilon^{(\nu)}, \theta^{(\nu)}) \to \tilde{u}(\cdot, \cdot, x^{(\infty)}, \epsilon^{(\infty)}, \theta)$ by continuity. Therefore, the distribution of $\tilde{u}_{(\nu)}(\cdot, \cdot, x^{(\nu)}, \epsilon^{(\nu)})$ converges weakly to the distribution of $\tilde{u}(\cdot, \cdot, x, \epsilon, \theta)$ by the continuous mapping theorem (e.g., van der Vaart and Wellner (2023, Theorem 1.11.1)). This uses the condition that $(x, \epsilon)$ is Euclidean, and thus distributions thereof are separable. □

*Proof of Lemma 12.* Suppose the first condition. Let $B$ be a closed set such that bd $A_0 \cap B = \emptyset$. First, suppose $A_0 \cap B = \emptyset$. Then, in any $T_4$ space, there are disjoint open sets $O_1$ and $O_2$ containing $A_0$ and $B$ respectively. Hence, if $A \subseteq O_1$ then $A \cap B = \emptyset$. Therefore, $\Pi^{(N)}(A \cap B = \emptyset) \geq \Pi^{(N)}(A \subseteq O_1) \to 1$. Second, suppose $A_0 \cap B \neq \emptyset$. Then, $x \in A_0 \cap B$. Since bd $A_0 \cap B = \emptyset$, $x \in$ int $A_0 \cap B$. Therefore, $\Pi^{(N)}(A \cap B \neq \emptyset) \geq \Pi^{(N)}(x \in A) \to 1$.

Suppose the second condition. Let $O$ be an open set such that $A_0 \subseteq O$. Then $A_0 \cap O^C = \emptyset$. Since $A_0$ is closed, this implies that bd $A_0 \cap O^C = \emptyset$, Therefore, $\Pi^{(N)}(A \subseteq O) = \Pi^{(N)}(A \cap O^C = \emptyset) \to 1$. Let $x \in$ int $A_0$. Then $\{x\}$ is a closed set in any $T_1$ space, such that bd $A_0 \cap \{x\} = \emptyset$. Therefore, $\Pi^{(N)}(x \in A) = \Pi^{(N)}(A \cap \{x\} \neq \emptyset) \to 1$.

In this proof, and other proofs in this section, the limits are valid whether interpreted as limits in probability or limits almost surely. In both cases, the proof repeatedly uses the "squeeze theorem" that if $g_1^{(N)} \leq g_2^{(N)} \leq g_3^{(N)}$ are random objects with $g_1^{(N)} \to c$ and $g_3^{(N)} \to c$ then also $g_2^{(N)} \to c$, either if the limits are in probability or almost surely. Often, implicitly, either $g_1^{(N)}$ or $g_3^{(N)}$ is the constant 0 or 1, since the limits always concern (posterior) probabilities that are necessarily bounded between 0 and 1. □

*Proof of Corollary 13.* Let $\{k_1, k_2, \ldots, k_d\}$ be the extreme points of $K$. Since $K \subseteq$ int $A_0$, also the extreme points are in int $A_0$. Therefore, $\Pi^{(N)}(\{k_1, k_2, \ldots, k_d\} \subseteq A) \to 1$ by Part 6.1 of Definition 6, and Lemma 12. Since the posterior has convex values, $\{k_1, k_2, \ldots, k_d\} \subseteq A$ implies that co$\{k_1, k_2, \ldots, k_d\} \subseteq A$. Therefore $\Pi^{(N)}(K \subseteq A) \to 1$. □

*Proof of Lemma 14.* **Proof that 1 implies 2:** Suppose $A_0$ is a singleton. Let $O$ be the open ball with radius $\frac{\epsilon}{2}$ centered at the singleton $A_0$. By construction, $\emptyset \neq A \subseteq O$ implies $d_H(A, A_0) < \epsilon$. By the condition in Part 1, and Lemma 12, $\Pi^{(N)}(d_H(A, A_0) < \epsilon) \geq \Pi^{(N)}(A \subseteq O) \to 1$.

Alternatively, suppose $A_0$ is a convex body.

Suppose $A \subseteq A_0^\epsilon = A_0 + \epsilon \mathcal{B} = \cup_{a \in A_0} \mathcal{B}_\epsilon(a)$, which is open since it is the union of open sets. $\mathcal{B}$ is the open unit ball centered at 0 and $\mathcal{B}_r(a)$ is the open ball of radius $r$ centered at $a$. Then, clearly $\sup_{a \in A} d(a, A_0) \leq \epsilon$. Thus, $\Pi^{(N)}(\sup_{a \in A} d(a, A_0) \leq \epsilon) \geq \Pi^{(N)}(A \subseteq A_0^\epsilon)$.

Since $A_0$ is a convex body, for any $\delta > 0$ there is a convex polytope $P \subseteq A_0$ such that $d_H(P, A_0) < \delta$. See for example Dudley (1974) and Bronshteyn and Ivanov (1975) for the main result, or the summaries in Gruber (1993) or Bronstein (2008) for more on such approximation results (i.e., the number of extreme points needed, and so forth). Suppose $k_\delta$ is the number of extreme points of $P$, which generically depends non-trivially on $\delta$. In general, the constructions in this literature appear (by design) to have extreme points on the boundary of the original convex body. However, we must avoid the boundary of $A_0$. Since $A_0$ is closed and convex, with non-empty interior, it is a regular closed set (i.e., it is the closure of the interior). See for example Rockafellar and Wets (2009, Theorem 2.33). Therefore, for each extreme point $p_i$ of $P$, there is a $q_i$ within a $\frac{\delta}{\sqrt{k_\delta}}$ neighborhood of $p_i$ such that $q_i$ is in the interior of $A_0$. Let $Q$ be the convex polytope with the $q_i$s as the extreme points.

Now consider $\sup_{x \in P} d(x, Q)$. Since $x \in P$, it holds that $x = \sum \lambda_i p_i$ with $\lambda_i \geq 0$ and $\sum \lambda_i = 1$. Consider the same weights applied to the extreme points $q_i$, and corresponding $z = \sum \lambda_i q_i$. Obviously, $z \in Q$. Note that each $p_i$ and $q_i$ are $d \times 1$ vectors, where $d$ is the dimension of the Euclidean space. Let $p$ and $q$ respectively be $d \times k_\delta$ stacked matrices. Let $\lambda$ be a $k_\delta \times 1$ stacked vector. Then, $d(x, y) = ||\sum \lambda_i (p_i - q_i)|| = ||(p - q)\lambda|| \leq ||p - q|| \times ||\lambda|| \leq ||p - q||$ where $||p - q||$ is understood to be the spectral norm of $p - q$. $||\lambda|| \leq 1$ since the $L_2$ norm is bounded above by the $L_1$ norm, and $\lambda$ has $L_1$ norm equal to 1 by construction. The spectral norm is bounded above by the Frobenius norm, which is $\sqrt{\sum_{i,j}(p_{i,j} - q_{i,j})^2}$. Since $q_i$ is in a $\frac{\delta}{\sqrt{k_\delta}}$ neighborhood of $p_i$, clearly $\sum_j (p_{i,j} - q_{i,j})^2 < \left(\frac{\delta}{\sqrt{k_\delta}}\right)^2$. Thus, $\sqrt{\sum_{i,j}(p_{i,j} - q_{i,j})^2} \leq \sqrt{k_\delta} \frac{\delta}{\sqrt{k_\delta}} = \delta$. Therefore, $d(x, y) \leq \delta$, so $\sup_{x \in P} d(x, Q) \leq \delta$. By symmetric arguments, $\sup_{y \in Q} d(P, y) \leq \delta$. Therefore, $d_H(P, Q) \leq \delta$. Therefore, by the triangle inequality, $d_H(Q, A_0) \leq d_H(Q, P) + d_H(P, A_0) \leq 2\delta$.

Hence, for any $\epsilon > 0$ there is a convex polytope $Q \subseteq A_0$ with extreme points in the interior of $A_0$ such that $d_H(Q, A_0) < \epsilon$ by setting $\delta < \frac{\epsilon}{2}$ in the above construction.

Suppose $Q \subseteq A$: $\sup_{a \in A_0} d(A, a) \leq \sup_{a \in A_0} d(Q, a) \leq d_H(Q, A_0) < \epsilon$ by construction of $Q$.

Therefore, overall, if $Q \subseteq A \subseteq A_0^\epsilon$, it follows $d_H(A, A_0) \leq \epsilon$. Thus, $\Pi^{(N)}(d_H(A, A_0) \leq \epsilon) \geq \Pi^{(N)}(Q \subseteq A \subseteq A_0^\epsilon) \to 1$, since previous results show $\Pi^{(N)}(Q \subseteq A) \to 1$ and $\Pi^{(N)}(A \subseteq A_0^\epsilon) \to 1$. The first follows from Corollary 13 and the second follows from Lemma 12.

**Proof that 2 implies 1:** Suppose $A_0 \cap B = \emptyset$. Since both $A_0$ and $B$ are closed, and $A_0$ is compact, the distance between $A_0$ and $B$ is strictly positive. Let that distance be $d(A_0, B) = \epsilon > 0$. Suppose $d_H(A, A_0) \leq \frac{\epsilon}{2}$. Suppose $x \in A$. Then, $d(x, A_0) \leq \frac{\epsilon}{2}$ by construction. Therefore, $\epsilon = d(A_0, B) \leq d(A_0, x) + d(x, B) \leq \frac{\epsilon}{2} + d(x, B)$, so $d(x, B) \geq \frac{\epsilon}{2}$ and thus $x \notin B$. Overall, this proves that $d_H(A, A_0) \leq \frac{\epsilon}{2}$ implies $A \cap B = \emptyset$. So, $\Pi^{(N)}(A \cap B = \emptyset) \geq \Pi^{(N)}(d_H(A, A_0) \leq \frac{\epsilon}{2}) \to 1$.

Suppose $A_0 \cap B \neq \emptyset$ and bd $A_0 \cap B = \emptyset$. Clearly, this cannot happen when $A_0$ is a singleton, so $A_0$ is a convex body. Let $x \in $ int $A_0 \cap B$. Let $\mathcal{B}$ be the open unit ball centered at 0. Since $x \in $ int $A_0$ there is $\epsilon > 0$ such that $\{x\} + \epsilon\mathcal{B} \subseteq A_0$. Supposing that $d_H(A, A_0) \leq \frac{\epsilon}{3}$, it follows that $A_0 \subseteq A + \frac{\epsilon}{2}\mathcal{B}$, by for example Aliprantis and Border (2006, Lemma 3.71). So, $\{x\} + \epsilon\mathcal{B} \subseteq A_0 \subseteq A + \frac{\epsilon}{2}\mathcal{B}$. Equivalently, $\{x\} + \frac{\epsilon}{2}\mathcal{B} + \frac{\epsilon}{2}\mathcal{B} \subseteq A_0 \subseteq A + \frac{\epsilon}{2}\mathcal{B}$. By the cancellation principle for convex sets (i.e., Rådström (1952, Lemma 1)), $\{x\} + \frac{\epsilon}{2}\mathcal{B} \subseteq A$. So, in particular, $x \in A \cap B$. So, $\Pi^{(N)}(A \cap B \neq \emptyset) \geq \Pi^{(N)}(A \cap \{x\} \neq \emptyset) \geq \Pi^{(N)}(d_H(A, A_0) \leq \frac{\epsilon}{3}) \to 1$. □

*Proof of Lemma 15.* **Part 15.1:** let $x \in $ int $A_0 \cap B$. Lemma 12 implies $\Pi^{(N)}(A \cap B \neq \emptyset) \geq \Pi^{(N)}(x \in A) \to 1$.

**Part 15.2:** first suppose $A_0 \cap B \neq \emptyset$. In the case that $A_0 \cap $ bd $B = \emptyset$ it must be that $A_0 \cap $ int $B \neq \emptyset$. Thus, it must be that $A_0 \cap $ int $B \neq \emptyset$. Let $x \in A_0 \cap $ int $B$. If $A_0$ is regular closed, $x$ is a point of closure of int $A_0$, and therefore every neighborhood of $x$ contains an element of int $A_0$. int $B$ is a neighborhood of $x$, so int $A_0 \cap $ int $B \neq \emptyset$. And thus let $x' \in $ int $A_0 \cap $ int $B$. Lemma 12 implies $\Pi^{(N)}(A \cap B \neq \emptyset) \geq \Pi^{(N)}(x' \in A) \to 1$. If $A_0$ is a singleton, then $A_0 \subseteq $ int $B$. So Lemma 12 shows $\Pi^{(N)}(A \cap B \neq \emptyset) \geq \Pi^{(N)}(A \subseteq $ int $B) \to 1$ given that $A \neq \emptyset$ with posterior probability 1 by assumption. Alternatively, suppose $A_0 \cap B = \emptyset$. Then certainly bd $A_0 \cap B = \emptyset$ so Part 6.1 of Definition 6 directly applies. □

*Proof of Theorem 16.* The theorem follows immediately from collecting the statements in Lemmas 12 and 14 into one statement. (The condition that $\Pi^{(N)}(A \neq \emptyset) \to 1$ if $A_0 \neq \emptyset$ is shared in Parts 16.1 and 16.2, so there is nothing to prove for that part of the equivalence.

Under the further conditions, the posterior for $A_0$ is assumed to have non-empty values, in which case the condition that $\Pi^{(N)}(A \neq \emptyset) \to 1$ if $A_0 \neq \emptyset$ is obviously true.) □

Appendix C. Graphical examples of ingredients in a CPDS

This part displays some graphical examples of some of the ingredients in a CPDS.

**Market entry probabilities:** Figure 1a displays an example of the upper bound and the lower bound on the counterfactual probability that at least one firm enters the market, as a function of the parameter value. This corresponds to $E_{P_{\tilde{u} \sim \tilde{F}_\theta(\tilde{u})}(\text{co } \mathcal{S}(\tilde{u}))}(\sup \omega(s))$ and $E_{P_{\tilde{u} \sim \tilde{F}_\theta(\tilde{u})}(\text{co } \mathcal{S}(\tilde{u}))}(\inf \omega(s))$ with $\omega(s) = s_{01} + s_{10} + s_{11}$, similarly to Equations 7 and 8. As an ingredient in a CPDS, the calculations in Figure 1a consider any possible value of the parameters. The actual bounds on the counterfactual probability that at least one firm enters the market would be the interval connecting the displayed upper bound and displayed lower bound, for any *given* value of the underlying parameter. Then, the overall identified set for the counterfactual probability that at least one firm enters the market would account for the identified set for the underlying parameter. Figure 1a uses the parametric specification of the utility functions in Equation 1.

Along the bottom of Figure 1a is the specification of values of $\alpha$ and $\beta$, which is assumed in this example to apply to both firms in the market. Both firms also share a fixed value of $\Delta = -1$. The restriction of equality of parameter values across firms, and the restriction of a fixed value of $\Delta = -1$, is made simply for the purposes of this example. These restrictions reduce the effective dimension of the parameter to 2, allowing the display of figures. The underlying computations easily accommodate the case that each firm has a different value of the parameter and an unrestricted value of $\Delta$. (And indeed the empirical application does not involve such a restriction.) The single observable determinant of utility for each firm is independently drawn from $\mathcal{N}(0.4375, (0.25)^2)$ and the single unobservable determinant of utility for each firm is independently drawn from $\mathcal{N}(0, (0.125)^2)$. These distributions are part of the specification of the counterfactual analysis. They could reflect the distribution in the real world, or a counterfactual specification, or something else, as discussed in Section 2.2.4.

The solution concept used in Figure 1a is the correlated equilibrium solution concept, but other solution concepts could also be used, per the discussion in Section 2.2.5.

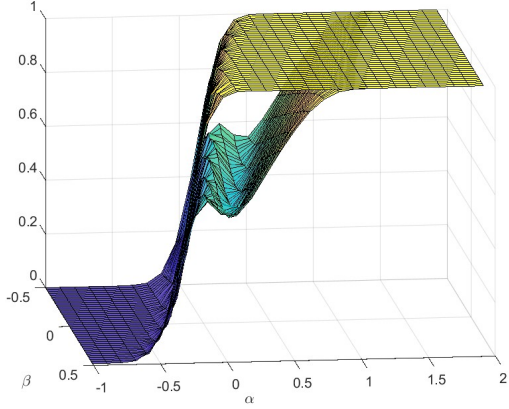
(A) Bounds on the probability that at least one firm enters the market

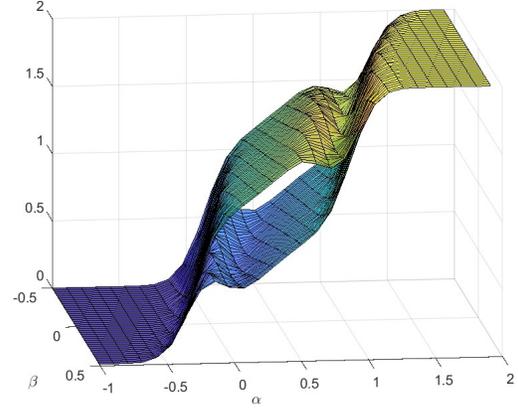
(B) Bounds on the expected number of entrants

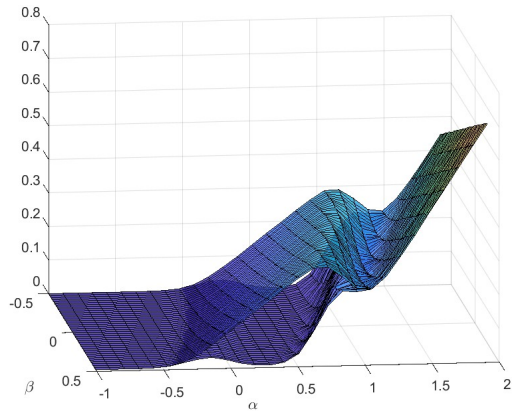
(C) Bounds on the expected profits

**Expected entrants:** Figure 1b provides another example for the expected number of market entrants. The same background information for Figure 1a applies to Figure 1b.

**Profits:** Figure 1c provides a final example for the total profits of both firms. Again, the same background information for Figure 1a applies to Figure 1c.

The figures exhibit some non-monotonicities. For example, in Figure 1c, there are non-monotonicities in the bounds on the expected profits as a function of $\alpha$. In a model with multiple DMs, increasing $\alpha$ has multiple effects that can result in non-monotonicities. As with models involving only a single DM, increasing $\alpha$ will "directly" increase the incentive for firms to enter the market, *taking as fixed the entry decisions of other firms.* However, in models with multiple DMs, increasing $\alpha$ will therefore "indirectly" *reduce* the incentive for firms to enter the market, because the other firms will now have a stronger incentive to enter the market. Because of the effects of competition on profits (i.e., negative $\Delta$), entry of the other firms will reduce the incentive to enter the market. Therefore, depending on the relative strengths of the various effects of $\alpha$, there can be a non-monotone effect of increasing $\alpha$. Another feature of the figures is that, for some values of the parameters, the lower bound (essentially) equals the upper bound. For instance, if $\alpha$ is large enough, the only market outcome is for both firms to enter the market, as displayed in Figure 1b.

As a technical note relating to footnote 1, note that there are both non-monotonicities and "flat spots" in these bounds as a function of the underlying model parameters. The "flat spots" happen in cases where the bounds are 0 or 1, and in other cases (in particular, where the non-monotonicities "happen"). Beyond all of the other issues already discussed, this complicates inference, because it precludes using various "standard" ideas for inference. In particular, abstracting away from many of the complications in this setting, one may think about the much simpler and standard problem of inference on a scalar-valued ("ordinary") function of a point identified parameter. It seems safe to say that the dominant approach involves using the delta method, either explicitly or implicitly; roughly, we would include here any method that involves using a linearization of the function of the parameter that involves the derivative. We should not expect anything like the "delta method" to apply in our setting, given the "flat spots" would translate to something along the lines of a zero derivative (if indeed the derivative even exists), which is ruled out by the delta method. Furthermore, the non-zero derivative condition that is used in the delta method at least somewhat conflicts with non-monotonicities in the function.[32] For example, it appears that Kaido, Molinari, and Stoye (2019) provides results for the most general available case of (frequentist) inference on a function of a partially identified parameter; and, their equation (G.31) involves such a "delta method" approach. Correspondingly, their results require a non-zero derivative in Kaido, Molinari, and Stoye (2019, Theorem 3.1, part III).

---

[32]The sharpest results on this topic are possible in cases where the underlying parameter is itself a scalar: A function from $\mathbb{R}$ to $\mathbb{R}$ with continuous, non-zero derivative is monotone. This is because continuous functions map connected sets to connected sets. Thus, if there were $a$ and $b$ such that $\frac{df}{dx}(a) < 0$ and $\frac{df}{dx}(b) > 0$, it would hold that $\frac{df}{dx}(\text{co }\{a,b\})$ maps to an interval that contains both a negative value and positive value, and hence contains zero, which would contradict the non-zero derivative condition. More generally, for a function $\mathbb{R}^d$ to $\mathbb{R}$, by the first-order necessary conditions for optimization, at any point where the function has a local minimum/maximum (which "tends" to arise with non-monotonicities), the derivative equals zero. Specifically, although it is a bit difficult to see in the figures, for example the lower bound in Figure 1a has a local minimum in the "trough" where $\alpha \approx 0.2$.

Appendix D. Conditions for uniform integrability in Theorem 4

**Lemma 17.** *Suppose the outcome satisfies* $\sup_{t \in \mathcal{T}(\tilde{\tilde{u}}) \cap K_\alpha^C} |t| \leq \sup_{w \in \mathcal{W}} |\sum_{i,a} w_{i,a} \tilde{\tilde{u}}_i(a)| 1(\sum_{i,a} w_{i,a} \tilde{\tilde{u}}_i(a) \in K_\alpha^C)$, *where* $\mathcal{W}$ *is a non-empty set of possible weights. Suppose there is* $\overline{W} < \infty$ *such that* $\sup_{w \in \mathcal{W}} ||w||_\infty \leq \overline{W}$. *Suppose* $\lim_{\alpha \to \infty} \sup_{\theta \in \Theta} E_{\tilde{\tilde{u}} \sim \tilde{F}_\theta(\tilde{\tilde{u}})}(|\tilde{\tilde{u}}_i(a)| 1(|\tilde{\tilde{u}}_i(a)| \geq \alpha)) = 0$. *Then it holds that* $\lim_{\alpha \to \infty} \sup_{\theta \in \Theta} E_{\tilde{\tilde{u}} \sim \tilde{F}_\theta(\tilde{\tilde{u}})}(\max\{\sup_{t \in \mathcal{T}(\tilde{\tilde{u}}) \cap K_\alpha^C} |t|, 0\}) = 0$.

The first condition on the outcome generally holds for various welfare outcomes, which are generally a weighted sum (not necessarily weighted average) of the utilities. For instance, the utility outcome of just a specific DM $i$ has this form when $w_{i,a}$ is the probability that action profile $a$ arises, and $w_{j,a} = 0$ for $j \neq i$. The sum of the utility outcomes for DM $i$ and DM $j$ has this form when $w_{i,a}$ and $w_{j,a}$ are the probability that action profile $a$ arises, and $w_{k,a} = 0$ for other $k$. The second condition requires a uniform upper bound on the possible weights $|w_{i,a}|$. In most cases, including the examples just above, the upper bound can be taken to be 1, reflecting that the welfare outcome places no more than weight 1 on the utility any particular DM gets from any particular action profile. The last condition is a standard uniform integrability condition for each utility component $\tilde{\tilde{u}}_i(a)$. By standard arguments, this uniform integrability condition follows from weak sufficient conditions (e.g., Gut (2006, Section 5.4)), like the condition that $\sup_{\theta \in \Theta} E_{\tilde{\tilde{u}} \sim \tilde{F}_\theta(\tilde{\tilde{u}})}(|\tilde{\tilde{u}}_i(a)|^{1+\epsilon})$ is finite for some $\epsilon > 0$. In turn, this would follow from the assumption that $E_{\tilde{\tilde{u}} \sim \tilde{F}_\theta(\tilde{\tilde{u}})}(|\tilde{\tilde{u}}_i(a)|^{1+\epsilon})$ depends continuously on $\theta$, and the assumption that $\Theta$ is compact.

Appendix E. The set $\mathcal{U}^e$ and location normalizations in utility

This concerns providing the details for the claim in the text that an ordinary density for the "un-normalized" terms of utility suffices for Assumption 5, assuming that $\mathcal{U}^e$ has Lebesgue measure zero. Although perhaps intuitively true, this claim is more subtle than it may immediately seem. Hence, this section. The subtle issue is that even if $\mathcal{U}^e$ has Lebesgue measure zero, it could be that there is a positive Lebesgue measure of the "un-normalized" utility terms that result in a discontinuity when combined with the 0 values for the "normalized" utility terms. Possibly, $\mathcal{S}$ might have discontinuities precisely at specifications of $\tilde{\tilde{u}}$ where the "normalized" utility terms are set to 0. This would indeed have Lebesgue measure zero with respect to Lebesgue measure over $\tilde{\tilde{u}}$ (and thus be consistent with the discussion of Assumption 3), while also implying that all utility functions actually generated from the specification involving the normalization would be part of $\mathcal{U}^e$.

This section proves actually this does not happen. Let $d_{\mathrm{n}} = \sum_{i=1}^{M} |\tilde{\mathcal{A}}_{-i}|$ be the number of location normalizations. Each normalization corresponds to subtracting off the "pre-normalization" value of $\tilde{\tilde{u}}_i(0, a_{-i})$ from $\tilde{\tilde{u}}_i(a)$, with corresponding vector-representation $\tilde{\tilde{u}}_{\mathrm{n}} \in \mathbb{R}^{d_{\mathrm{n}}}$ that collects all these subtracted values.

Let $f(\tilde{\tilde{u}}_{\mathrm{u}}, \tilde{\tilde{u}}_{\mathrm{n}}) = (\tilde{\tilde{u}}_{\mathrm{u}}, 0) + (g(\tilde{\tilde{u}}_{\mathrm{n}}), \tilde{\tilde{u}}_{\mathrm{n}})$ be the function that "adds back" a specification $\tilde{\tilde{u}}_{\mathrm{n}}$ for the normalized utility terms to a specification $\tilde{\tilde{u}}_{\mathrm{u}}$ for the unnormalized utility terms (the utility terms that are not normalized to 0). Specifically, corresponding to the element of $\tilde{\tilde{u}}_u$ that is $\tilde{\tilde{u}}_i(a_i, a_{-i})$, $g(\cdot)$ picks out the element of $\tilde{\tilde{u}}_n$ that is $\tilde{\tilde{u}}_i(0, a_{-i})$; in particular, $g(\cdot)$ is a linear transform and $g(0) = 0$. In this representation, the normalized terms of utility are re-arranged to be the last elements of the overall vector of utility for notational convenience.

Let $\mathcal{U}_u^e$ be the set of values of $\tilde{\tilde{u}}_u^*$ where there is a discontinuity in $\mathcal{S}$ (defined in this section to be failure of lower hemi-continuity) for some value of $\tilde{\tilde{u}}_n^*$. We really only care about what happens when $\tilde{\tilde{u}}_n^* = 0$, but as the next paragraph discusses, actually that restriction does not change the set of discontinuities (and that plays a key part of the proof).

Given that only differences in utility matter, in particular $\mathcal{S}(f(\tilde{\tilde{u}}_{\mathrm{u}}^*, \tilde{\tilde{u}}_{\mathrm{n}}^*)) = \mathcal{S}(f(\tilde{\tilde{u}}_{\mathrm{u}}^*, \tilde{\tilde{u}}_{\mathrm{n}}))$. Therefore, if $f(\tilde{\tilde{u}}_{\mathrm{u}}^*, \tilde{\tilde{u}}_{\mathrm{n}}^*)$ is a point of discontinuity in $\mathcal{S}$, then $f(\tilde{\tilde{u}}_{\mathrm{u}}^*, \tilde{\tilde{u}}_{\mathrm{n}})$ for all $\tilde{\tilde{u}}_n \in \mathbb{R}^{d_n}$ are all points of discontinuity in $\mathcal{S}$. The details of this claim are: Suppose there is a discontinuity in $\mathcal{S}$ at $f(\tilde{\tilde{u}}_{\mathrm{u}}^*, \tilde{\tilde{u}}_{\mathrm{n}}^*)$. Then there is an open set $\mathcal{M}$ that meets $\mathcal{S}(f(\tilde{\tilde{u}}_{\mathrm{u}}^*, \tilde{\tilde{u}}_{\mathrm{n}}^*))$ such that there is no neighborhood $\mathcal{O}^*$ of $f(\tilde{\tilde{u}}_{\mathrm{u}}^*, \tilde{\tilde{u}}_{\mathrm{n}}^*)$ such that $\mathcal{S}(\tilde{\tilde{u}})$ meets $\mathcal{M}$ for all $\tilde{\tilde{u}} \in \mathcal{O}^*$. By strategic equivalence, $\mathcal{S}(f(\tilde{\tilde{u}}_{\mathrm{u}}^*, \tilde{\tilde{u}}_{\mathrm{n}})) = \mathcal{S}(f(\tilde{\tilde{u}}_{\mathrm{u}}^*, \tilde{\tilde{u}}_{\mathrm{n}}^*))$. Thus, $\mathcal{M}$ also meets $\mathcal{S}(f(\tilde{\tilde{u}}_{\mathrm{u}}^*, \tilde{\tilde{u}}_{\mathrm{n}}))$. To prove a contradiction, suppose that $\mathcal{S}$ were continuous at $f(\tilde{\tilde{u}}_{\mathrm{u}}^*, \tilde{\tilde{u}}_{\mathrm{n}})$. Then there is a neighborhood $\mathcal{O}$ of $f(\tilde{\tilde{u}}_{\mathrm{u}}^*, \tilde{\tilde{u}}_{\mathrm{n}})$ such that $\mathcal{S}(\tilde{\tilde{u}})$ meets $\mathcal{M}$ for all $\tilde{\tilde{u}} \in \mathcal{O}$. Then define $\mathcal{O}' = \mathcal{O} + f(\tilde{\tilde{u}}_{\mathrm{u}}^*, \tilde{\tilde{u}}_{\mathrm{n}}^*) - f(\tilde{\tilde{u}}_{\mathrm{u}}^*, \tilde{\tilde{u}}_{\mathrm{n}})$, which is hence a neighborhood of $f(\tilde{\tilde{u}}_{\mathrm{u}}^*, \tilde{\tilde{u}}_{\mathrm{n}}^*)$. Since $g$ is linear, $\mathcal{O}' = \mathcal{O} + (g(\tilde{\tilde{u}}_n^* - \tilde{\tilde{u}}_n), \tilde{\tilde{u}}_n^* - \tilde{\tilde{u}}_n)$. By construction of $\mathcal{O}$, $S(\tilde{\tilde{u}})$ meets $\mathcal{M}$ for all $\tilde{\tilde{u}} \in \mathcal{O}$. Thus, $\mathcal{S}(\tilde{\tilde{u}}')$ meets $\mathcal{M}$ for all $\tilde{\tilde{u}}' \in \mathcal{O}'$ since $\mathcal{S}(\tilde{\tilde{u}}') = \mathcal{S}(\tilde{\tilde{u}}' - (g(\tilde{\tilde{u}}_n^* - \tilde{\tilde{u}}_n), \tilde{\tilde{u}}_n^* - \tilde{\tilde{u}}_n))$ by strategic equivalence. This contradicts the hypothesis of a discontinuity in $\mathcal{S}$ at $f(\tilde{\tilde{u}}_{\mathrm{u}}^*, \tilde{\tilde{u}}_{\mathrm{n}}^*)$.

Let $N = \mathbb{R}^{d_{\mathrm{n}}}$ be the set of possible values for $\tilde{\tilde{u}}_{\mathrm{n}}$. Then, $\mathcal{E} = \{(\tilde{\tilde{u}}_u + g(\tilde{\tilde{u}}_n), \tilde{\tilde{u}}_n) : \tilde{\tilde{u}}_u \in \mathcal{U}_u^e, \tilde{\tilde{u}}_n \in N\} \subseteq \tilde{\mathcal{U}}^e$ is a set of discontinuities in $\mathcal{S}$. Here, $\tilde{\mathcal{U}}^e \subseteq \mathcal{U}^e$ is the set of discontinuities in $\mathcal{S}$. In fact, the same argument shows that $\mathcal{E} = \tilde{\mathcal{U}}^e$.

We want to prove $\mathcal{U}_u^e$ has Lebesgue measure zero, assuming that $\tilde{\mathcal{U}}^e$ has Lebesgue measure zero, because that implies that $P((\tilde{\tilde{u}}_u, 0) \in \tilde{\mathcal{U}}^e) \le P(\tilde{\tilde{u}}_u \in \mathcal{U}_u^e) = 0$, where the last equality uses the assumption that $\tilde{\tilde{u}}_u$ has an ordinary density. That is the claim in the text.

With sections $\mathcal{E}_{\tilde{\tilde{u}}_n^*} = \{\tilde{\tilde{u}}_u + g(\tilde{\tilde{u}}_n^*) : \tilde{\tilde{u}}_u \in \mathcal{U}_u^e\}$ for given $\tilde{\tilde{u}}_n^*$, and using Tonelli's theorem (e.g., Tao (2011, Corollary 1.7.19)), $\mu(\mathcal{E}) = \int \mu(\mathcal{E}_{\tilde{\tilde{u}}_n^*}) d\mu(\tilde{\tilde{u}}_n^*) = \mu(\mathcal{E}_0)\mu(N)$, where $\mu(\cdot)$ is Lebesgue measure on the relevant Euclidean space. This uses the fact that $\mu(\mathcal{E}_{\tilde{\tilde{u}}_n^*}) = \mu(\mathcal{E}_0)$ for all $\tilde{\tilde{u}}_n^*$, by translation invariance of Lebesgue measure. Thus, $\mu(\mathcal{E}_0) = 0$, since $\mu(\mathcal{E}) = 0$. And thus $\mu(\mathcal{U}_u^e) = 0$, since $\mathcal{U}_u^e = \mathcal{E}_0$. Further, under the condition that $\tilde{\mathcal{U}}^e$ is Borel (as in the discussion surrounding Assumption 3), $\mathcal{E}_{\tilde{\tilde{u}}_n^*}$ is Borel since it is a section of Borel $\mathcal{E}$ (e.g., Halmos (1974, page 141, Theorem A) or Tao (2011, Exercise 1.7.18)), and thus $\mathcal{U}_u^e = \mathcal{E}_0$ is Borel.